\documentclass[a4paper]{jpconf}
\usepackage{amsmath}
\usepackage{graphicx}
\pdfoutput=1
\begin{document}
\title{Time dispersion in quantum mechanics}

\author{John Ashmead}

\address{Visiting Scholar, University of Pennsylvania, USA}

\ead{jashmead@seas.upenn.edu}

\begin{abstract}
In quantum mechanics the time dimension is treated as a parameter,
while the three space dimensions are treated as observables. This
assumption is both untested and inconsistent with relativity. From
dimensional analysis, we expect quantum effects along the time axis
to be of order an attosecond. Such effects are not ruled out by current
experiments. But they are large enough to be detected with current
technology, if sufficiently specific predictions can be made. To supply
such we use path integrals. The only change required is to generalize
the usual three dimensional paths to four. We predict a large variety
of testable effects. The principal effects are additional dispersion
in time and full equivalence of the time/energy uncertainty principle
to the space/momentum one. Additional effects include interference,
diffraction, and entanglement in time. The usual ultraviolet divergences
do not appear: they are suppressed by a combination of dispersion
in time and entanglement in time. The approach here has no free parameters;
it is therefore falsifiable. As it treats time and space with complete
symmetry and does not suffer from the ultraviolet divergences, it
may provide a useful starting point for attacks on quantum gravity.
\end{abstract}

\begin{quotation}
``Wheeler's often unconventional vision of nature was grounded in
reality through the principle of radical conservatism, which he acquired
from Niels Bohr: Be conservative by sticking to well-established physical
principles, but probe them by exposing their most radical conclusions.''
--  Kip Thorne \cite{Thorne:2009wa}.

``You can have as much junk in the guess as you like, provided that
the consequences can be compared with experiment.'' -- Richard P. Feynman \cite{Feynman:1965jb} 
\end{quotation}

\section{Introduction\label{sec:intro}}

\subsection{Should the wave function extend in time as it does in space?\label{subsec:intro-idea}}

In relativity, time and space enter on a basis of formal equivalence.
In special relativity, the time and space coordinates rotate into
each other under Lorentz transformations. In general relativity, the
time and the radial coordinate change places at the Schwarzschild
radius (for instance in Adler \cite{Adler:1965fv}). In wormholes and other
exotic solutions to general relativity, time can even curve back on
itself as in G\"{o}del or Thorne \cite{Godel:1949lx,Thorne:1994pl}.

But in quantum mechanics ``time is a parameter not an operator''
(Hilgevoord \cite{Hilgevoord:1996bh,Hilgevoord:1998qu}). This is
clear in the Schr\"{o}dinger equation:

\begin{equation}
\imath\frac{d}{d\tau}\psi_{\tau}\left(\vec{x}\right)=\hat{H}\psi_{\tau}\left(\vec{x}\right)
\end{equation}
Here the wave function is indexed by time: if we know the wave function
at time $\tau$ we can use this equation to compute the wave function
at time $\tau+\epsilon$. The wave function has in general non-zero
dispersion in space, but is always treated as having zero dispersion
in time. This would appear to be inconsistent with special relativity.

\begin{figure}[h]
\begin{centering}
\includegraphics[width=10cm]{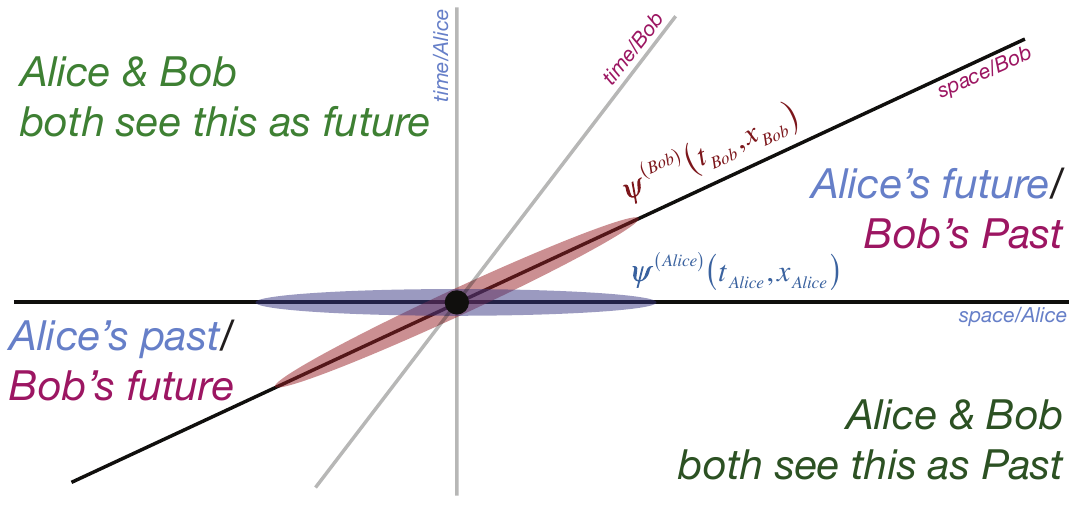} 
\par\end{centering}
\caption{Wave functions for Alice and Bob\label{fig:wf-alice-bob-2}}
\end{figure}

Consider Alice in her laboratory, her co-worker Bob jetting around
like a fusion powered mosquito. Both are studying the same system
but with respective wave functions:

\begin{equation}
\begin{array}{c}
{\psi^{\left({Alice}\right)}}\left({{t_{Alice}},{x_{Alice}}}\right)\hfill\\
{\psi^{\left({Bob}\right)}}\left({{t_{Bob}},{x_{Bob}}}\right)\hfill
\end{array}
\end{equation}

Alice and Bob center their respective wave functions on the particle:

\begin{equation}
\begin{array}{c}
\left\langle {t_{Alice}}\right\rangle =\left\langle {t_{Bob}}\right\rangle =0\hfill\\
\left\langle {x_{Alice}}\right\rangle =\left\langle {x_{Bob}}\right\rangle =0\hfill
\end{array}
\end{equation}

These are distinct wave functions but give the same predictions for
all observables. And do so to an extremely high degree of reliability.

But at Alice\textsc{'}s time zero, Bob\textsc{'}s wave function extends
into her past and future. And at Bob\textsc{'}s time zero her wave
function extends into his past and future.

There are at least two problems here.

One is that in quantum mechanics there is a strict ``plane
of the present''. The quantum mechanical wave function is non-localized
in space but is strictly localized in time. What if Alice decided
to work with Bob\textsc{'}s wave function, rather than her own? She
will get by hypothesis all the same predictions, but will be using
a wave function that from her point of view slops into past and future.

The other is that from the point of view of special relativity, there
should not be a strict ``plane of the present'' in the first place.
We should be able to rotate between the four dimensional references
frames of Alice and Bob as easily as we rotate between references
frames for the three space dimensions.

What happens if we shift to four dimensional wave functions?

\begin{equation}
\psi\left(\vec{x}\right)\to\psi\left(t,\vec{x}\right)
\end{equation}

Assume the coordinate systems for Alice and Bob are related by a Lorentz
transformation $\Lambda$:

\begin{equation}
{x^{\left({Bob}\right)}}=\Lambda{x^{\left({Alice}\right)}}
\end{equation}

Then their wave functions can be related by a Lorentz transformation
of their coordinates: 
\begin{equation}
{\psi^{\left({Bob}\right)}}\left({x^{\left({Bob}\right)}}\right)={\psi^{\left({Alice}\right)}}\left({\Lambda{x^{\left({Alice}\right)}}}\right)
\end{equation}
and matters are much more straightforward.

We make this our basic hypothesis: \textit{the quantum mechanical
wave function should be extended in the time direction on the same
basis as it is extended along the three space dimensions}.

We are playing a ``game of if'' here: we will push the idea as hard
as we can and see what breaks. We are not going to argue that this
is or is not true. We are going to look for experimental tests and
then let the experimentalists decide the question.

There are two principal effects: 
\begin{enumerate}
\item Dispersion in time appears on same basis as dispersion in space. Physical
wave functions are always a bit spread out in space; they will now
also be a bit spread out in time. 
\item The uncertainty principle for time/energy is treated on same basis
as the uncertainty principle for space/momentum. If a particle\textsc{'}s
position in time is well-defined, its energy will highly uncertain
and vice versa. 
\end{enumerate}

\subsubsection{Dispersion in time\label{subsec:intro-idea-dispersion-in-time}}

If the wave functions normally have an extension in time then every
time-specific measurement should show additional dispersion in time.

Suppose we are measuring the time-of-arrival of a particle at a detector.
Define the average time-of-arrival as:

\begin{equation}
\left\langle {\tau^{\left({TOA}\right)}}\right\rangle \equiv\int\limits _{-\infty}^{\infty}{d\tau}\tau{p^{\left({TOA}\right)}}\left(\tau\right)
\end{equation}
with an associated uncertainty:

\begin{equation}
\left\langle {\Delta{\tau^{\left({TOA}\right)}}}\right\rangle \equiv\sqrt{\int\limits _{-\infty}^{\infty}{d\tau}{\tau^{2}}{p^{\left({TOA}\right)}}\left(\tau\right)-{{\left\langle {\tau^{\left({TOA}\right)}}\right\rangle }^{2}}}
\end{equation}

The probability distribution for the particle will normally be spread
out in space, so its arrival times will also be spread out, depending
on the velocity of the particle and its dispersion in space.

But if it also has a dispersion in time, then part of the wave function
will reach the detector – thanks to the fuzziness in time – a
bit sooner and also a bit later than otherwise expected. There will
be an additional dispersion in the time-of-arrival due to the dispersion
in time.

As we will see (subsection \ref{subsec:free-toa}), at non-relativistic
speeds, the dispersion in the time-of-arrival is dominated by the
dispersion in space, so this effect may be hard to pick out. At relativistic
speeds, the contributions of the space and time dispersions can be
comparable.

\subsubsection{Uncertainty principle for time and energy\label{subsec:intro-idea-uncertainty-principle}}

Of particular significance for this work are differences in the treatment
of the uncertainty principle for time/energy as opposed to that for
space/momentum.

In the early days of quantum mechanics, these were treated on same
basis. See for instance the discussions between Bohr and Einstein
of the famous clock-in-a-box experiment \cite{Schilpp:1949oz} or
the comments of Heisenberg in \cite{Heisenberg:1930kb}.

In later work this symmetry was lost. As Busch \cite{Busch-2001}
puts it ``\ldots{} different types of time energy uncertainty can
indeed be deduced in specific contexts, but \ldots{} there is no unique
universal relation that could stand on equal footing with the position-momentum
uncertainty relation.'' See also Pauli, Dirac, and Muga \cite{Pauli:1980wd,Dirac:1958ty,Muga:2002ft,Muga:2008vv}.

\textit{There does not appear to be any experimental test of this
or observational evidence for it; it is merely the way the field has
developed.}

That is not to say that there are not uncertainties with respect to
time, but they are side effects of other uncertainties in quantum
mechanics. 
For instance,
if a particle is spread out in space,
moving to the right,
and going towards a detector at a fixed position,
its time-of-arrival will have a dispersion in time.
But this is a side-effect
of the dispersion in space.

Now consider a particle going through a narrow slit in time, for instance
a camera shutter. Its wave function will be clipped in time. If the
wave function is not extended in time, then the wave function will
merely be clipped: the resulting dispersion in time at detector will
be \textsl{reduced}.

But if the wave function is extended in time and the Heisenberg uncertainty
principle applies in time/energy on the same basis as with space/momentum,
then an extremely fast camera shutter will give a small uncertainty in time
at the gate:
\begin{equation}
\Delta t\to0
\end{equation}
causing the uncertainty in energy to become arbitrarily great:
\begin{equation}
\Delta E\geq\frac{1}{{\Delta t}}\Rightarrow\Delta E\to\infty
\end{equation}
which will in turn cause the wave function to be diffracted, to fan
out in time, and the dispersion in time-of-arrival to become arbitrarily
great.

\subsubsection{A necessary hypothesis\label{subsec:intro-idea-hypothesis}}

This question does not appear to have been attacked directly. As noted,
the assumption that the wave function is not extended in time seems
to have crept into the literature of its own, without experimental
test or observational evidence.

To make an experimental test of this question we have to develop predictions
for both branches: 
\begin{enumerate}
\item Assume the wave function is not extended in time. Make predictions
about time-of-arrival and the like. 
\item Assume the wave function is extended in time. Make equivalent predictions. 
\item Compare. 
\end{enumerate}
We have to develop both branches in a way that makes the comparison
straightforward.

Further, to make the results falsifiable we have to develop the extended-in-time
branch in a way that is clearly correct. A null result should show
that the wave function is \textit{not} extended in time.

These objectives drive what follows.

\subsubsection{Literature\label{subsec:intro-idea-literature}}

The literature for special relativity and for quantum mechanics is
vast. Our focus is on the critical intersection of the two. References
of particular interest here include: 
\begin{itemize}
\item Stueckelberg \& Feynman\textsc{'}s original papers: \cite{Stueckelberg:1941-2,Stueckelberg:1941la,Feynman:1948,Feynman:1949sp,Feynman:1949uy,Feynman:1950rj}. 
\item Reviews of the role of time, Schulman, Zeh, Muga, Callender: \cite{Schulman:1997fd,Zeh:2001xb,Muga:2002ft,Muga:2008vv,Callender:2017rt}. 
\item Block universe picture: Parmenides, Barbour, Price: \cite{Kirk:1983vn,Price:1996mr,Barbour:2000hl}. 
\item Cramer\textsc{'}s transactional interpretation: \cite{Cramer:1986aq,Cramer:1988wh,Kastner:2013fk,Cramer:2016eu}. 
\item The time symmetric quantum mechanics of Aharonov and Reznik: \cite{Aharonov:2005fg,Reznik:1995hj}. 
\item The relativistic dynamics program of Horwitz, Piron, Land, Collins,
and others: \cite{Horwitz:1973ys,Fanchi:1978aa,Fanchi:1993aa,Fanchi:1993ab,Land:1996aj,Horwitz:2005ix,Fanchi:2011aa,Horwitz:2015jk}.
This program is a natural outgrowth of Stueckelberg \& Feynman\textsc{'}s
work.
\end{itemize}

%TQM is RD weaponized for falsifiability

Our approach here may be understood as falling within the relativistic
dynamics framework, but with the emphasis placed on the ``coordinate
time'' rather than the ``evolution parameter'' aspects of that
program.  We will discuss this in detail once an appropriate foundation has been laid.

\subsection{Order of magnitude estimate\label{subsec:estimated-scale}}

Has this hypothesis has already been falsified?
Quantum mechanics has been tested with extraordinary precision. Should
associated effects have been seen already, even if not looked for?

Consider the atomic scale given by the Bohr radius $5.3\ 10^{-11}m$.
We take this as an estimate of the uncertainty in space.

We assume the maximum symmetry possible between time and space. We
therefore infer that the uncertainty in time should be of order the
uncertainty in space (in units where $c=1$).

Dividing the Bohr radius by the speed of light we get the Bohr radius
in time $a_{0}=.177\ 10^{-18}s$, or less than an attosecond. $.177as$
is therefore our starting estimate of the uncertainty in time.

Therefore from strictly dimensional and symmetry arguments, the effects
will be small, of order attoseconds. This is sufficient to explain
why such effects have not been seen.

At the same time, the time scales we can look at experimentally are
now getting down to the attosecond range. A recent paper by Ossiander
et al \cite{Ossiander:2016fp} reports results at the sub-attosecond
level.

Therefore if we can provide the experimentalists with a sufficiently
well-defined target, the hypothesis should be falsifiable in practice.

\subsection{Plan of attack\label{subsec:intro-plan}}
\begin{quotation}
\textit{Look, I don't care what your theory of time is. Just give
me something I can prove wrong.} -- experimentalist at the 2009 Feynman
Festival in Olomouc 
\end{quotation}

\subsubsection{Primary objective is falsifiability}

It is not enough to extend quantum mechanics to include time. It is
necessary to do so in a way that can be proved wrong. The approach
has to be so strongly and clearly constrained that if it is proved
wrong, the whole project of extending quantum mechanics to include
time is falsified.

Our requirements are therefore that we have: 
\begin{enumerate}
\item the most complete possible equivalence in the treatment of time and
space -- manifest covariance at every point at a minimum, 
\item consistency with existing experimental and observation results, 
\item and consistency between the single particle and multiple particle
domains. 
\end{enumerate}
These requirements leave us with no free parameters. And having no
free parameters means in turn that our hypothesis is falsifiable in
principle.

To get to falsifiable in practice, we will look for the simplest cases
that make a direct comparison possible. 

We will also look at points
of principle that need to be addressed, to ensure that the approach is not ruled out by, say, violations of unitarity.

We will use the acronym SQM for standard quantum mechanics. We will
use the acronym TQM for temporal quantum mechanics. By TQM we mean
SQM with time treated on the system basis as space: time just as much
an observable as the three space dimensions.

We do \textit{not} mean by ``temporal quantum mechanics'' that time
itself comes in small chunks or quanta! For instance, there has been
speculation that time is granular at the scale of the Planck time:
${t_{Planck}}\equiv\sqrt{\frac{{\hbar G}}{{c^{5}}}}\approx5.39116x{10^{-44}}s=5.39116x{10^{-26}}as$.
Maybe it is, maybe it isn't. But as this is 25 orders of magnitude
smaller than the times we are considering here, it is reasonable for
us to take time as continuous. And since space is treated by SQM as
continuous, and since the defining assumption of TQM is the maximum
symmetry between time and space, we are required to take time
in TQM continuous.

\subsubsection{Extruding quantum mechanics in time}

If you have worked with a CAD/CAM or 3D drawing program you are familiar
with the ``extrude'' operation, where you take a circle or square
and extrude it into the third dimension\footnote{See Abbott's classic Flatland \cite{Abbott:1884fn} for a delightful
example.}. That is all we are going to do here. We are going to use
manifest covariance as the rule for going from three dimensions to
four -- so that ${s^{2}}={t^{2}}-{x^{2}}-{y^{2}}-{z^{2}}$ is the
invariant rather than ${r^{2}}={x^{2}}+{y^{2}}+{z^{2}}$ -- but the
principle is the same. 

The advantage is that with this approach we have no free parameters.
The moment we introduce a free parameter we put falsifiability at
risk since we are thereby creating a ``fudge factor''. And any fudge
factor might let us say ``well perhaps no effect was seen at scale
$x$, but at smaller scale $y$, then we shall see!''. We want our
experimentalists to be able to say, ``not seen, therefore not there!''

It might be that we should have done the extrude at a slight angle,
extruding our circle into, say, an oblate or prolate spheroid rather
than a perfect one. But this does not seriously impact falsifiability
-- such an oblate or prolate spheroid should still have the same
overall scale of a Bohr radius in time, it will still give our experimentalist
a fixed target. (We discuss possible meanings of ``a slight angle''
in subsection \ref{subsec:lsa-frame} and again in the discussion).

We will use path integrals as our defining formalism. With the path
integral approach we can extrude the paths from three to four dimensions
in a straightforward way, while leaving the rest of the machinery
essentially untouched. With other formalisms, it is less clear what
``extrude'' means. But once we have the meaning of extrude worked out for path
integrals, we can, as it were, rotate our path integral formalism
into other formalisms, seeing what ``extruding quantum mechanics
in time'' looks like as a Schr\"{o}dinger equation or in quantum field
theory. 

Our plan:

\subsubsection{Single particle case}

In the single particle case we will: 
\begin{enumerate}
\item Generalize path integrals to include time as an observable. 
\item Derive the corresponding Schr\"{o}dinger equation as the short time limit
of the path integral. 
\item Develop the free solutions. We will estimate the initial wave function,
let it evolve in time, and detect it. We will compute the dispersions
of time-of-arrival measurements in SQM and in TQM. In general the
differences are real but small. 
\item Analyze the single and double slit experiments. The single slit in
time experiment provides the decisive test of temporal quantum mechanics.
In SQM, the narrower the slit, the \textit{less} the dispersion in
subsequent time-of-arrival measurements. In TQM, the narrower the
slit, the \textit{greater} the dispersion in subsequent time-of-arrival
measurements. \textsl{In principle, the difference may be made arbitrarily
great.} 
\end{enumerate}

\subsubsection{Multiple particle case}

We will then extend TQM to include the multiple particle case, i.e.
field theory. We will show, using a toy model, that: 
\begin{enumerate}
\item We can extend the usual path integral approach to include time as
an observable. The basis functions in Fock space extend in a natural
way from three to four dimensions, the Lagrangian is unchanged, and
the usual Feynman diagram expansions appear. 
\item For each Feynman diagram in SQM we can compute the TQM equivalent.
Therefore any problem that can be solved using Feynman diagrams in
SQM can be solved in TQM. 
\item \textit{The usual ultraviolet divergences do not appear} (the combination of
dispersion in time and entanglement in time contain them). 
\item And that there are a large number of additional experimental effects
to be seen, including: 
\begin{enumerate}
\item wave functions anti-symmetric in time, 
\item correlations, entanglement, and interference in time, 
\item and forces of anticipation and regret. 
\end{enumerate}
\end{enumerate}

\subsubsection{Overall conclusions}

With this done, we will argue in the discussion: 
\begin{enumerate}
\item that TQM is not ruled out a priori. 
\item that TQM is falsifiable. And given experimental work like Ossiander\textsc{'}s,
probably with current technology. 
\item that TQM is a source of interesting experiments. Every foundational
experiment in SQM has an ``in time'' variant. 
\item that TQM is a potential starting point for attacks on the quantum
gravity problem, since TQM is manifestly covariant and untroubled
by the ultraviolet divergences. 
\item that as TQM is a straight-forward extrapolation of quantum mechanics
and special relativity, experiments that falsify TQM are likely to
require modification of our understanding of either quantum mechanics
or special relativity or both. Something will have to break. 
%(We suspect our Olomouc experimentalist will not much care which, so long as he gets to do the breaking.) 
\end{enumerate}

\subsubsection{Limits of the current investigation}

We do no more here than is required to establish that TQM is well-defined, self-consistent, and falsifiable.

It will be clear from the development that the basic idea could be extended in a number of directions;
we review some of these in the concluding discussion.

The conventions used in this work are spelled out in \ref{sec:app-conv}.

\section{Path integrals\label{sec:paths}}

\subsection{Overview}

To extend quantum mechanics to include time we will take as our starting
point Feynman\textsc{'}s path integral approach to quantum mechanics
\cite{Feynman:2010bt,Schulman:1981um,Rivers:1987ma,Swanson:1992ju,Khandekar:1993ci,Kashiwa:1997xt,Grosche:1998uj,Zinn-Justin:2005nx,Kleinert:2009hw,Zee:2010oy}.

With the path integral approach, the only change we will need to make
is to generalize the paths from varying in three dimensions to varying
in four.

To make clear what this means, consider the case of Alice walking
her dog, say from her front door to Bob\textsc{'}s.

Alice will take the shortest (classical) path from door to door.

But her dog will dart from side to side, now investigating a mailbox
to the left, now checking out a lamppost to the right. In fact, as
a quantum dog he will investigate all such paths simultaneously. While
he will start at the same time and place as Alice, and finish at the
same time and place as Alice, in between he will travel simultaneously
along all possible paths.

But -- in SQM -- only along paths in space. At each tick of Alice\textsc{'}s
digital watch, her dog will be found off to the left or right, jumping
up or digging down, further along the path to Bob\textsc{'}s, or holding
back for an important investigation.

But in TQM, the quantum dog can -- and therefore will -- advance
into the future and drop back into the past. So that tick by tick
of Alice\textsc{'}s watch, her dog\textsc{'}s paths will have to tracked
in four dimensions rather than three.

This is harder to visualize, being out of our normal experience. So
we develop the analysis a bit formally, letting math take the place
of an as yet undeveloped intuition.

Path integrals, as the name suggests, are done by summing over all
paths from starting point to finish, weighting each path by the integral
of the Lagrangian (the action) along it:

\begin{equation}
{\psi_{T}}\left({x_{T}}\right)=\int{\mathcal{D}{x_{\tau}}\exp\left({\imath\int\limits _{0}^{T}{d\tau L\left[{x,\dot{x}}\right]}}\right){\psi_{0}}\left({x_{0}}\right)}
\end{equation}

Piece by piece: 
\begin{enumerate}
\item ${\psi_{0}}\left({x_{0}}\right)$ is the initial wave function. We
will be breaking these down into sums over Gaussian test functions
using Morlet wavelet analysis. 
\item $\tau$ is the clock time as given by Alice\textsc{'}s digital watch.
We will break up the paths into the bits from one clock tick to the
next. 
\item $\mathcal{D}{x_{\tau}}$ represents the paths. Each path is defined
by its coordinates at each clock tick. In SQM, these are the values
of $x_{\tau},y_{\tau},z_{\tau}$ at each clock tick. In TQM these
are values of $t_{\tau},x_{\tau},y_{\tau},z_{\tau}$ at each clock
tick. 
\item $L\left[{x,\dot{x}}\right]$ is a suitable Lagrangian. We will be
using one that works equally well for both SQM and TQM. 
\item ${\imath\int\limits _{0}^{T}{d\tau L\left[{x,\dot{x}}\right]}}$ is
the action, the integral over the Lagrangian taken path by path. 
\item And ${\psi_{T}}\left({x_{T}}\right)$ is the final wave function,
the amplitude for the dog to arrive at Bob\textsc{'}s door step. 
\end{enumerate}
We will look at: 
\begin{enumerate}
\item What do we mean by $\tau$ the clock time? 
\item What do we mean by the coordinate time $t$ in $t,x,y,z$? 
\item How do we define the initial wave function in a way that does not
potentially bias the outcome? 
\item What Lagrangian shall we use? 
\item How do we get the sums to converge? 
\item Having gotten the sums to converge, how do we normalize them? 
\item And what do all the pieces look like when we put them back together? 
\end{enumerate}

\subsection{Laboratory time\label{subsec:paths-clock}}

We consider the action, the integral
of the Lagrangian over time:
\begin{equation}
\int\limits _{0}^{T}{d\tau\mathcal{L}\left[{x,\dot{x}}\right]}
\end{equation}

In classical mechanics, we are free to take the parameter $\tau$ as any monotonically increasing
variable. We will get the same classical equations of motion in any
case.

A typical choice is to select $\tau$ as the proper time of the particle
in question. However this makes it difficult to extend the work to
the multiple particle case, where there are many particles and therefore
many proper times in play.

Here we choose to use the time as shown on a laboratory clock. We
take the term laboratory time from Busch \cite{Busch-2001}. We will
use the terms clock time and laboratory time interchangeably.

It is useful to visualize this clock as a metronome, breaking up the
clock time into a series of $N$ ticks each of length $\epsilon$.
If $\tau=0$ at the source, and $\tau=T$ at the detector, we have:
\begin{equation}
\epsilon\equiv\frac{T}{N}
\end{equation}

We will take the limit as $N\to\infty$ as the final step in the calculation.

\subsection{Coordinate time\label{subsec:Coordinate-time}}

We visualize a four dimensional coordinate system coordinates $t,x,y,z$.
We will refer to $t$ as coordinate time by analogy with the three
coordinate space dimensions: coordinate $x$, coordinate $y$, and
coordinate $z$.

Paths are defined with reference to this coordinate system. If the
time by Alice\textsc{'}s watch is $\tau$, then each path $\pi$ will
have a location at $\tau$ given by:
\begin{equation}
{\pi_{\tau}}\left({t,x,y,z}\right)
\end{equation}

It may help to think of the coordinates as laid out on a piece of
four dimensional graph paper. At a specific clock tick $n$, a specific
path $\pi$ will be represented by a dot on a specific vertex on the
four dimensional graph paper. If we want to see the progress of the
path with respect to clock time, we can flip the series of pieces
of graph paper like one of those old time flip movies.

If our graph paper has $M$ grid lines in each direction, the number
of vertices on a page is $M^{4}$, and the number of paths total is
$M^{4N}$. Each different sequence of grid points counts as a distinct
path.

The path integral measure ${\cal D}x$ is usually defined by assigning
a weight of one to each distinct path, and then taking the limit as
the spacing goes to zero.

Since coordinate time is on the same footing as the three space coordinates
it has a corresponding energy operator:
\begin{equation}
{p_{x}}\equiv-\imath\frac{\partial}{{\partial x}}\Rightarrow E\equiv\imath\frac{\partial}{{\partial t}}
\end{equation}

We will refer to this as coordinate energy. It is not positive definite
or bounded from below. Since $p_{x}$ can be positive or negative,
by our controlling requirement of covariance $E$ can be positive
or negative.

We discuss the relationship between clock time and coordinate time in \ref{sec:app-time}.

\subsection{Initial wave function\label{subsec:paths-wf}}

We need a starting set of wave functions $\psi_{0}$ at clock time
$\tau=0$. We will need wave functions that extend in both coordinate
time and space. The usual choices would be $\delta$ functions or
plane waves.

In coordinate time these might be:
\begin{equation}
\delta(t-t_{0})
\end{equation}
\begin{equation}
e^{-\imath E\left(t-t_{0}\right)}
\end{equation}
or in space:
\begin{equation}
\delta(x-x_{0})
\end{equation}
\begin{equation}
e^{\imath p\left(x-x_{0}\right)}
\end{equation}
But neither $\delta$ functions nor plane waves are physical. Their
use creates a risk of artifacts.

More physical would be Gaussian test functions, for instance in coordinate
time:
\begin{equation}
\varphi\left(t\right)\equiv\sqrt[4]{\frac{1}{\pi\sigma_{t}^{2}}}e^{-\imath E\left(t-t_{0}\right)-\frac{\left(t-t_{0}\right)^{2}}{2\sigma_{t}^{2}}}
\end{equation}
or in space:
\begin{equation}
\varphi\left(x\right)\equiv\sqrt[4]{\frac{1}{\pi\sigma_{x}^{2}}}e^{\imath p\left(x-x_{0}\right)-\frac{\left(x-x_{0}\right)^{2}}{2\sigma_{x}^{2}}}
\end{equation}
But while Gaussian test functions are physically reasonable they are
not completely general.

We can achieve both generality and physical reasonableness by using
a basis of Morlet wavelets \cite{Morlet:1982mw,Chui:1992be,Meyer:1992hl,Kaiser:1994ph,Berg:1999mo,Bratteli:2002ar,Addison:2002jk,Visser:2003ga,Antoine:2004rm,Ashmead:2012kx}.

\begin{figure}[h]
\begin{centering}
\includegraphics[width=10cm]{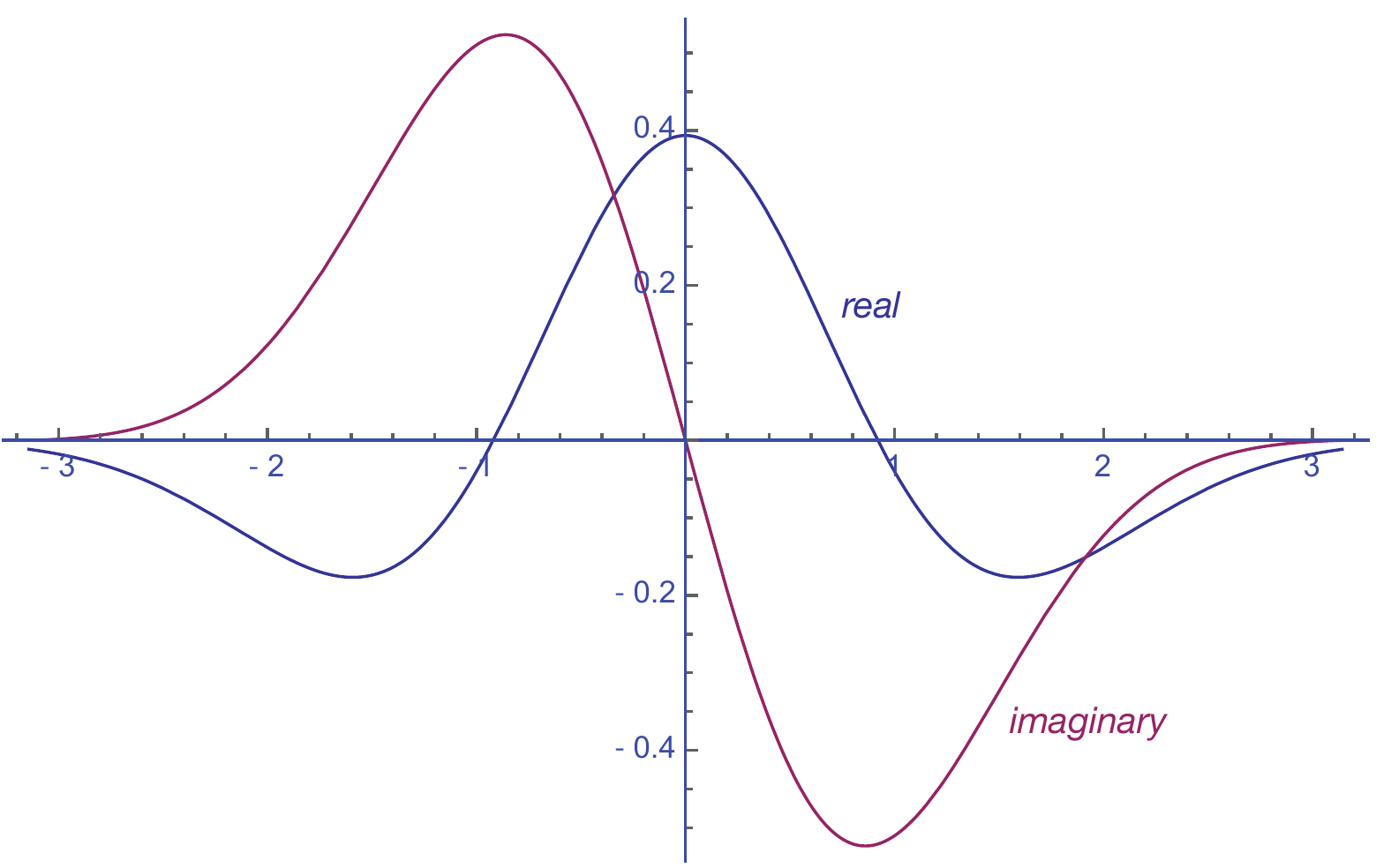} 
\par\end{centering}
\caption{Mother Morlet wavelet\label{fig:Morlet-wavelet}}
\end{figure}

Morlet wavelets are derived by starting with a ``mother'' wavelet:
\begin{equation}
\phi^{\left(mother\right)}\left(t\right)\equiv\left(e^{-\imath t}-\frac{1}{\sqrt{e}}\right)e^{\left(-\frac{t^{2}}{2}\right)}
\end{equation}
and scaling and displacing it with the replacement $t\to\frac{t-l}{s}$:
\begin{equation}
\phi_{sl}\left(t\right)=\frac{1}{\sqrt{\left|s\right|}}\left(e^{-\imath\left(\frac{t-l}{s}\right)}-\frac{1}{\sqrt{e}}\right)e^{-\frac{1}{2}\left(\frac{t-l}{s}\right)^{2}}
\end{equation}
Any normalizable function $f$ can be broken up into wavelet components
$\hat{f}$ using:
\begin{equation}
\hat{f}_{sl}=\int\limits _{-\infty}^{\infty}dt\phi_{sl}^{*}\left(t\right)f\left(t\right)
\end{equation}
And then recovered using the inverse Morlet wavelet transform:
\begin{equation}
f\left(t\right)=\frac{1}{C}\int\limits _{-\infty}^{\infty}\frac{dsdl}{s^{2}}\phi_{sl}\left(t\right)\hat{f}_{sl}
\end{equation}

The value of $C$ is worked out in \cite{Ashmead:2012kx}.

Therefore we can write any physically reasonable wave function $f$ in
time in terms of Morlet wavelets.

%\begin{equation}
%f\left(t\right)=\frac{1}{C}\int{\frac{{dsdl}}{{s^{2}}}{{\hat{f}}_{sl}}}{\phi_{sl}}\left(t\right)
%\end{equation}

We may include space by using products of Morlet wavelets:
\begin{equation}
\begin{array}{rcl}
  {{\hat f}_{{s_t}{l_t}{s_x}{l_x}}} &=& \int {dtdx\phi _{{s_t}{l_t}}^*\left( t \right)} \phi _{{s_x}{l_x}}^*\left( x \right)f\left( {t,x} \right) \hfill \\
  f\left( {t,x} \right) &=& \frac{1}{{{C^2}}}\int {\frac{{d{s_t}d{l_t}}}{{s_t^2}}\frac{{d{s_x}d{l_x}}}{{s_x^2}}{\phi _{{s_t}{l_t}}}\left( t \right)} {\phi _{{s_x}{l_x}}}\left( x \right){{\hat f}_{{s_t}{l_t}{s_x}{l_x}}} \hfill \\ 
\end{array} 
\end{equation}

Clearly it would be cumbersome to track four dimensional Morlet wavelets
at every step.

Fortunately we do not need to perform the Morlet wavelet
analyses; we merely need the ability to do so. As each Morlet wavelet
may be written as a sum of a pair of Gaussians, Morlet wavelet analysis
lets us write any physically reasonable wave function as a sum over
Gaussians. Provided we are dealing only with linear operations --
the case throughout here -- we can work directly with Gaussian test
functions. By Morlet wavelet analysis the results will then be valid
for any physically reasonable wave functions.

\subsection{Lagrangian\label{subsec:paths-lag}}

To sum over the paths -- to construct the path integral -- we will
need to weight each path by the exponential of the action, where the
action is defined as the integral of the Lagrangian over the laboratory
time:
\begin{equation}
e^{\imath\int\limits _{0}^{T}d\tau L\left(x^{\mu},\dot{x}^{\mu}\right)}
\end{equation}

We require a Lagrangian which: 
\begin{enumerate}
\item Is manifestly covariant, 
\item Produces the correct classical equations of motion, 
\item And gives the correct Schr\"{o}dinger equation. 
\end{enumerate}
We would further prefer a Lagrangian which is the same for both SQM
and TQM. This will let us argue that we are treating SQM and TQM with
the most complete possible equality.

Somewhat surprisingly such a Lagrangian exists. In Goldstein\textsc{'}s
well-known text on classical mechanics \cite{Goldstein:1980ce} we
find:
\begin{equation}
L\left(x^{\mu},\dot{x}^{\mu}\right)=-\frac{1}{2}m\dot{x}^{\mu}\dot{x}_{\mu}-q\dot{x}^{\mu}A_{\mu}\left(x\right)\label{eq:paths-lag-bridge}
\end{equation}

The potentials are not themselves functions of the laboratory time
$\tau$. The mass $m$ is the rest mass of the particle, an invariant.

This Lagrangian is unusual in that it uses four independent variables
(the usual three space coordinates plus a time variable) but still
gives the familiar classical equations of motion (see \ref{sec:app-classical}).

This Lagrangian therefore provides a natural bridge from a three to
a four dimensional picture.

The classical equations of motion are still produced if we add a dimensionless
scale $a$ and an additive constant $b$ to the Lagrangian:
\begin{equation}
-\frac{1}{2}am\dot{x}^{\mu}\dot{x}_{\mu}-aq\dot{x}^{\mu}A_{\mu}\left(x\right)-ab\frac{m}{2}\label{eq:paths-lag-lag}
\end{equation}

The Lagrangian is therefore only determined up to $a$ and $b$. The
requirement that we match the SQM results will fix $a$ and $b$ (subsection
\ref{subsec:lsa-constants}).

\subsection{Convergence\label{subsec:Convergence-of-the-sums}}

How do we get the integrals over the paths to converge without breaking covariance?

We compute the path integral for the kernel by slicing the clock time
into an infinite number of intervals and integrating over each:
\begin{equation}
K_{BA}=\lim_{N\to\infty}C_{N}\int\prod\limits _{n=1}^{N-1}dt_{n}d\vec{x}_{n}e^{\imath\varepsilon\sum\limits _{j=1}^{N}L_{j}}
\end{equation}
with $C_{N}$ an appropriate normalization factor.

Consider the discrete form of the Lagrangian. We use a tilde to mark
the coordinate time part and an overbar to mark the space part:
\begin{equation}
L_{j}\equiv\tilde{L}_{j}^{t}+\bar{L}_{j}^{\vec{x}}+L_{j}^{m}
\end{equation}
\begin{equation}
\tilde{L}_{j}^{t}\equiv-a\frac{m}{2}\left(\frac{t_{j}-t_{j-1}}{\varepsilon}\right)^{2}-qa\frac{t_{j}-t_{j-1}}{\varepsilon}\frac{\Phi\left(x_{j}\right)+\Phi\left(x_{j-1}\right)}{2}
\end{equation}
\begin{equation}
\bar{L}_{j}^{\vec{x}}\equiv a\frac{m}{2}\left(\frac{\vec{x}_{j}-\vec{x}_{j-1}}{\varepsilon}\right)^{2}+qa\frac{\vec{x}_{j}-\vec{x}_{j-1}}{\varepsilon}\cdot\frac{\vec{A}\left(x_{j}\right)+\vec{A}\left(x_{j-1}\right)}{2}
\end{equation}
\begin{equation}
L_{j}^{m}\equiv-ab\frac{m}{2}
\end{equation}
We are using the mid-point rule, averaging the scalar and the vector
potentials over the start and end points of the step, by analogy with
the rule for three dimensions (Schulman \cite{Schulman:1981um}, Grosche and Steiner \cite{Grosche:1998uj}).

Now look at a single step for the free case, vector potential $A_{\mu}$
zero:
\begin{equation}
\int\limits _{-\infty}^{\infty}\int\limits _{-\infty}^{\infty}dt_{j}d\vec{x}e^{-\imath am\frac{\left(t_{j}-t_{j-1}\right)^{2}}{2\varepsilon}+\imath am\frac{\left(\vec{x}_{j}-\vec{x}_{j-1}\right)^{2}}{2\varepsilon}}
\end{equation}
The formal tricks normally used to ensure convergence do not work
here (e.g. Kashiwa or Zinn-Justin \cite{Kashiwa:1997xt,Zinn-Justin:2005nx}).
Perhaps the most popular of these is the use of Wick rotation to shift
to a Euclidean time\label{paths-sums-euclidean}: 
\begin{equation}
\tau\to-\imath\tau
\end{equation}

This causes integrals to converge rapidly going into the future, but
makes the past inaccessible. For instance, factors of $\exp\left({-\imath\omega\tau}\right)$
-- which spring up everywhere in path integrals -- converge going
into the future, but blow up going into the past. If we are to treat
time on the same footing as space -- our central assumption -- then
past and future must be treated as symmetrically as left and right.

Another approach is to add a small convergence factor at a cleverly
chosen spot in the arguments of the exponentials. But if we attach
a convergence factor to $t$ and $x$ separately, we break manifest
covariance. If we attach our convergence factor to both, the fact
that the $t$ and $x$ parts enter with opposite sign means any convergence
factor that works for one will fail for the other. We could try attaching
one to the mass $m$, but this also fails. For instance if $a>0$
and we subtract a small factor of $\imath\delta$ from the mass:
\begin{equation}
m\to m-\imath\delta
\end{equation}
the $t$ integral converges but the $x$ integral diverges.

We recall the kernel has meaning only when applied to a specific physical
wave function. If we break the incoming wave up into Morlet wavelets
and then into Gaussian test functions, we see that each integral converges
by inspection, the factor $e^{-\frac{1}{2}\left(\frac{t-l}{s}\right)^{2}}$
ensures this.

So for physically significant wave functions, there is no problem
in the first place. Effectively we are taking seriously the point
that the path integral kernel is a distribution, only meaningful with
respect to specific wave functions.

\subsection{Normalization\label{subsec:Normalization}}

Now that we have our path integrals converging, we have to normalize
them. If we start from the Schr\"{o}dinger equation, the normalization
is wired in. But in path integrals we are a bit at sea.

We will here deal with the free case, verifying the normalization
is correct in the general case in \ref{sec:app-unitarity}.

The normalization factor for $N$ steps we will call $C_{N}$. The
defining requirement is that, if the initial wave function is normalized
to one, then with the inclusion of $C_{N}$, the final wave function
will be normalized to one as well:
\begin{equation}
\int{d{t_{0}}d{{\vec{x}}_{0}}\left|{{\psi_{0}}{{\left({{t_{0}},{{\vec{x}}_{0}}}\right)}^{2}}}\right|=1}\to\int{d{t_{N}}d{{\vec{x}}_{N}}\left|{{\psi_{N}}{{\left({{t_{N}},{{\vec{x}}_{N}}}\right)}^{2}}}\right|=1}
\end{equation}
If $C_{N}$ depends on the particular $\psi_{0}$, then we have failed.

We now compute the factor of $C_{N}$.

\subsubsection{Normalization in time}

We start with the coordinate time dimension only. Consider a Gaussian
test function centered on an initial position in coordinate time $\bar{t}_{0}$:
\begin{equation}
\tilde{\varphi}_{0}\left(t_{0}\right)\equiv\sqrt[4]{\frac{1}{\pi\sigma_{t}^{2}}}e^{-\imath E_{0}t_{0}-\frac{\left(t_{0}-\bar{t}_{0}\right)^{2}}{2\sigma_{t}^{2}}}\label{eq:wf-time-0}
\end{equation}
We write the kernel for the time part as:
\begin{equation}
\tilde{K}_{\tau}\left(t_{N};t_{0}\right)\sim\int\prod\limits _{j=1}^{N-1}dt_{j}e^{-\imath am\sum\limits _{k=1}^{N}\frac{\left(t_{k}-t_{k-1}\right)^{2}}{2\varepsilon}}
\end{equation}
The wave function after the initial integral over $t_{0}$ is:
\begin{equation}
\tilde{\varphi}_{\varepsilon}\left(t_{1}\right)=\int dt_{0}e^{-\imath\frac{am}{2\varepsilon}\left(t_{1}-t_{0}\right)^{2}}\tilde{\varphi}_{0}\left(t_{0}\right)
\end{equation}
or:
\begin{equation}
{\tilde{\varphi}}_{\varepsilon}\left(t_{1}\right)=\sqrt{\frac{2\pi\varepsilon}{\imath am}}\sqrt[4]{\frac{1}{\pi\sigma_{t}^{2}}}\sqrt{\frac{1}{f_{\varepsilon}^{\left(t\right)}}}e^{-\imath E_{0}t_{1}+\imath\frac{E_{0}^{2}}{2am}\varepsilon-\frac{1}{2\sigma_{t}^{2}f_{\varepsilon}^{\left(t\right)}}\left(t_{1}-\bar{t}_{0}-\frac{E_{0}}{am}\varepsilon\right)^{2}}
\end{equation}
with the dispersion factor $f_{\tau}^{\left(t\right)}\equiv1-\imath \frac{\tau}{am\sigma_{t}^{2}}$.

The normalization requirement is:
\begin{equation}
1=\int dt_{1}\tilde{\varphi}_{\varepsilon}^{*}\left(t_{1}\right)\tilde{\varphi}_{\varepsilon}\left(t_{1}\right)
\end{equation}
The first step normalization is correct if we multiply the kernel
by a factor of $\sqrt{\frac{\imath am}{2\pi\varepsilon}}$. Since
this normalization factor does not depend on the laboratory time the
overall normalization for $N$ infinitesimal kernels is the product
of $N$ of these factors:
\begin{equation}
C_{N}\equiv\sqrt{\frac{\imath am}{2\pi\varepsilon}}^{N}
\end{equation}
Note also that the normalization does not depend on the specifics
of the Gaussian test function (the values of $E_{0}$, $\sigma_{t}^{2}$,
and $\bar{t}_{0}$) so it is valid for an arbitrary sum of Gaussian
test functions as well. And therefore, by Morlet wavelet decomposition,
for an arbitrary wave function.

As noted, the phase is arbitrary. If we were working the other way,
from Schr\"{o}dinger equation to path integral, the phase would be determined
by the Schr\"{o}dinger equation itself. The specific phase choice we are
making here has been chosen to help ensure the four dimensional Schr\"{o}dinger
equation is manifestly covariant, see below. We may think of the phase
choice as a choice of gauge (see \ref{sub:app-gauge}).

Therefore the expression for the free kernel in coordinate time is
(with $t^{\prime\prime}\equiv t_{N}$, $t^{\prime}\equiv t_{0}$):
\begin{equation}
\tilde{K}_{\tau}\left(t^{\prime\prime};t^{\prime}\right)=\int dt_{1}dt_{2}\ldots dt_{N-1}\sqrt{\frac{\imath am}{2\pi\varepsilon}}^{N}e^{-\imath\sum\limits _{j=1}^{N}\left(\frac{am}{2\varepsilon}\left(t_{j}-t_{j-1}\right)^{2}\right)}
\end{equation}
Doing the integrals we get:
\begin{equation}
\tilde{K}_{\tau}\left(t^{\prime\prime};t^{\prime}\right)=\sqrt{\frac{\imath am}{2\pi\tau}}e^{-\imath am\frac{\left(t^{\prime\prime}-t^{\prime}\right)^{2}}{2\tau}}
\end{equation}
and free wave functions in coordinate time:
\begin{equation}
{{\tilde{\varphi}}_{\tau}}\left(t\right)=\sqrt[4]{{\frac{1}{{\pi\sigma_{t}^{2}}}}}\sqrt{\frac{1}{{f_{\tau}^{\left(t\right)}}}}{e^{-\imath{E_{0}}t-\frac{1}{{2\sigma_{t}^{2}f_{\tau}^{\left(t\right)}}}{{\left({t-{{\bar{t}}_{0}}-\frac{{E_{0}}}{am}\tau}\right)}^{2}}+\imath\frac{{E_{0}^{2}}}{{2am}}\tau}}
\end{equation}

\subsubsection{Normalization in space}

We redo the analysis for coordinate time for space. We use the correspondences:
\begin{equation}
t\to x,m\to-m,\bar{t}_{0}\to\bar{x}_{0},E_{0}\to-p_{0},\sigma_{t}^{2}\to\sigma_{x}^{2}
\end{equation}
With these we can write down the equivalent set of results by inspection.
Since we will need the results below, we do this explicitly. We get
the initial Gaussian test function:
\begin{equation}
\bar{\varphi}_{0}\left(x_{0}\right)=\sqrt[4]{\frac{1}{\pi\sigma_{x}^{2}}}e^{\imath p_{0}x_{0}-\frac{\left(x_{0}-\bar{x}_{0}\right)^{2}}{2\sigma_{x}^{2}}}\label{eq:wf-space-0}
\end{equation}
free kernel:
\begin{equation}
\bar{K}_{\tau}\left(x^{\prime\prime};x^{\prime}\right)\sim\int dx_{1}dx_{2}\ldots dx_{N-1}e^{\imath\sum\limits _{j=1}^{N}\frac{am}{2\varepsilon}\left(x_{j}-x_{j-1}\right)^{2}}
\end{equation}
and normalized kernel:
\begin{equation}
\bar{K}_{\tau}\left(x^{\prime\prime};x^{\prime}\right)=\sqrt{-\frac{\imath am}{2\pi\tau}}e^{\imath am\frac{\left(x^{\prime\prime}-x^{\prime}\right)^{2}}{2\tau}}\label{eq:free-kernel-3D-space}
\end{equation}
The kernel matches the usual (non-relativistic) kernel \cite{Feynman:1965bh,Schulman:1981um}
if $a=1$.

The wave function is:
\begin{equation}
\bar{\varphi}_{\tau}\left(x\right)=\sqrt[4]{\frac{1}{\pi\sigma_{x}^{2}}}\sqrt{\frac{1}{f_{\tau}^{\left(x\right)}}}e^{\imath p_{0}x-\frac{1}{2\sigma_{x}^{2}f_{\tau}^{\left(x\right)}}\left(x-\bar{x}_{0}-\frac{p_{0}}{am}\tau\right)^{2}-\imath\frac{p_{0}^{2}}{2am}\tau}\label{eq:wf-space-tau-1}
\end{equation}
with the definition of the dispersion factor $f_{\tau}^{\left(x\right)}=1+\imath \frac{\tau}{am\sigma_{x}^{2}}$
parallel to that for coordinate time (but with opposite sign for the
imaginary part).

\subsubsection{Normalization in time and space}

The full kernel is the product of the coordinate time kernel, the
three space kernels, and the constant term $e^{-\imath\frac{abm}{2}\tau}$.
We understand $x$ to refer to coordinate time and all three space
dimensions:
\begin{equation}
K_{\tau}\left(x^{\prime\prime};x^{\prime}\right)=-\imath\frac{a^{2}m^{2}}{4\pi^{2}\tau^{2}}e^{-\frac{\imath am}{2\tau}\left(x^{\prime\prime}-x^{\prime}\right)^{2}-\imath\frac{abm}{2}\tau}\label{eq:free-kernel}
\end{equation}
We have done the analysis only for Gaussian test functions, but by
Morlet wavelet decomposition it is completely general.

\subsection{Formal expression\label{subsec:paths-full}}

We now have the full path integral\label{final-path-integral}:
\begin{equation}
K_{\tau}\left(x^{\prime\prime};x^{\prime}\right)=\lim\limits _{N\to\infty}\int\mathcal{D}xe^{\imath\sum\limits _{j=1}^{N}L_{j}\epsilon}\label{eq:path-integral}
\end{equation}
with the measure:
\begin{equation}
\mathcal{D}x\equiv\left(-\imath a^{2}\frac{m^{2}}{4\pi^{2}\varepsilon^{2}}\right)^{N}\prod\limits _{n=1}^{N-1}d^{4}x_{n}
\end{equation}
and the discretized Lagrangian at each step:
\begin{equation}
L_{j}\equiv-am\frac{\left(x_{j}-x_{j-1}\right)^{2}}{2\varepsilon^{2}}-aq\frac{x_{j}-x_{j-1}}{\epsilon}\frac{A\left(x_{j}\right)+A\left(x_{j-1}\right)}{2}-ab\frac{m}{2}\label{eq:lagrangian}
\end{equation}

We will show that $a=b=1$ in the next section.

\section{Schr\"{o}dinger equation\label{sec:seqn}}

\subsection{Overview}

The path integral and Schr\"{o}dinger equation views are complementary.
We need both to fully understand either.

We derive the Schr\"{o}dinger equation from the path integral by taking
the short time limit of the path integral form.

By comparing the result to the Klein-Gordon equation -- and making
a reasonable assumption about the long time evolution of the wave
functions -- we are able to fix the additive and scale constants
in the Lagrangian.

The resulting equation looks like the Klein-Gordon equation over short
times but shows some drift over longer times. We use some heuristic
arguments to estimate the scale of the long term drift as of order
picoseconds, a million times longer than the attosecond scale of the
time dispersions we are primarily concerned with here. We will therefore
be able to largely ignore this drift.

With $a,b$ defined, we look at a further problem. We have done the
derivation of path integral and Schr\"{o}dinger equation forms from Alice's
perspective. But what of Bob, jetting around like a fusion powered
mosquito?

We resolve this conflict by arguing that we can find a natural rest
frame that both can use. Starting with an argument of Weinberg\textsc{'}s,
we argue we can associate an energy-momentum tensor with spacetime.
This means we can associate a local rest frame with spacetime. And
this local rest frame can provide the neutral and agreed defining
frame for TQM.

This will complete the formal development of TQM.

Before turning to applications, we will then look at the relationship of TQM to the relativistic dynamics program.
We will argue that TQM may be understood as a member of that program
aggressively specialized to achieve falsifiability.

\subsection{Derivation of the Schr\"{o}dinger equation}

Normally the path integral expression is derived from the Schr\"{o}dinger
equation. But because for us the path integral provides the defining
formulation we need to run the analysis in the ``wrong'' direction.

Our starting point is a derivation of the path integral from the Schr\"{o}dinger
equation by Schulman \cite{Schulman:1981um}. We run his derivation
in reverse and with one extra dimension\footnote{This derivation is done for the free case in Fanchi \cite{Fanchi:1993ab}. }.

We start with the discrete form of the path integral. We consider
a single step of length $\epsilon$, taking $\epsilon\to0$ at the
end. Because of this, only terms first order in $\epsilon$ are needed.

Following Schulman, we define the coordinate difference:
\begin{equation}
\xi\equiv x_{j}-x_{j+1}
\end{equation}
We rewrite the functions of $x_{j}$ as functions of $\xi$ and $x_{j+1}$.
We expand the vector potential:
\begin{equation}
A_{\nu}\left(x_{j}\right)=A_{\nu}\left(x_{j+1}\right)+\left(\xi^{\mu}\partial_{\mu}\right)A_{\nu}\left(x_{j+1}\right)+\ldots
\end{equation}
and the wave function:
\begin{equation}
\psi_{\tau}\left(x_{j}\right)=\psi_{\tau}\left(x_{j+1}\right)+\left(\xi^{\mu}\partial_{\mu}\right)\psi_{\tau}\left(x_{j+1}\right)+\frac{1}{2}\xi^{\mu}\xi^{\nu}\partial_{\mu}\partial_{\nu}\psi_{\tau}\left(x_{j+1}\right)+\ldots
\end{equation}
giving:
\begin{equation}
\begin{array}{c}
\psi_{\tau+\varepsilon}\left(x_{j+1}\right)=\sqrt{\frac{\imath am}{2\pi\varepsilon}}\sqrt{-\frac{\imath am}{2\pi\varepsilon}}^{3}\int d^{4}\xi e^{-\frac{\imath am\xi^{2}}{2\varepsilon}-\imath ab\frac{m}{2}\varepsilon}\\
\times e^{\imath aq\xi^{\nu}\left(A_{\nu}\left(x_{j+1}\right)+\frac{1}{2}\left(\xi^{\mu}\partial_{\mu}\right)A_{\nu}\left(x_{j+1}\right)+\ldots\right)}\\
\times\left(\psi_{\tau}\left(x_{j+1}\right)+\left(\xi^{\mu}\partial_{\mu}\right)\psi_{\tau}\left(x_{j+1}\right)+\frac{1}{2}\xi^{\mu}\xi^{\nu}\partial_{\mu}\partial_{\nu}\psi_{\tau}\left(x_{j+1}\right)+\ldots\right)
\end{array}
\end{equation}
We now expand in powers of $\xi\sim\sqrt{\epsilon}$. We do not need
more than the second power:
\begin{equation}
\begin{array}{c}
\psi_{\tau+\varepsilon}\left(x_{j+1}\right)=\sqrt{\frac{\imath am}{2\pi\varepsilon}}\sqrt{-\frac{\imath am}{2\pi\varepsilon}}^{3}\int d^{4}\xi e^{-\frac{\imath am\xi^{2}}{2\varepsilon}}\\
\times\left(1+\imath aq\xi^{\nu}A_{\nu}\left(x_{j+1}\right)+\frac{\imath aq}{2}\xi^{\nu}\xi^{\mu}\partial_{\mu}A_{\nu}\left(x_{j+1}\right)-\frac{a^{2}q^{2}}{2}\xi^{\mu}A_{\mu}\left(x_{j+1}\right)\xi^{\nu}A_{\nu}\left(x_{j+1}\right)-\imath\frac{abm\varepsilon}{2}\right)\\
\times\left(\psi_{\tau}\left(x_{j+1}\right)+\left(\xi^{\mu}\partial_{\mu}\right)\psi_{\tau}\left(x_{j+1}\right)+\frac{1}{2}\xi^{\mu}\xi^{\nu}\partial_{\mu}\partial_{\nu}\psi_{\tau}\left(x_{j+1}\right)+\ldots\right)
\end{array}
\end{equation}
The term zeroth order in $\xi$ gives:
\begin{equation}
\sqrt{\frac{\imath am}{2\pi\varepsilon}}\sqrt{-\frac{\imath am}{2\pi\varepsilon}}^{3}\int d\xi^{4}e^{-\frac{\imath am\xi^{2}}{2\varepsilon}}=1
\end{equation}
This is not surprising; the normalization above was chosen to do this.

Terms linear in $\xi$ give zero when integrated. 

The terms second
order in $\xi$ (first in $\epsilon$) are:

\begin{equation}
\left(\begin{array}{c}
-\imath\frac{abm\varepsilon}{2}+\imath aq\left(\xi^{\nu}A_{\nu}\right)\left(\xi^{\mu}\partial_{\mu}\right)+\frac{\imath aq}{2}\xi^{\nu}\xi^{\mu}\left(\partial_{\mu}A_{\nu}\right)\hfill\\
-\frac{a^{2}q^{2}}{2}\left(\xi^{\mu}A_{\mu}\right)\left(\xi^{\nu}A_{\nu}\right)+\frac{1}{2}\xi^{\mu}\xi^{\nu}\partial_{\mu}\partial_{\nu}\hfill
\end{array}\right){\psi_{\tau}}
\end{equation}
Integrals over off-diagonal powers of order $\xi^{2}$ give zero.
Integrals over diagonal $\xi^{2}$ terms give:

\begin{equation}
\begin{array}{c}
\sqrt{\frac{\imath am}{2\pi\varepsilon}}\int d\xi_{0}e^{-\frac{\imath am\xi_{0}^{2}}{2\varepsilon}}\xi_{0}^{2}=\frac{\varepsilon}{\imath am}\\
\sqrt{-\frac{\imath am}{2\pi\varepsilon}}\int d\xi_{i}e^{\frac{\imath am\xi_{i}^{2}}{2\varepsilon}}\xi_{i}^{2}=-\frac{\varepsilon}{\imath am}
\end{array}
\end{equation}
The expression for the wave function is therefore:
\begin{equation}
\psi_{\tau+\varepsilon}=\psi_{\tau}-\frac{\imath abm\varepsilon}{2}\psi_{\tau}+\frac{q\varepsilon}{m}\left(A^{\mu}\partial_{\mu}\right)\psi_{\tau}+\frac{q}{2m}\varepsilon\left(\partial^{\mu}A_{\mu}\right)\psi_{\tau}+\frac{\imath aq^{2}\varepsilon}{2m}A^{2}\psi_{\tau}-\frac{\imath\varepsilon}{2am}\partial^{2}\psi_{\tau}
\end{equation}
Taking the limit $\epsilon\to0$ and multiplying by $\imath$, we
get the Schr\"{o}dinger equation for TQM:
\begin{equation}
\imath\frac{\partial\psi_{\tau}}{\partial\tau}=ab\frac{m}{2}\psi_{\tau}+\frac{\imath q}{m}\left(A^{\mu}\partial_{\mu}\right)\psi_{\tau}+\frac{\imath q}{2m}\left(\partial^{\mu}A_{\mu}\right)\psi_{\tau}-\frac{aq^{2}}{2m}A^{2}\psi_{\tau}+\frac{1}{2ma}\partial^{2}\psi_{\tau}
\end{equation}
or\label{Schrodinger-equation}:

\begin{equation}
\imath\frac{\partial\psi_{\tau}}{\partial\tau}\left(t,\vec{x}\right)=-\frac{1}{2ma}\left(\left(\imath\partial_{\mu}-aqA_{\mu}\left(t,\vec{x}\right)\right)\left(\imath\partial^{\mu}-aqA^{\mu}\left(t,\vec{x}\right)\right)-a^{2}bm^{2}\right)\psi_{\tau}\left(t,\vec{x}\right)\label{eq:schrodinger-equation-ab}
\end{equation}
If we make the customary identifications $\imath\frac{\partial}{\partial t}\to E$,
$-\imath\vec{\nabla}\to\hat{\vec{p}}$ or $\imath\partial_{\mu}\to p_{\mu}$
we have:
\begin{equation}
\imath\frac{\partial\psi_{\tau}}{\partial\tau}=-\frac{1}{2ma}\left(\left(p_{\mu}-aqA_{\mu}\right)\left(p^{\mu}-aqA^{\mu}\right)-a^{2}bm^{2}\right)\psi_{\tau}
\end{equation}

\subsection{Long, slow approximation\label{subsec:lsa-constants}}

We can now fix the scale and additive constants by looking at the
behavior of the Schr\"{o}dinger equation over longer times.

In his development of quantum mechanics from a time-dependent perspective
\cite{Tannor:2007qz}, Tannor used a requirement of constructive
interference in time to derive the Bohr condition for the allowed
atomic orbitals. We use a similar approach here.

If we average over a sufficiently long period of time, the results
will be dominated by the components with:
\begin{equation}
\imath\frac{\partial\psi_{\tau}\left(x\right)}{\partial\tau}=0
\end{equation}

The argument here is not that the typical variation from the long,
slow solution is small, but rather that over time interactions with
the system in question will tend to be dominated by interactions with
the stabler, slower moving components. Interactions with more rapidly
varying components will tend to average to zero.

Accepting this, then the right side looks like the Klein-Gordon equation.
To complete this identification, first look at the case with the vector
potential $A$ zero:
\begin{equation}
\left(\hat{p}^{2}-a^{2}bm^{2}\right)\psi=0\to a^{2}b=1
\end{equation}
Now when $A$ is not zero we have:
\begin{equation}
\left(\left(\hat{p}-aqA\right)^{2}-m^{2}\right)\psi=0\to a=1\to b=1
\end{equation}
We will refer to this as the ``long, slow approximation''.

In the free case, the long, slow approximation picks out the on-shell
components:
\begin{equation}
\left(\hat{p}^{2}-m^{2}\right)\psi=0
\end{equation}
And more generally the solutions of the Klein-Gordon equation with
the minimal substitution $p\to p-qA$:
\begin{equation}
\left(\left(\hat{p}-qA\right)^{2}-m^{2}\right)\psi=0
\end{equation}

The two constants are now fixed. 
The full Schr\"{o}dinger equation is:
\begin{equation}
\imath\frac{\partial\psi_{\tau}}{\partial\tau}\left(t,\vec{x}\right)=-\frac{1}{2m}\left(\left(\imath\partial_{\mu}-qA_{\mu}\left(t,\vec{x}\right)\right)\left(\imath\partial^{\mu}-qA^{\mu}\left(t,\vec{x}\right)\right)-m^{2}\right)\psi_{\tau}\left(t,\vec{x}\right)\label{eq:schroedinger-equation}
\end{equation}
and in momentum space:
\begin{equation}
\imath\frac{\partial\psi_{\tau}}{\partial\tau}=-\frac{1}{{2m}}\left({\left({{p_{\mu}}-q{A_{\mu}}}\right)\left({{p^{\mu}}-q{A^{\mu}}}\right)-m^{2}}\right){\psi_{\tau}}
\end{equation}

The free Schr\"{o}dinger equation is:
\begin{equation}
2m\imath\frac{\partial\psi_{\tau}}{\partial\tau}\left({t,\vec{x}}\right)=\left({{\partial_{\mu}}{\partial^{\mu}}+{m^{2}}}\right){\psi_{\tau}}\left({t,\vec{x}}\right)\label{eq:seqn-coord}
\end{equation}
and in momentum space:
\begin{equation}
2m\imath\frac{\partial\psi_{\tau}}{\partial\tau}=-\left({{p_{\mu}}{p^{\mu}}-{m^{2}}}\right){\psi_{\tau}}\left({t,\vec{x}}\right)\label{eq:seqn-mom}
\end{equation}

We establish in \ref{sec:app-unitarity} this equation
is unitary. Therefore if a wave function is normalized at $\tau=0$:
\begin{equation}
1=\int{dtd\vec{x}\psi_{0}^{*}\left({t,\vec{x}}\right){\psi_{0}}\left({t,\vec{x}}\right)}
\end{equation}
then at any later clock time $\tau>0$ we will have:
\begin{equation}
1=\int{dtd\vec{x}\psi_{\tau}^{*}\left({t,\vec{x}}\right){\psi_{\tau}}\left({t,\vec{x}}\right)}
\end{equation}

We next estimate the time scales over which we expect the
long, slow approximation to be valid.

\subsection{How long and how slow?\label{subsec:lsa-decoherence}}

We have argued that we can fix the scaling and additive constants
by looking at the behavior of the Schr\"{o}dinger equation over long times.
What do we mean by long times?

To see the relevant scale, we estimate the clock frequency $f$:
\begin{equation}
f\sim-\frac{{{E^{2}}-{{\vec{p}}^{2}}-{m^{2}}}}{{2m}}
\end{equation}

In the non-relativistic case $E$ is of order mass plus kinetic
energy:
\begin{equation}
E\sim m+\frac{{{\vec{p}}^{2}}}{{2m}}
\end{equation}

So we have:
\begin{equation}
{E^{2}}-{{\vec{p}}^{2}}-{m^{2}}\sim{\left({m+\frac{{{\vec{p}}^{2}}}{{2m}}}\right)^{2}}-{{\vec{p}}^{2}}-{m^{2}}={\left({\frac{{{\vec{p}}^{2}}}{{2m}}}\right)^{2}}
\end{equation}

This is just the kinetic energy, squared. In an atom the kinetic energy
is of order the binding energy:
\begin{equation}
\frac{{{\vec{p}}^{2}}}{{2m}}\sim eV
\end{equation}

So the numerator is of order $eV$ squared. But the denominator is
of order $MeV$. So we can estimate the clock frequency $f$ as:
\begin{equation}
f\sim\frac{{e{V^{2}}}}{{MeV}}\sim{10^{-6}}eV
\end{equation}

Energies of millionths of an electron volt ${10^{-6}eV}$ correspond
to times of order millions of attoseconds $10^{6}as$ or picoseconds,
a million times longer than the natural time scale of the effects
we are looking at. So the long, slow approximation is reasonable. 

The physical picture that emerges is of a particle that is extended in time on the same basis as space,
with its wave function extend in all four dimensions,
and -- if we think in terms of path integrals -- the associated paths wandering around in all four dimensions,
not just the traditional three.

It may be amusing to note that from  this perspective there is no such thing as an onshell particle;
in momentum space the onshell part of the wave function is always a set of measure zero.
It is only over scales of picoseconds and greater that 
an onshell description of the particle may give an acceptable approximation.

\subsection{Observer independent choice of frame\label{subsec:lsa-frame}}

There remains one piece of the puzzle; we need to establish that the use of the clock time $\tau$
does not itself violate covariance.

If we did the above derivation for Bob rather than Alice, we would
see Alice's clock time replaced by Bob's clock time $\tau\to\gamma\tau$ ($\gamma  \equiv \frac{1}{{\sqrt {1 - {v^2}} }}$)
where $v$ is his velocity relative to her.

We therefore have one free parameter left to fix before we can declare
our analysis free of free parameters.

If Bob is not going that quickly (relative to Alice) the errors
created by ambiguities with respect to $\tau$ will introduce only small corrections;
of only second order and therefore not relevant for falsifiability.

Even if we were prepared to accept that, 
establishing frame independence is interesting as a point
of principle. 

This may be done in a natural way by making use of an
observation from Weinberg \cite{Weinberg:1972un}. Per Weinberg,
we may treat the Einstein field equation for general relativity as
representing conservation of energy-momentum when exchanges of energy
momentum with spacetime are included.

Consider the Einstein field equations:
\begin{equation}
{G_{\mu\nu}}\equiv{R_{\mu\nu}}-\frac{1}{2}{g_{\mu\nu}}R=-8\pi G{T_{\mu\nu}}
\end{equation}

Rewrite as:
\begin{equation}
{\left({{G^{\mu\nu}}+8\pi G{T^{\mu\nu}}}\right)_{;\nu}}=0
\end{equation}

We may use this to associate an energy momentum tensor ($t_{\mu\nu}$
in Weinberg\textsc{'}s notation) with local space time. Define:
\begin{equation}
{g_{\mu\nu}}={\eta_{\mu\nu}}+{h_{\mu\nu}}
\end{equation}
where $h_{\mu\nu}$ vanishes at infinity but is not assumed small.
The part of the Ricci tensor linear in $h$ is:
\begin{equation}
R_{\mu\nu}^{\left(1\right)}\equiv\frac{1}{2}\left({\frac{{{\partial^{2}}h_{\lambda}^{\lambda}}}{{\partial{x^{\mu}}\partial{x^{\nu}}}}-\frac{{{\partial^{2}}h_{\mu}^{\lambda}}}{{\partial{x^{\lambda}}\partial{x^{\nu}}}}-\frac{{{\partial^{2}}h_{\nu}^{\lambda}}}{{\partial{x^{\lambda}}\partial{x^{\mu}}}}+\frac{{{\partial^{2}}{h_{\mu\nu}}}}{{\partial{x^{\lambda}}\partial{x_{\lambda}}}}}\right)
\end{equation}
The exact Einstein equations may be written as: 
\begin{equation}
R_{\mu\nu}^{\left(1\right)}-\frac{1}{2}{\eta_{\mu\nu}}R_{\lambda}^{\left(1\right)\lambda}=-8\pi G\left({{T_{\mu\nu}}+{t_{\mu\nu}}}\right)
\end{equation}
where $t_{\mu\nu}$ is defined by: 
\begin{equation}
{t_{\mu\nu}}\equiv\frac{1}{{8\pi G}}\left({{R_{\mu\nu}}-\frac{1}{2}{g_{\mu\nu}}R_{\lambda}^{\lambda}-R_{\mu\nu}^{\left(1\right)}+\frac{1}{2}{\eta_{\mu\nu}}R_{\lambda}^{\left(1\right)\lambda}}\right)
\end{equation}

Weinberg then argues we may interpret $t_{\mu\nu}$ as the energy-momentum
of the gravitational field itself.

Accepting this, we change to a coordinate frame in which $t_{\mu\nu}$ is diagonalized.
We will refer to this as the rest frame of the vacuum or the $V$
frame. We can treat this $V$ frame as the defining frame for the
four dimensional Schr\"{o}dinger equation. As the $V$ frame is invariant
(up to rotations in three-space) we now have an invariant definition
of the four dimensional Schr\"{o}dinger equation.

This is obviously going to be a free-falling frame. So Alice and Bob
-- if they are working in a terrestrial laboratory -- will have
to adjust their calculations to include a correction for the upwards
force the laboratory floor exerts against them. If their colleague
Vera is working in an orbiting laboratory, she will be able to calculate
without correction. 

Here after, unless stated to contrary, we will assume we are working in the ``rest frame of the vacuum''.

\subsection{Relationship between TQM and the Relativistic Dynamics program \label{subsec:tqm-and-rd}}
\begin{quotation}
``In other words, we are trying to prove ourselves
wrong as quickly as possible, because only in that way can we find
progress.'' -- Richard P. Feynman \cite{Feynman:1965jb}
\end{quotation}
We have now fully defined the path integral expression and the Schr\"{o}dinger
equation.

It is appropriate therefore to pause to look at relationship of TQM
to the relativistic dynamics program. (A helpful overview of the relativistic dynamics program is provided
in Fanchi \cite{Fanchi:1993aa}). 

\subsubsection{TQM as part of the Relativistic Dynamics program}

We start with Feynman \cite{Feynman:1948}, who uses the Lagrangian:
\begin{equation}
{L_{Feynman}}\left({{x^{\mu}},{{\dot{x}}^{\mu}}}\right)=-{{\dot{x}}^{\mu}}{{\dot{x}}_{\mu}}-q{{\dot{x}}^{\mu}}{A_{\mu}}\left(x\right)
\end{equation}
giving:
\begin{equation}
\imath\frac{\partial}{{\partial s}}{\psi_{s}}=\frac{{\left({{p^{\mu}}-{q}{A^{\mu}}}\right)\left({{p_{\mu}}-{q}{A_{\mu}}}\right)}}{2}{\psi_{s}}
\end{equation}

However with this approach, as Feynman notes, each particle would
need its own evolution parameter $s$. This would make extension to
the multiple particle case problematic.

There are many variations on this in the literature. Fanchi
\cite{Fanchi:1993ab} derives:
\begin{equation}
\imath\frac{\partial}{{\partial s}}{\psi_{s}}=\frac{{\left({{p^{\mu}}-{q}{A^{\mu}}}\right)\left({{p_{\mu}}-{q}{A_{\mu}}}\right)}}{2m}{\psi_{s}}
\end{equation}

This differs from Feynman's by a factor of $\frac{1}{m}$.
Fanchi's evolution parameter $s$ is for all practical purposes the negative
of our clock time $\tau$. 

Land and Horwitz \cite{Land:1996aj} give:
\begin{equation}
\left({\imath\frac{\partial}{{\partial s}}+{q}{a_{5}}}\right){\psi_{s}}=\frac{{\left({{p^{\mu}}-{q}{A^{\mu}}}\right)\left({{p_{\mu}}-{q}{A_{\mu}}}\right)}}{{2m}}{\psi_{s}}
\end{equation}
This is Fanchi's with the addition of the $a_{5}$, a gauge term. 

If we take $s\to-\tau$ and make the choice of gauge (see \ref{sub:app-gauge}):
\begin{equation}
{a_{5}}\Rightarrow\frac{{m}}{{2q}}
\end{equation}
we get:
\begin{equation}
\left({\imath\frac{\partial}{{\partial\tau}}}\right){\psi_{s}}=-\frac{{\left({{p^{\mu}}-q{A^{\mu}}}\right)\left({{p_{\mu}}-q{A_{\mu}}}\right)-{m^{2}}}}{{2m}}{\psi_{s}}
\end{equation}
which is the same as our Schr\"{o}dinger equation (equation \ref{eq:schroedinger-equation}).

At this point we may argue that as the right hand-side is the Klein-Gordon
equation, and the Klein-Gordon equation is strongly confirmed by experimental
evidence we have:
\begin{equation}
0\approx\frac{{\left({{p^{\mu}}-q{A^{\mu}}}\right)\left({{p_{\mu}}-q{A_{\mu}}}\right)-{m^{2}}}}{{2m}}{\psi_{s}}
\end{equation}

So the combination of the mass gauge ($a_{5}=\frac{m}{2q}$) and the
comparison with the Klein-Gordon equation give us the long slow approximation. 

We have therefore placed our Schr\"{o}dinger equation within the context
of the relativistic program. This will let us use results from that
program in the appropriate limit.

\subsubsection{TQM and falsifiability}

TQM may be thought of as a specialization of the relativistic dynamics
program aimed at making the hypothesis ``time should be treated on
the same basis as space in quantum mechanics'' falsifiable. This
means in turn that we have to be able to rule out the \textit{class}
of such theories, not just one in particular.

To do this we need to extend quantum mechanics to include time in
a way that depends only on covariance. We can admit no free or adjustable
parameters; any such risk falsifiability. 

We may understand each step of the program here from this point of
view. 
\begin{enumerate}
\item We used clock time as the ``evolution parameter''. This is defined
operationally -- by the laboratory clock -- and therefore introduces
no free parameters.
\item We used path integrals as the defining representation; with these
the only change we need to make is to have the paths extend in four
rather than three dimensions.
\item We could not use plane waves or delta functions as the initial wave
functions: as they are not physical, they introduce the risk of inducing
mathematical artifacts and again threatening falsifiability.
\item We needed to treat general wave functions, not just a specialized
subset. Otherwise the results might be conditional on the subset chosen.
\item We needed to ensure that our path integrals are convergent without
the use of tricks. Otherwise the tricks can bring the results into
question. The use of Morlet wavelet decomposition solved this and
the previous two problems.
\item The resulting Schr\"{o}dinger equation depends on the clock time. To eliminate
such dependence (to lowest order) we showed that the time scales associated
with the clock time are of order picoseconds, while the scales associated
with dispersion in time are of order attoseconds -- a million times
smaller. So long as we work at the attosecond scale we should be able
to ignore effects associated with clock time.
\item The clock time in turn depends on the choice of the laboratory frame.
To further control for effects of the specific choice of laboratory
frame, we showed we can define an invariant frame (the ``rest frame
of the vacuum''). By defining all calculations with reference to this
invariant frame we eliminate the dependence of the clock time on the
choice of a specific laboratory frame.
\end{enumerate}
While we have fully defined the path integral expression and the Schr\"{o}dinger
equation, there are still some further questions that need to be addressed
if we are to achieve falsifiability:
\begin{enumerate}
\item How do we estimate the initial wave functions in time in a robust
and frame independent way (subsection \ref{subsec:free-initial})?
\item How do we understand the evolution of the wave function with respect
to clock time: with the long, slow approximation we appear to be saying
the wave function is frozen in time, that $\imath\frac{\partial}{{\partial\tau}}\psi\approx0$,
which is clearly far from the case (subsection \ref{subsec:free-evolution})?
\item How do we define the rules for detection in a way which is manifestly
covariant (subsection \ref{subsec:free-toa})?
\item How do we produce at least one experiment with which we can unambiguously
show that the wave function should not be extended in time (section
\ref{sec:slits})?
\item Realistic experiments will need to run at ultra-short times and therefore
high energies. At these energies, particles may spring into existence
from the vacuum. How do we extend this analysis to the multiple particle
case in a way which is consistent with the work done in the single
particle case (section \ref{sec:Multiple-particle-case})?
\end{enumerate}

At attosecond times, we expect that the effects of dispersion in time will dominate. 
Over picosecond and longer times we expect that the effects of the clock time will need to be included.

Therefore, at attosecond times, we expect TQM will function as a kind of lowest common denominator for the relativistic dynamics program.
At picosecond times, we may need to include contributions from the evolution parameter and therefore be able to discriminate among various branches of the relativistic dynamics program.

\section{Free solutions\label{sec:free}}

We now have the Schr\"{o}dinger equation. What do its free solutions look
like?

We examine in turn the birth, life, and death of a free particle.
The calculations are straightforward. But each stage will present
problems specific to TQM.

For convenience, we assemble the solutions of the free Schr\"{o}dinger\textsc{'}s equation in \ref{sec:app-free}.

\subsection{Initial wave function\label{subsec:free-initial}}

What do our wave functions look like at start?

We have a chicken and egg problem here. Any initial wave function
had itself to come from somewhere. How can we estimate the initial
wave functions without first knowing them?

We look specifically at the Klein-Gordon equation for a static electric potential with no magnetic field:
\begin{equation}
{A_{0}}=\Phi\left({\vec{x}}\right),\vec{A}=0
\end{equation}
with potential:
\begin{equation}
V\equiv q \Phi
\end{equation}

This includes attractive potentials, scattering potentials, and the free case as the special case when $\Phi=0$.

In SQM the Klein-Gordon equation is:
\begin{equation}
\left({{{\left({\imath{\partial_{\tau}}-V\left({\vec{x}}\right)}\right)}^{2}}-{\nabla^{2}}-{m^{2}}}\right){\bar{\psi}_{\tau}}\left({\vec{x}}\right)=0
\end{equation}

In TQM the equivalent is the four dimensional Schr\"{o}dinger equation:
\begin{equation}
\imath\frac{\partial}{{\partial\tau}}{\psi_{\tau}}\left({t,\vec{x}}\right)=-\frac{1}{{2m}}\left({{{\left({\imath{\partial_{t}}-V\left({\vec{x}}\right)}\right)}^{2}}-{\nabla^{2}}-{m^{2}}}\right){\psi_{\tau}}\left({t,\vec{x}}\right)
\end{equation}

The long, slow approximation picks out the solutions with:
\begin{equation}
\left({{{\left({\imath{\partial_{t}}-V\left({\vec{x}}\right)}\right)}^{2}}-{\nabla^{2}}-{m^{2}}}\right){\psi_{\tau}}\left({t,\vec{x}}\right)=0
\end{equation}

We will assume we are already in possession of solutions to the SQM
version of the problem:
\begin{equation}
\left(\left(E-V\left(\vec{x}\right)\right)^{2}-\vec{p}^{2}-m^{2}\right){\bar{\psi}_{E}}\left({\vec{x}}\right)=0
\end{equation}

This is necessarily in a specific frame, since
SQM solutions are always (from our perspective) taken with reference
to a specific frame. 

We leverage the SQM solutions in two different
ways\footnote{This problem is also solved in Fanchi \cite{Fanchi:1993ab}, using different
methods.} to get the TQM solution: 
\begin{enumerate}
\item \textit{Separation of variables.} Each SQM solution induces a TQM
solution, which is the SQM solution with a plane wave bolted
on. This is technically correct, but unphysical.
\item \textit{Maximum entropy. }We can use the long, slow approximation
to estimate the mean and uncertainty of the coordinate energy. With
these we can use the method of Lagrange multipliers to get the maximum
entropy solution. Maximum entropy solutions tend to be robust: even
if we are wrong about the details, the order of magnitude should be
correct. This will provide our preferred starting point. 
\end{enumerate}
As a quick check on the sanity of all this we will use the virial
theorem to estimate the TQM version of the atomic wave functions.
The width in time/energy of these matches the initial order of magnitude
estimate we gave in the introduction. 

We will also show that while we have chosen a specific frame in which
to estimate the dispersion in time/energy we can get the estimate
in an invariant way.

\subsubsection{Solution by separation of variables\label{subsubsec:separation-of-variables}}

We solve the TQM Klein-Gordon equation using separation of variables, looking for solutions
of the form:
\begin{equation}
\varphi_{n}\left(t,\vec{x}\right)=\tilde{\phi}{}_{n}\left(t\right)\bar{\varphi}_{n}\left(\vec{x}\right)
\end{equation}
where $\bar{\varphi}_{n}\left(\vec{x}\right)$ is a solution of the
SQM Klein-Gordon equation:
\begin{equation}
\left(\left(\bar{E}_{n}-V\left(\vec{x}\right)\right)^{2}-\vec{p}^{2}-m^{2}\right)\bar{\varphi}_{n}\left(\vec{x}\right)=0
\end{equation}
Then the coordinate time part is:
\begin{equation}
\tilde{\phi}{}_{n}\left(t\right)=\frac{1}{\sqrt{2\pi}}e^{-\imath\bar{E}_{n}t}
\end{equation}
Because the potential is constant in time each use of the operator
$E$ turns into a constant $E_{n}$ via:
\begin{equation}
E\to\imath\frac{\partial}{\partial t}\to\bar{E}_{n}
\end{equation}
We get immediately:
\begin{equation}
\left(\left(\bar{E}_{n}-V\left(\vec{x}\right)\right)^{2}-\vec{p}^{2}-m^{2}\right)\varphi_{n}\left(t,\vec{x}\right)=0
\end{equation}

So every solution of the Klein-Gordon equation in SQM generates a
corresponding solution in TQM.

We could accomplish the same thing, formally, by taking $\tau\to t$.
Since we expect in general that $\left\langle t\right\rangle \approx\tau$
(discussed further in the next section) this will often a give reasonable
first approximation.

However this is not entirely satisfactory. We have a solution which
is ``fuzzy'' in space, but ``crisp'' in time. A more realistic,
if more complex solution, would include off-shell components. Even
if our wave function started out as a simple plane wave in time, internal
decoherence would rapidly turn it into something a bit more cloud-like.
While mathematically acceptable, our solution is not physically plausible.

\subsubsection{Solution by maximum entropy}

\label{subsubsec:free-initial-maximum-entropy}

The long, slow approximation picks out a single solution. But in practice
we expect there would be a great number of solutions, with the one
given by the long slow approximation merely the most typical.

Such a sum will have an associated probability density function. We
can get a reasonable first estimate of this by defining appropriate
constraints and then using the method of Lagrange multipliers to pick
out the distribution with maximum entropy.

From the probability density, we will infer the wave function.

\paragraph{Estimate of the probability density}

We start with the free case, as representing the simplest case of
a constant potential. We treat the bound case below. 

We assume we
are given the SQM wave function in three dimensions.  
We use this to compute
the expectation of the energy,
and the expectation of the energy squared.
These give us our constraints:
\begin{equation}
\begin{array}{l}
\left\langle 1\right\rangle =1\\
\bar{E}\equiv\left\langle E\right\rangle =\sqrt{m^{2}+\left\langle \vec{p}\right\rangle ^{2}}\\
\left\langle E^{2}\right\rangle =\left\langle m^{2}+\vec{p}^{2}\right\rangle =m^{2}+\left\langle \vec{p}^{2}\right\rangle 
\end{array}\label{eq:constraints-1}
\end{equation}

The uncertainty in energy is defined as:
\begin{equation}
\Delta E\equiv\sqrt{\left\langle {E^{2}}\right\rangle -{{\left\langle E\right\rangle }^{2}}}
\end{equation}

The expectations are defined as integrals over the probability density:
\begin{equation}
\left\langle f\right\rangle \equiv\int d\vec{p}\bar{\rho}\left(\vec{p}\right)f\left(\vec{p}\right)
\end{equation}
The constraints imply $\Delta E=\Delta p$.

We will work with box normalized energy eigenfunctions\label{free-initial-box-normal-in-time}:
\begin{equation}
\phi_{n}\left(t\right)\equiv\frac{1}{\sqrt{2T}}e^{-\imath E_{0}nt}
\end{equation}
\begin{equation}
E_{0}\equiv\frac{\pi}{T}
\end{equation}
Here $n$ is an integer running from negative infinity to positive
and the eigenfunctions are confined to a box extending $T$ seconds
into the future and $T$ seconds into the past, where $T$ is much
larger than any time of interest to us. (We will use a similar approach
in developing the four dimensional Fock space below, subsection \ref{subsec:mul-fock-4d}).

A general wave function can be written as:
\begin{equation}
\psi\left(t\right)=\sum\limits _{n=-\infty}^{\infty}c_{n}\phi_{n}\left(t\right)
\end{equation}
The coefficients $c$ only appear as the square:
\begin{equation}
\rho_{n}\equiv c_{n}^{*}c_{n}
\end{equation}
Expressed in this language we have the constraints:
\begin{equation}
\begin{array}{c}
C_{0}\equiv\sum\limits _{n=-\infty}^{\infty}\rho_{n}-1=0\\
C_{1}\equiv E_{0}\sum\limits _{n=-\infty}^{\infty}n\rho_{n}-\left\langle E\right\rangle =0\\
C_{2}\equiv E_{0}^{2}\sum\limits _{n=-\infty}^{\infty}n^{2}\rho_{n}-\left\langle E^{2}\right\rangle =0
\end{array}
\end{equation}

We would like to find the solution that maximizes the entropy:
\begin{equation}
S\equiv\sum\limits _{n=-\infty}^{\infty}-\rho_{n}\ln\left(\rho_{n}\right)
\end{equation}
We form the Lagrangian from the sum of the entropy and the constraints,
with Lagrange multipliers $\lambda_{0}$, $\lambda_{1}$, and $\lambda_{2}$:
\begin{equation}
L\equiv S+\lambda_{0}C_{0}+\lambda_{1}C_{1}+\lambda_{2}C_{2}
\end{equation}
To locate the configuration of maximum entropy, we take the derivative
with respect to the $\rho_{n}$:
\begin{equation}
\frac{\partial L}{\partial\rho_{n}}=0
\end{equation}
getting:
\begin{equation}
-\ln\rho_{n}-1+\lambda_{0}+E_{0}n\lambda_{1}+E_{0}^{2}n^{2}\lambda_{2}=0
\end{equation}
Therefore the distribution of the $\rho_{n}$ is given by an exponential
with zeroth, first, and second powers of the energy:
\begin{equation}
\rho\left(E\right)\sim e^{-a-bE-cE^{2}}
\end{equation}
The constraints force the constants:
\begin{equation}
\rho\left(E\right)=\frac{1}{\sqrt{2\pi\Delta E^{2}}}e^{-\frac{\left(E-\bar{E}\right)^{2}}{2\left(\Delta E\right)^{2}}}
\end{equation}

\paragraph{Estimate of the wave function}

We therefore have the probability density in energy; now we wish to
estimate the corresponding wave function in energy (or equivalently
time). 

The simplest -- and therefore best -- way to do this is to write the total wave function as the direct product of a
wave function in energy times the postulated wave function in momentum:
\begin{equation}
\hat{\varphi}\left(E,p\right)\sim\hat{\tilde{\varphi}}\left(E\right)\hat{\bar{\varphi}}\left(p\right)
\end{equation}

We then use the Gaussian test function that matches the derived density function for the energy factor:
\begin{equation}
\hat{\tilde{\varphi}}_{0}\left(E\right)=\sqrt[4]{\frac{1}{\pi\sigma_{E}^{2}}}e^{\imath\left(E-E_{0}\right)t_{0}-\frac{\left(E-E_{0}\right)^{2}}{2\sigma_{E}^{2}}}
\end{equation}
with values:
\begin{equation}
\sigma_{E}^{2}=\sigma_{p}^{2}
\end{equation}
and:
\begin{equation}
E_{0}=\sqrt{m^{2}+\bar{p}^{2}}
\end{equation}

Taking the Fourier transform of the energy part we have:
\begin{equation}
\tilde{\varphi}_{0}\left(t\right)=\sqrt[4]{\frac{1}{\pi\sigma_{t}^{2}}}e^{-\imath E_{0}\left(t-t_{0}\right)-\frac{t^{2}}{2\sigma_{t}^{2}}}
\end{equation}
\begin{equation}
\sigma_{t}^{2}=\frac{1}{\sigma_{E}^{2}}
\end{equation}
We set $t_{0}=0$ as the overall phase is already supplied by the
space/momentum part.

The full wave functions are the products of the coordinate time and
space (or energy and momentum) parts:
\begin{equation}
\varphi_{0}\left(t,x\right)=\tilde{\varphi}_{0}\left(t\right)\bar{\varphi}_{0}\left(x\right)
\end{equation}
\begin{equation}
\hat{\varphi}_{0}\left(E,p\right)=\hat{\tilde{\varphi}}_{0}\left(E\right)\hat{\bar{\varphi}}_{0}\left(p\right)
\end{equation}

\subsubsection{Bound state wave functions\label{subsec:free-initial-bound}}

We extend this approach to estimate the dispersion of a bound wave
function in time.

In the case of a Coulomb potential we can estimate $\Delta p$ from
the virial theorem:
\begin{equation}
\left\langle \frac{\vec{p}^{2}}{2m}\right\rangle =-\frac{1}{2}\left\langle V\right\rangle 
\end{equation}
which implies:
\begin{equation}
\left\langle \frac{\vec{p}^{2}}{2m}\right\rangle +\left\langle V\right\rangle =\bar{E}_{n}\to\left\langle \frac{\vec{p}^{2}}{2m}\right\rangle =-\bar{E}_{n}
\end{equation}
Since the average momentum is zero, we have the estimate of the uncertainty in energy as:
\begin{equation}
\Delta E_{n}=\sqrt{-2m\bar{E}_{n}}
\end{equation}
Substituting the mass of the electron and the Rydberg constant, we
have for the hydrogen ground state:
\begin{equation}
\Delta E_{1}=\sqrt{2\cdot13.6\textrm{eV}\cdot\left(0.511\cdot10^{6}\right)\textrm{eV}}=3728\textrm{ eV}
\end{equation}
And the corresponding dispersion in coordinate time is:
\begin{equation}
\Delta t=\frac{\hbar}{\Delta E_{n}}=.1766\textit{as}
\end{equation}

This matches the order of magnitude estimate we started with (subsection
\ref{subsec:estimated-scale}). The numerical closeness is coincidental,
but does increase confidence the order of magnitude is correct. 

We have therefore as the initial estimate of the TQM wave function for a hydrogen atom:
\begin{equation}
\begin{array}{c}
  {{\hat \psi }_n}\left( {E,\vec p} \right) = {{\hat \varphi }_n}\left( E \right){{\hat {\bar {\psi }}}_n}\left( {\vec p} \right) \hfill \\
  {{\hat {\tilde \varphi }}_n}\left( E \right) = \sqrt[4]{{\frac{1}{{2\pi \hat \sigma _n^2}}}}{e^{\imath \left( {E - {{\bar E}_n}} \right){t_0} - \frac{{{{\left( {E - {{\bar E}_n}} \right)}^2}}}{{2\hat \sigma _n^2}}}} \hfill \\
  \hat \sigma _n^2 =  - 4m{{\bar E}_n} \hfill \\ 
\end{array}
\end{equation}

\subsubsection{Frame independence of the estimate\label{subsubsec:free-initial-free-wave-functions}}

We have given a reasonable estimate of the initial wave function in
time/energy. 
However, we have not yet established that the estimate is independent
of the initial choice of frame. 

For both the bound and free cases, we choose our initial frame as the rest frame.
We make our estimate, via maximum entropy in this frame, then to get the wave function in a arbitrary frame
Lorentz transform to that frame:
\begin{equation}
\psi '\left( {x'} \right) = \psi \left( {\Lambda x} \right)
\end{equation}
Since the rest frame is an invariant, the resulting initial wave functions are well-defined.

\subsection{Evolution of the free wave function\label{subsec:free-evolution}}

Now that we know what our starting wave functions look like, how do
they evolve over time?

For a first examination we work with the non-relativistic case; we
will extend to the relativistic case below (section \ref{sec:Multiple-particle-case}).
We will first look at the SQM case, then the TQM case, 
then compare the two.

\subsubsection{Evolution in SQM}

We start with the familiar problem of the evolution of the non-relativistic
wave function with respect to clock time. We work with two dimensions
-- $\tau,x$ -- since the extension to $y,z$ is straightforward.

The non-relativistic Schr\"{o}dinger equation is:
\begin{equation}
\imath\frac{\partial}{{\partial\tau}}\bar{\psi}=\frac{{p^{2}}}{{2m}}\bar{\psi}=-\frac{{\partial_{x}^{2}}}{{2m}}\bar{\psi}
\end{equation}

At clock time zero we start with a Gaussian test function in momentum
with average position $x_{0}$, average momentum $p_{0}$, and dispersion
in momentum $\sigma_{p}$. To reduce clutter we use $p$ for
$p_{x}$:
\begin{equation}
{{\hat{\bar{\varphi}}}_{0}}\left(p\right)=\sqrt[4]{{\frac{1}{{\pi\sigma_{p}^{2}}}}}{e^{-\imath p{x_{0}}-\frac{{{\left({p-{p_{0}}}\right)}^{2}}}{{2\sigma_{p}^{2}}}}}
\end{equation}

In momentum space the problem is trivial. The solution is:
\begin{equation}
{{\hat{\bar{\varphi}}}_{\tau}}\left(p\right)=\sqrt[4]{{\frac{1}{{\pi\sigma_{p}^{2}}}}}{e^{-\imath p{x_{0}}-\frac{{{\left({p-{p_{0}}}\right)}^{2}}}{{2\sigma_{p}^{2}}}-\imath\frac{{p^{2}}}{{2m}}\tau}}
\end{equation}

In coordinate space we get:
\begin{equation}
\bar{\varphi}_{\tau}\left(x\right)=\sqrt[4]{\frac{1}{\pi\sigma_{x}^{2}}}\sqrt{\frac{1}{f_{\tau}^{\left(x\right)}}}e^{\imath p_{0}x-\frac{1}{2\sigma_{x}^{2}f_{\tau}^{\left(x\right)}}\left(x-{x}_{0}-\frac{p_{0}}{m}\tau\right)^{2}-\imath\frac{p_{0}^{2}}{2m}\tau}
\end{equation}
with:
\begin{equation}
{\sigma_{x}}=\frac{1}{\sigma_{p}}
\end{equation}
\begin{equation}
f_{\tau}^{\left(x\right)}=1+\imath\frac{\tau}{m\sigma_{x}^{2}}
\end{equation}

\subsubsection{Evolution in TQM}

In two dimensions the Schr\"{o}dinger equation for TQM is:
\begin{equation}
\imath\frac{\partial\psi_{\tau}}{\partial\tau}\left({t,\vec{x}}\right)=-\frac{{{E^{2}}-{{p}^{2}}-{m^{2}}}}{{2m}}{\psi_{\tau}}\left({t,\vec{x}}\right)=\left({\frac{{\partial_{t}^{2}}}{{2m}}-\frac{{\partial_{x}^{2}}}{{2m}}+\frac{m}{2}}\right){\psi_{\tau}}\left({t,\vec{x}}\right)
\end{equation}

We start in energy momentum space. The momentum part is as above.
We start with a Gaussian test function in energy, with average time
at start $t_{0}$, average energy $E_{0}$, and dispersion in energy
$\sigma_{E}$:
\begin{equation}
{{\hat{\tilde{\varphi}}}_{0}}\left(E\right)\equiv\sqrt[4]{{\frac{1}{{\pi\sigma_{E}^{2}}}}}{e^{\imath E{t_{0}}-\frac{{{\left({E-{E_{0}}}\right)}^{2}}}{{2\sigma_{E}^{2}}}}}
\end{equation}

Usually we will take $t_{0}=\tau_{0}=0$.

As a function of clock time we get:

\begin{equation}
{{\hat{\psi}}_{\tau}}\left({E,p}\right)={{\hat{\tilde{\varphi}}}_{0}}\left(E\right){{\hat{\bar{\varphi}}}_{0}}\left({p}\right)\exp\left({-\imath\frac{{{E^{2}}-{p^{2}}-{m^{2}}}}{{2m}}\tau}\right)
\end{equation}

We divide up the pieces of the clock time part, assigning the $\frac{E^{2}}{2m}$
to the energy part, the $\frac{p^{2}}{2m}$ to the momentum part,
and keeping the third part outside:
\begin{equation}
{{\hat{\psi}}_{\tau}}\left({E,p}\right)={{\hat{\tilde{\varphi}}}_{\tau}}\left(E\right){{\hat{\bar{\varphi}}}_{\tau}}\left(p\right)\exp\left({\imath\frac{m}{2}\tau}\right)\label{eq:free-evolution-tq-wf}
\end{equation}

Now the energy part works in parallel to the momentum part:
\begin{equation}
{{\hat{\tilde{\varphi}}}_{\tau}}\left(E\right)\equiv\sqrt[4]{{\frac{1}{{\pi\sigma_{E}^{2}}}}}{e^{\imath E{t_{0}}-\frac{{{\left({E-{E_{0}}}\right)}^{2}}}{{2\sigma_{E}^{2}}}-\imath\frac{{E_{0}^{2}}}{{2m}}\tau}}\label{eq:free-evolution-wf-E}
\end{equation}

And in coordinate space:
\begin{equation}
\tilde{\varphi}_{\tau}\left(t\right)=\sqrt[4]{\frac{1}{\pi\sigma_{t}^{2}}}\sqrt{\frac{1}{f_{\tau}^{\left(t\right)}}}e^{-\imath E_{0}t+\imath\frac{E_{0}^{2}}{2m}\tau-\frac{1}{2\sigma_{t}^{2}f_{\tau}^{\left(t\right)}}\left(t-t_{0}-\frac{E_{0}}{m}\tau\right)^{2}}\label{eq:free-evolution-wf-t}
\end{equation}
with the usual ancillary definitions: 
\begin{equation}
{\sigma_{E}}=\frac{1}{{\sigma_{t}}}
\end{equation}
\begin{equation}
f_{\tau}^{\left(t\right)}\equiv1-\imath\frac{\tau}{m\sigma_{t}^{2}}
\end{equation}
and with the expectation for coordinate time:
\begin{equation}
\bar{t}_{\tau}={t}_{0}+\frac{E_{0}}{m}\tau\label{eq:expect-time}
\end{equation}
implying a velocity for coordinate time with respect to laboratory
time:
\begin{equation}
\gamma=\frac{E_{0}}{m}
\end{equation}

In the non-relativistic case, $\gamma\approx1$. 
So the expectation
of the coordinate time advances at the traditional one-second-per-second
rate relative to the clock time.

%While we are focused on the non-relativistic case the wave function
%here is fully relativistic and may be written, if we wish, in a covariant
%form, appendix \ref{subsubsec:app-free-Four-space}.

\subsubsection{Comparison of TQM to SQM\label{subsec:free-comparison}}

The TQM and SQM approaches develop in close parallel. In both, the
wave functions are centered on the classical trajectory as given by
the solution of the Euler-Lagrange equations. 

There is however one peculiarity.

We are relying on the long, slow approximation. Especially over short
times, this means that the total wave function appears to be relatively
static with respect to evolution in clock time:
\begin{equation}
\imath\frac{\partial}{{\partial\tau}}\psi\approx0
\end{equation}

In momentum space:
\begin{equation}
f_{p}\equiv-\frac{{{E^{2}}-{p^{2}}-{m^{2}}}}{{2m}}\approx0
\end{equation}

However, real wave functions are not static with respect to clock
time.

The resolution is that most of the clock time dependence is carried
by the coordinate time:
\begin{equation}
\frac{d}{{d\tau}}\psi=\frac{\partial}{{\partial\tau}}\psi+\frac{{dt}}{{d\tau}}\frac{\partial}{{\partial t}}\psi\approx\frac{{dt}}{{d\tau}}\frac{\partial}{{\partial t}}\psi
\end{equation}

And in the non-relativistic case we have:
\begin{equation}
\frac{{dt}}{{d\tau}}=\frac{E}{m}\approx1
\end{equation}

So we get as a rough approximation:
\begin{equation}
\frac{d}{{d\tau}}\psi\approx\frac{\partial}{{\partial t}}\psi
\end{equation}

We see that the expectation of the coordinate time is about equal
to the clock time. While Alice\textsc{'}s dog is always getting ahead
of and behind her, on average his position is about equal to hers:
\begin{equation}
\left\langle t\right\rangle \approx\tau
\end{equation}

The Schr\"{o}dinger equation gives the partial derivative with respect
to clock time; the total derivative behaves as expected.

\subsection{Time of arrival measurements\label{subsec:free-toa}}

\begin{figure}[h]
\begin{centering}
\includegraphics[width=11cm]{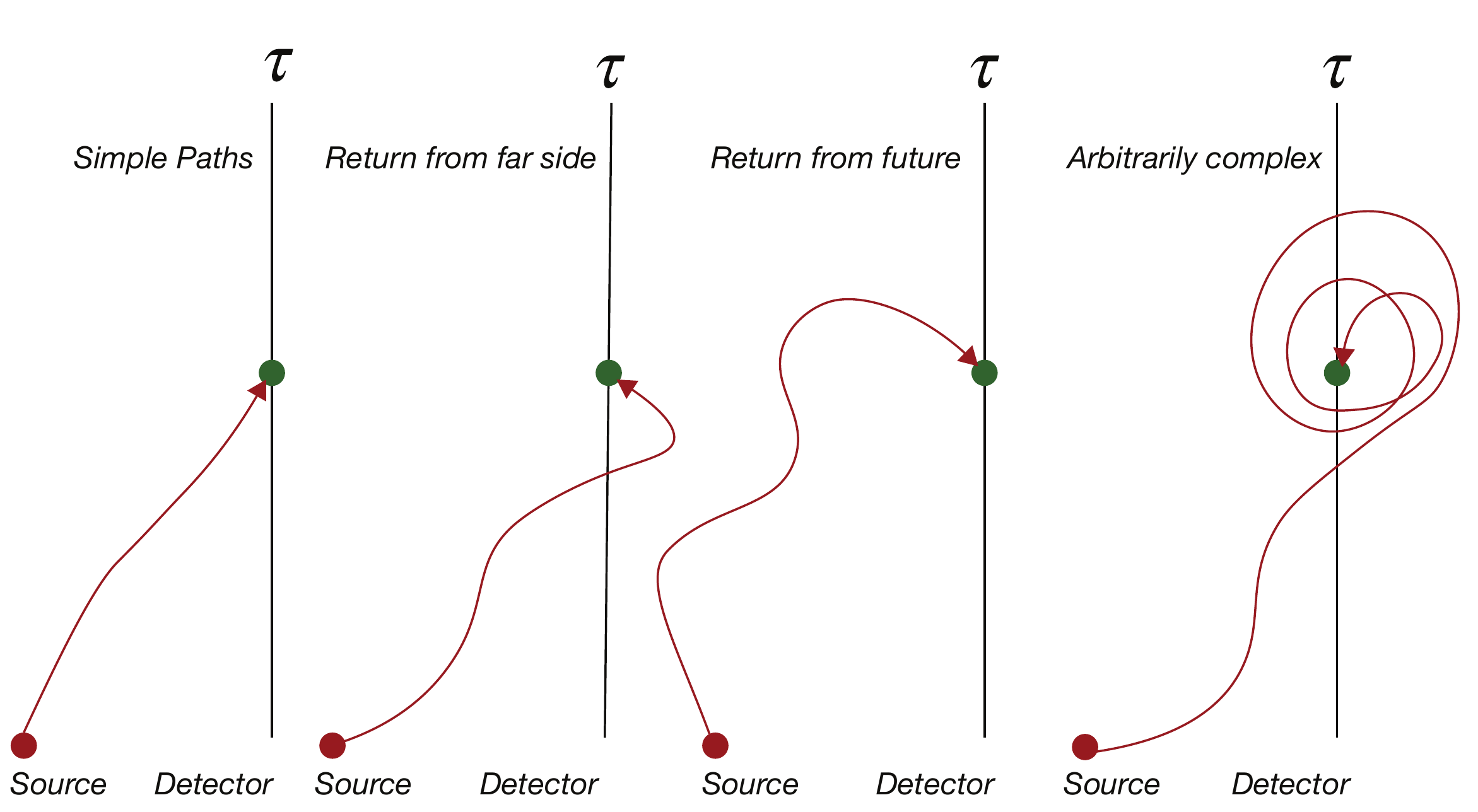} 
\par\end{centering}
\caption{Complex paths near the detector\label{fig:free-toa-complex-paths}}
\end{figure}

So we know what our wave function looks like at start and how it evolves
with time. To complete the analysis of the free case we look at how it
is detected.

We look specifically at the measurement of time-of-arrival. We assume
we have a particle going left to right, starting at $x=0$. We place
a detector at position $x=L$. It records when it detects the particle.
The metric we are primarily interested in is the dispersion in time-of-arrival
at the detector.

In SQM, if a detector located at position $X$ registers a hit by
a particle we take the particle\textsc{'}s position in space as also
$x=X$. Therefore in TQM, if detector active at laboratory time $T$
registers a hit by a particle we \textit{must} take the particle\textsc{'}s
position in time as $t=T$. This is required by our principle of maximum
symmetry between time and space.

By the same token, in SQM if an emitter located at position $X$ emits
a particle, we take the start position of the path as $x=X$. Therefore
in TQM, if an emitter active at laboratory time $T$ emits a particle,
we \textit{must} take the start position in coordinate time as $t=T$.

In a practical treatment we would replace the phrases ``at X'' or
``at T'' with ``within the range $X\pm\frac{\Delta X}{2}$'' and
``within the range $T\pm\frac{\Delta T}{2}$''.

If we know both source and detector positions in space and time then
all corresponding paths are clamped at both ends. In between source
and detector the paths can examine all sorts of interesting times
and spaces but each path is clamped at the endpoints. Alice and her
dog leave from the same starting point in space time and arrive at
the same ending point in space time, but while classical Alice takes
the shortest path between the start and end points, \textit{the quantum
dog explores all paths}.
%All paths in space for SQM; all paths in space and time for TQM.

\paragraph{Meaning of ``all paths''}

Paths in TQM are much more complex than those in
SQM. If a detector is a camera shutter, open for a fraction of a second,
then any paths that arrive early or late will merely be ``eaten''
by the closed shutter. But what if our apparatus can somehow be toggled
from transparent to absorptive and back, as via ``electromagnetically
induced transparency'' \cite{RevModPhys.77.633}? Then the paths
can arrive early but then circle back, or arrive late but circle forward,
or even perform a drunkard\textsc{'}s walk around the detector till
they choose to fall into it.

Since we are primarily interested in comparisons of TQM to SQM, rather
than in fully exploring the elaborations of TQM, we will focus on
the camera shutter model. Paths that arrive before the shutter is
open or after the shutter is closed again will be silently absorbed
by the camera itself. We defer to a later investigation examination
of more complex paths.

We employ the rest frame of the camera.

\subsubsection{Metrics}

With that dealt with, to compute the dispersion in time, we log how
many hits we get in each time interval (``clicks per tick''):
\begin{equation}
\rho\left(\tau\right)
\end{equation}
then calculate the average:
\begin{equation}
\left\langle \tau\right\rangle \equiv\int\limits _{-\infty}^{\infty}{d\tau\rho\left(\tau\right)}
\end{equation}
and the uncertainty:
\begin{equation}
{\left\langle {\Delta\tau}\right\rangle ^{2}}\equiv\int\limits _{-\infty}^{\infty}{d\tau{\tau^{2}}\rho\left(\tau\right)}-{\left\langle \tau\right\rangle ^{2}}
\end{equation}

\subsubsection{Time of arrival in SQM\label{subsec:free-toa-sqm}}

\begin{figure}[h]
\begin{centering}
\includegraphics[width=10cm]{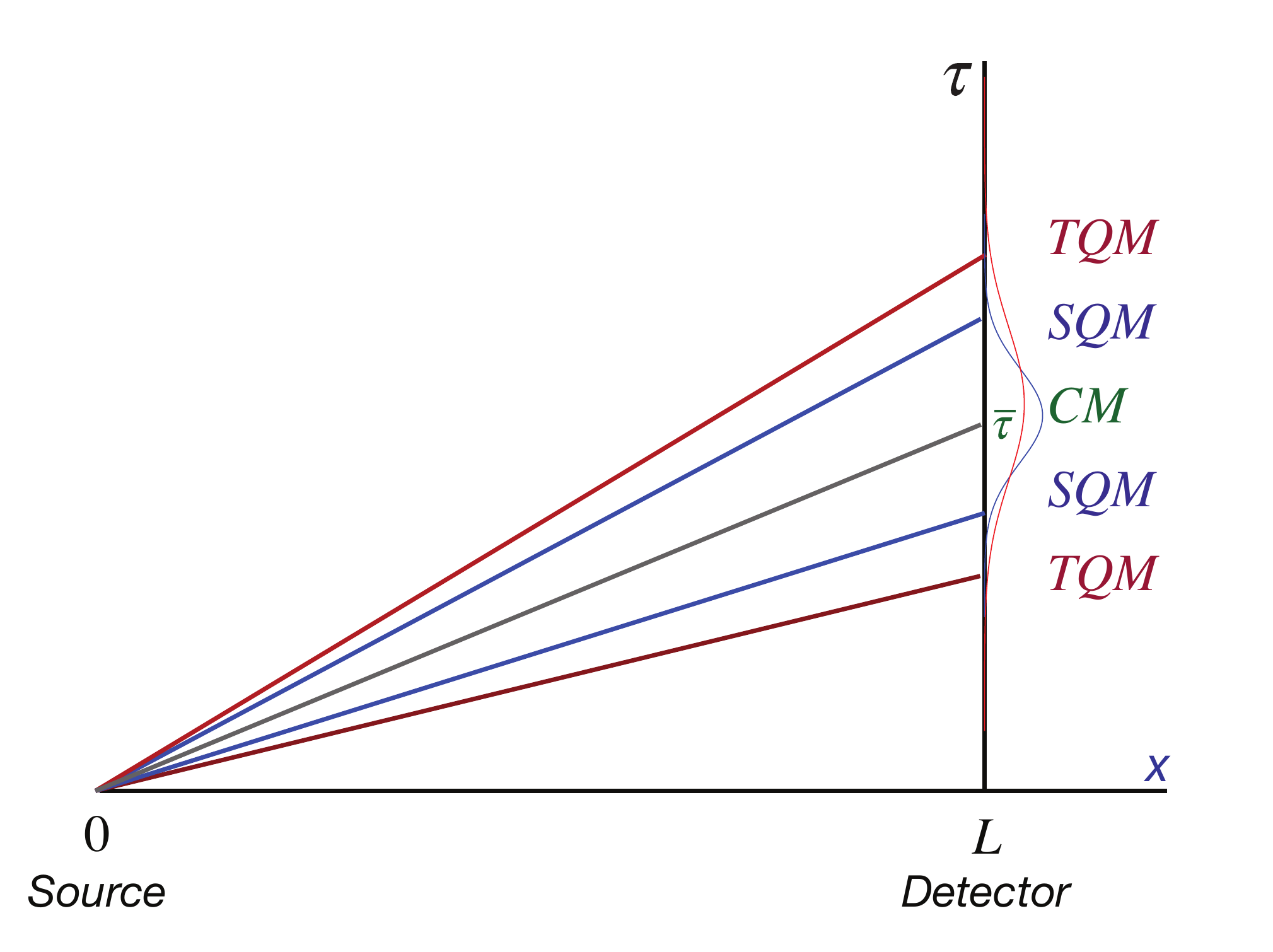} 
\par\end{centering}
\caption{Time of arrival\label{fig:free-toa-toa}}
\end{figure}

We start with a particle with initial position $x=0$ and with average
momentum (in the $x$ direction) of $p_{0}$. We will assume that
the initial dispersion in momentum is small. We have the wave function
from above. The probability density is then:
\begin{equation}
\bar{\rho}_{\tau}\left(x\right)\equiv\left|\bar{\varphi}_{\tau}\left(x\right)\right|^{2}=\sqrt{\frac{1}{\pi\sigma_{x}^{2}\left|f_{\tau}^{\left(x\right)}\right|^{2}}}e^{-\frac{\left(x-\frac{p_{0}}{m}\tau\right)^{2}}{\sigma_{x}^{2}\left|f_{\tau}^{\left(x\right)}\right|^{2}}}
\end{equation}

We assume there is a detector at position $x=L$.  
If the particle is released at time $\tau=0$ the average time of arrival is:
\begin{equation}
\bar{\tau}_{D}\equiv\frac{mL}{p_{0}}
\end{equation}
We can write the difference from the average time as $\delta\tau\equiv\tau-\bar{\tau}$
We are interested in the uncertainty in time of arrival at the detector or the expectation of ${\delta\tau}^{2}$.

If we were looking at measurements of the $x$ position we would know how to proceed.  
We would compute:
\begin{equation}
\Delta x = \sqrt {\int {dx{{\left( {x - \bar x} \right)}^2}} {\rho _\tau }\left( x \right)} 
\end{equation}

We would like something not much more complex for the uncertainty in time.
And which makes sense for both SQM and TQM.

We will take an ad hoc approach here 
but then check it against the results of a more detailed analysis by Muga and Leavens\cite{Muga:2000nx}.

Normally we think of a wave function as something that evolves in time;
it is first a function of $\tau$, then of $x$.
But here we are not interested in the probability to be at $x$ at a specific time $\tau$;
we are interested in the probability to be at a specific clock time $\tau$ for a fixed $x=L$.
So we will rewrite the probability density as a function of $\tau$.

We will assume that we are dealing with a reasonably well-focused particle so we may use a paraxial approximation.
We can write $x = v\tau$ or $x = L + \delta x$.  We use this to rewrite the density function as a function of $\delta \tau$.
We take:
\begin{equation}
L + \delta x = v\left( {{{\bar \tau }_D} + \delta \tau } \right) \to \delta x =   v\delta \tau 
\end{equation}
We keep only terms up through second order in ${\delta\tau}$.

We rewrite the numerator in the density function in terms of $\delta\tau$:

\begin{equation}
\bar{\rho}_{\tau}\left(L\right)\approx\sqrt{\frac{1}{\pi\sigma_{x}^{2}\left|f_{\tau}^{\left(x\right)}\right|^{2}}}e^{-\frac{\left(v\delta\tau\right)^{2}}{\sigma_{x}^{2}\left|f_{\tau}^{\left(x\right)}\right|^{2}}}
\end{equation}
Since the numerator is already only of second order in $\delta\tau$ we need only 
keep the zeroth order in $\delta\tau$ in the denominator:
\begin{equation}
\sigma_{x}^{2}\left|f_{\tau}^{\left(x\right)}\right|^{2}=\sigma_{x}^{2}+\frac{\tau^{2}}{m^{2}\sigma_{x}^{2}}=\sigma_{x}^{2}+\frac{\left(\bar{\tau}+\delta\tau\right)^{2}}{m^{2}\sigma_{x}^{2}}\approx\frac{\left(\bar{\tau}+\delta\tau\right)^{2}}{m^{2}\sigma_{x}^{2}}\approx\frac{\bar{\tau}^{2}}{m^{2}\sigma_{x}^{2}}\label{eq:probdens-space}
\end{equation}
giving:
\begin{equation}
\bar{\rho}_{\delta\tau}\approx\sqrt{\frac{v^{2}m^{2}\sigma_{x}^{2}}{\pi\bar{\tau}^{2}}}e^{-\frac{v^{2}m^{2}\sigma_{x}^{2}}{\bar{\tau}^{2}}\left(\delta\tau\right)^{2}}
\end{equation}
We define an effective dispersion in time:
\begin{equation}
\bar{\sigma}_{\tau}\equiv\frac{1}{mv\sigma_{x}}\bar{\tau}\label{eq:bar-sigma-tau}
\end{equation}
And the probability of detection as:
\begin{equation}
\bar{\rho}_{\delta\tau}=\sqrt{\frac{1}{\pi\bar{\sigma}_{\tau}^{2}}}e^{-\frac{\left(\delta\tau\right)^{2}}{\bar{\sigma}_{\tau}^{2}}}\label{eq:probdens-space-1}
\end{equation}

This is normalized to one, centered on $\tau=\bar{\tau}$, and with  uncertainty:
\begin{equation}
\Delta\tau=\frac{1}{\sqrt{2}}\bar{\sigma}_{\tau}
\end{equation}

Particularly important is the inverse dependence on the velocity.
Intuitively if we have a slow moving (non-relativistic) particle,
it will take a long time to pass through by the $x$ position of the detector,
causing the associated uncertainty in time to be relatively large.

\paragraph{Comparison to a time-of-arrival operator\label{sec:Comparison-to-TOA}}

As a cross-check, we compare our treatment to the time-of-arrival
operator analysis in Muga and Leavens (who are
following Kijowski \cite{Kijowski:1974hi}). They give a probability
density in time of:
\begin{equation}
\rho\left(\tau\right)={\left|{\int\limits _{0}^{\infty}dp\sqrt{\frac{p}{m}}{e^{-\imath\frac{{{p^{2}}\tau}}{{2m}}}}\hat{\bar{\varphi}}\left(p\right)\exp\left({\imath pL}\right)}\right|^{2}}+{\left|{\int\limits _{-\infty}^{0}dp\sqrt{\frac{{-p}}{m}}{e^{-\imath\frac{{{p^{2}}\tau}}{{2m}}}}\hat{\bar{\varphi}}\left(p\right)\exp\left({-\imath pL}\right)}\right|^{2}}
\end{equation}
where $\hat{\bar{\varphi}}$ is an arbitrary momentum space wave function
normalized to one.

Assume:
\begin{equation}
{p_{0}}\gg{\sigma_{p}}
\end{equation}
so:
\begin{equation}
p\approx{p_{0}}
\end{equation}

If our wave functions are closely centered on $p$, the term with
negative momentum can be dropped or even flipped in sign without effect
on the value of the integral. Further, by comparison to the exponential
part, the term under the square root is roughly constant:
\begin{equation}
\sqrt{\frac{p}{m}}\approx\sqrt{\frac{p_{0}}{m}}
\end{equation}

So we may replace Muga and Leaven\textsc{'}s expression by the simpler\label{toa-positive-freq}:
\begin{equation}
\rho\left(\tau\right)\approx\frac{{p_{0}}}{m}{\left|{\int\limits _{-\infty}^{\infty}dp{e^{-\imath\frac{{{p^{2}}\tau}}{{2m}}}}\hat{\bar{\varphi}}\left(p\right)\exp\left({\imath pL}\right)}\right|^{2}}
\end{equation}

As the contents of the integral are the Fourier transform of the (clock)
time dependent momentum space wave function, it is the clock time
dependent space wave function at $x=L$:
\begin{equation}
\rho\left(\tau\right)\approx\frac{{p_{0}}}{m}{\left|{{{\bar{\varphi}}_{\tau}}\left(L\right)}\right|^{2}}
\end{equation}

We are now in $x$ space.
To get to $\tau$ space, we take:
\begin{equation}
\int {\frac{{{p_0}}}{m}dx}  = \int {vdx}  = \int {d\tau } 
\end{equation}
and the probability density from Muga and Leavens is now identical to ours.

\subsubsection{Time of arrival in TQM\label{subsec:free-toa-tqm}}

It is striking that there is considerable uncertainty in time even
when time is treated classically. Our hypothesized uncertainty in
time will be added to this pre-existing uncertainty.

Using the time wave function from equation \ref{eq:free-evolution-wf-t}
we have for the probability density in time:

\begin{equation}
\tilde{\rho}_{\tau}\left(t\right)=\sqrt{\frac{1}{\pi\sigma_{t}^{2}\left|f_{\tau}^{\left(t\right)}\right|^{2}}}e^{-\frac{1}{\sigma_{t}^{2}\left|f_{\tau}^{\left(t\right)}\right|^{2}}\left(t-\tau\right)^{2}}
\end{equation}
We multiply by the space part from above to get the full probability
density:
\begin{equation}
\rho_{D}\left(t,L\right)=\tilde{\rho}_{D}\left(t\right)\bar{\rho}_{D}\left(L\right)
\end{equation}
If $t$ were replaced by space dimension $y$, we would have no doubt
as to how to proceed. To get the overall uncertainty in $y$ we would
integrate over clock time:
\begin{equation}
\left(\Delta y\right)^{2}=\int dy\left(y-\bar{y}\right)^{2}\int d\tau\rho_{\tau}\left(y\right)\bar{\rho}_{\tau}\left(L\right)
\end{equation}
Therefore we write (taking $y\to t$):

\begin{equation}
\left(\Delta t\right)^{2}=\int dt\left(t-\bar{\tau}\right)^{2}\int d\tau\tilde{\rho}_{\tau}\left(t\right)\bar{\rho}_{\tau}\left(L\right)\label{eq:free-toa-convolute}
\end{equation}
This is a convolution of clock time with coordinate time. To solve
we first invoke the same approximations as above:
\begin{equation}
\begin{array}{c}
\sigma_{t}^{2}\left|f_{\tau}^{\left(t\right)}\right|^{2}=\sigma_{t}^{2}+\frac{\tau^{2}}{m^{2}\sigma_{t}^{2}}=\sigma_{t}^{2}+\frac{\left(\bar{\tau}+\delta\tau\right)^{2}}{m^{2}\sigma_{t}^{2}}\approx\frac{\left(\bar{\tau}+\delta\tau\right)^{2}}{m^{2}\sigma_{t}^{2}}\approx\frac{\bar{\tau}^{2}}{m^{2}\sigma_{t}^{2}}\\
\tilde{\sigma}_{\tau}\equiv\frac{\bar{\tau}}{m\sigma_{t}}\\
\tilde{\rho}_{\tau}\left(t\right)\approx\sqrt{\frac{1}{\pi\tilde{\sigma}_{\tau}^{2}}}e^{-\frac{1}{\tilde{\sigma}_{\tau}^{2}}\left(t-\tau\right)^{2}}
\end{array}
\end{equation}
So we have for the full probability distribution in $t$:
\begin{equation}
\begin{array}{c}
\rho_{\bar{\tau}}\left(t\right)\equiv\int d\tau\sqrt{\frac{1}{\pi\tilde{\sigma}_{\tau}^{2}}}e^{-\frac{1}{\tilde{\sigma}_{\tau}^{2}}\left(t-\tau\right)^{2}}\sqrt{\frac{1}{\pi\bar{\sigma}_{\tau}^{2}}}e^{-\frac{1}{\bar{\sigma}_{\tau}^{2}}\left(\tau-\bar{\tau}\right)^{2}}\\
\left(\Delta t\right)^{2}=\int dt\left(t-\bar{\tau}\right)^{2}\rho_{\bar{\tau}}\left(t\right)
\end{array}
\end{equation}
The convolution over $\tau$ is trivial giving:
\begin{equation}
{{\rho_{\bar{\tau}}}\left(t\right)=\sqrt{\frac{1}{{\pi\sigma_{\tau}^{2}}}}{e^{-\frac{{{\left({t-\bar{\tau}}\right)}^{2}}}{{\sigma_{\tau}^{2}}}}}}
\end{equation}
with the total dispersion in clock time being the sum of the dispersions in coordinate time and in space:
\begin{equation}
{\sigma_{\tau}^{2}\equiv\tilde{\sigma}_{\tau}^{2}+\bar{\sigma}_{\tau}^{2}}
\end{equation}
This is intuitively reasonable. 

The uncertainty is:
\begin{equation}
\Delta\tau=\frac{1}{{\sqrt{2}}}\sqrt{\tilde{\sigma}_{\tau}^{2}+\bar{\sigma}_{\tau}^{2}}
\end{equation}
We collect the
definitions for the two dispersions:
\begin{equation}
\begin{array}{c}
\bar{\sigma}_{\tau}^{2}=\frac{{{\bar{\tau}}^{2}}}{{{m^{2}}{v^{2}}\sigma_{x}^{2}}}\hfill\\
\tilde{\sigma_{t}}^{2}\approx\frac{{{\bar{\tau}}^{2}}}{{{m^{2}}\sigma_{t}^{2}}}\hfill
\end{array}
\end{equation}

From the long, slow approximation, we would expect particle wave functions
to have initial dispersions in energy/time comparable to their dispersions
in momentum/space. $\sigma_{t}\sim\sigma_{x}$. But the conventional
contribution has an additional $\frac{1}{v}$ in it. Since in the
non-relativistic case, $v\ll1$ the total uncertainty will be dominated
by the space part.

This helps to explain why dispersion in time has not been seen by
accident. It also motivates an exploration of the relativistic case,
where the effects of dispersion in time should be at least comparable
to the effects of dispersion in space (section \ref{sec:Multiple-particle-case}).

\section{Single and double slit experiments\label{sec:slits}}

There is already a significant literature on the ``in time'' versions
of the single and double slit experiments. The investigation of this
problem started over sixty years ago with Moshinsky \cite{Moshinsky:1951qu,Moshinsky:1952zt}
and continues, with a recent review by Gerhard and Paulus \cite{Gerhard-G.-Paulus:2009zr}.
Particularly interesting for our purposes are treatments of scattering
of wave functions, as Umul \cite{Umul:2008xy} and Marchewka and
Schuss \cite{Marchewka:1998ci,Marchewka:1999nl}.

Neither single nor double slit has, to the best of our knowledge, been solved exactly.
Approximations are necessary. Further the SQM and TQM branches
have to be approximated in a way that lets us compare the two branches
directly. We will treat the single slit first, then the double.

\subsection{Single slit in time}

The single slit in time experiment provides
the decisive test of temporal quantum mechanics. 

In SQM, the narrower
the slit, the \textit{less} the dispersion in subsequent time-of-arrival
measurements. In TQM, the narrower the slit, the \textit{greater}
the subsequent dispersion in subsequent time-of-arrival measurements.
\textsl{In principle, the difference may be made arbitrarily great.}

This distinction follows directly from the fundamental principles
of quantum mechanics. Picture a quantum wave function going through
a gate in space. If the gate is wide, diffraction by the edges is
minimal and the subsequent broadening of the wave function minimal.
The gate will clip the beam around the edges and that will be about
it. But if the gate is narrow, then the wave function will spread
in a nearly circular pattern and the subsequent broadening will be
arbitrarily great.

In terms of the uncertainty principle, the gate represents a measurement
of the position. The narrower the gate, the less the uncertainty at
the gate. If $\Delta{y}$ is small, then $\Delta p_{y}$ must be correspondingly
large and the resulting spread greater at the detector. As $\Delta y\to0\Rightarrow\Delta{p_{y}}\to\infty$.
But a large $\Delta p_{y}$ implies -- with a bit of time -- a large
spread at the detector.

We translate this from space to time. We will start with a beam moving
from left to right in the $x$ direction, going through an extremely fast
camera shutter, and arriving at a detector. In both SQM and TQM, the
faster the shutter the smaller $\Delta t$. But in SQM the beam is
clipped and the dispersion at the detector correspondingly reduced.
While in TQM the smaller the $\Delta t$ the correspondingly greater the
$\Delta E$. The greater the $\Delta E$, the greater the dispersion in
velocities and the greater the dispersion in time-of-arrival at the
detector.

The distribution of detections in clock time will give us the dispersion
in time-of-arrival, which is the key measurement. 

We will ignore paths that loop back and forth through the gate. Most
elementary treatments make this assumption of a single passage (for
an analysis of the effects of multiple passages of a gate see Yabuki,
Raedt, and Sawant \cite{Yabuki-1986,PhysRevA.85.012101,Sawant:2013yu}).

We will also, following Feynman and Hibbs \cite{Feynman:1965bh},
take the gate as having a Gaussian shape, rather then turning on and
off instantly. This is more realistic and avoids some distracting
mathematical complexities that result from using hard-edged gates.

Note also, in space we can make the gate entirely perpendicular to
the beam. Beam traveling in $x$, gate in $y$. The beam can start
with zero momentum in the $y$ direction, letting the $x$ momentum
act as a carrier. But there is no such thing as a particle that does
not have at least some momentum in time, i.e. energy.  
Therefore it is difficult
to achieve a complete 
separation between the measurement of $p_{x}$ and $E$.
Here we will use increased dispersion in the time-of-arrival measurements as the test of TQM.
(Below we will discuss briefly a way to separate the measurements of uncertainty in space and in time:  subsubsection
\ref{paragraph:mul-abs-two-different}.)

As usual we look first at SQM, then TQM.
\subsubsection{Single slit in SQM\label{subsec:single-sqm}}

\begin{figure}[h]
\begin{centering}
\includegraphics[width=10cm]{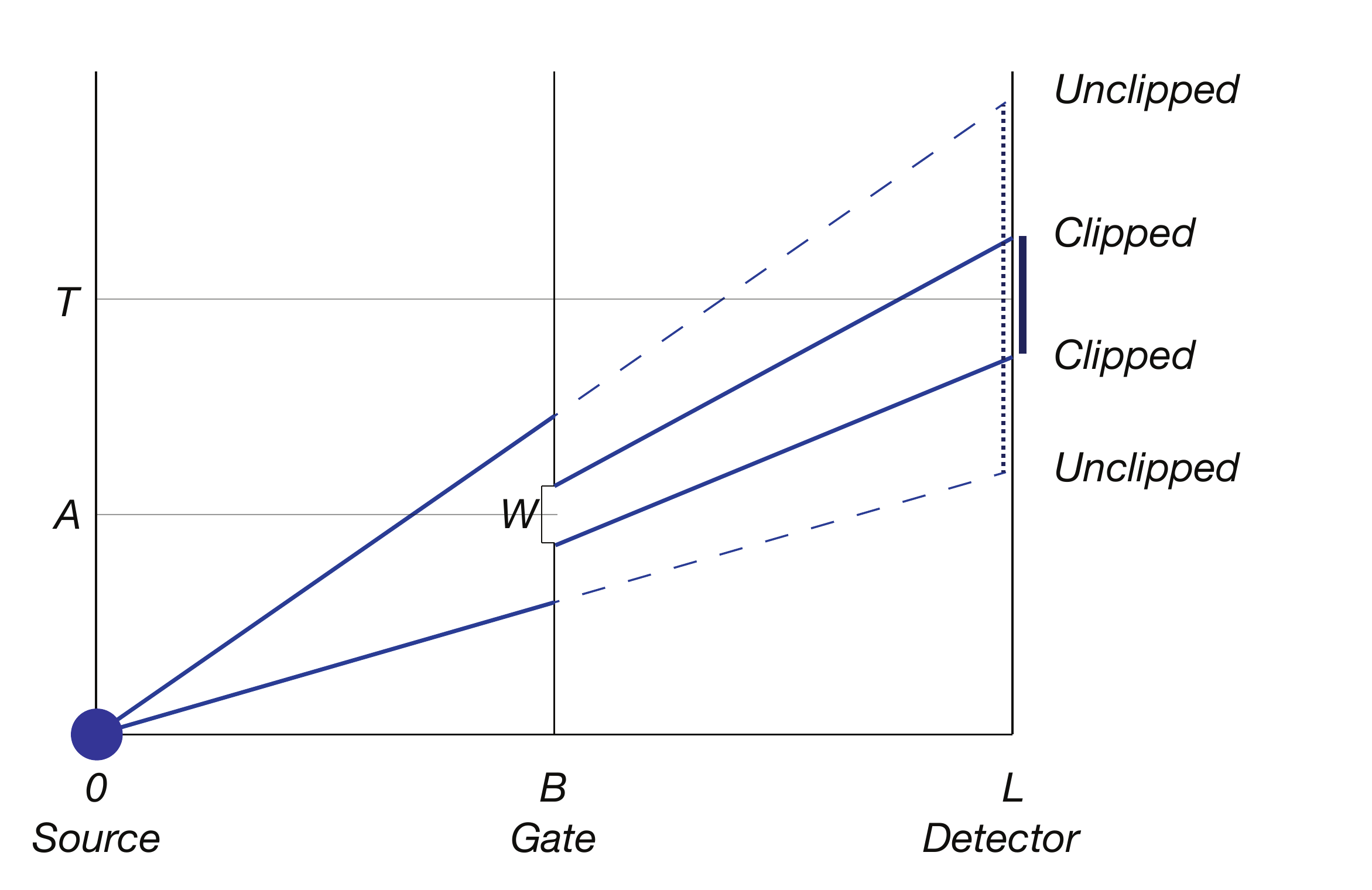} 
\par\end{centering}
\caption{Single slit in SQM\label{fig:free-toa-sqm}}
\end{figure}

We start by calculating the effects of a single slit in time on the
dispersion of time-of-arrival measurements in SQM.

The gate is located at $x=B$, centered on clock time $A$, with width
in time $W$:

\begin{equation}
\overline{G}_{\tau}=e^{-\frac{\left(\tau-A\right)^{2}}{2W^{2}}}
\end{equation}

We start with a wave function in $x$:
\begin{equation}
\bar{\varphi}_{\tau}\left(x\right)=\sqrt[4]{\frac{1}{\pi\sigma_{x}^{2}}}\sqrt{\frac{1}{f_{\tau}^{\left(x\right)}}}e^{\imath p_{0}x-\frac{1}{2\sigma_{x}^{2}f_{\tau}^{\left(x\right)}}\left(x-x_{0}-\frac{p_{0}}{m}\tau\right)^{2}-\imath\frac{p_{0}^{2}}{2m}\tau}
\end{equation}
or in $p$:
\begin{equation}
{{\hat{\bar{\varphi}}}_{\tau}}\left(p\right)=\sqrt[4]{{\frac{1}{{\pi\sigma_{p}^{2}}}}}{e^{-\imath p{x_{0}}-\frac{{{\left({p-{p_{0}}}\right)}^{2}}}{{2\sigma_{p}^{2}}}-\imath\frac{{p^{2}}}{{2m}}\tau}}
\end{equation}

We take the initial position $x_{0}=0$, initial velocity $v\equiv\frac{p_{0}}{m}$. We assume the particle
is non-relativistic so that $v\ll1.$

We assume that the incoming wave function can be treated as a sum
of $p$ rays. This worked in the time-of-arrival case, has the merit
of simplicity, and lets us make a direct comparison between TQM and
SQM. 

The evolution of the wave function is simplest in the $p$ basis,
but the gate is defined in the $\tau$ basis.   
Therefore we will need to shift back and forth between the $p$ and the $\tau$ basis.
To do this we use:
\begin{equation}
p=m\frac{x}{\tau}
\end{equation}
with the ancillary definitions of $\delta p$ and $\delta \tau$:
\begin{equation}
\begin{array}{c}
p=p_{0}+\delta p=mv+\delta p\hfill\\
\tau=\bar{\tau}+\delta\tau\hfill
\end{array}
\end{equation}

For simplicity we center the particle beam on the gate. If $\bar{\tau}_{G}$
is the average time at which the particle reaches the gate, we arrange
it so that:
\begin{equation}
\overline{\tau}_{G}=A
\end{equation}

With the detector at position $x=L$, we define $T$
as the average time at which the particle is detected. 
This gives:
\begin{equation}
v=\frac{B}{A}=\frac{L}{T}
\end{equation}

We assume the beam is well-focused and use a paraxial
approximation. To quadratic order:
\begin{equation}
\begin{array}{c}
\delta p=m\frac{x}{{\bar{\tau}+\delta\tau}}-m\frac{x}{{\bar{\tau}}}\approx mv\left({-\frac{{\delta\tau}}{{\bar{\tau}}}+\frac{{\delta{\tau^{2}}}}{{{\bar{\tau}}^{2}}}}\right)\hfill\\
\delta\tau=m\frac{x}{{p_{0}+\delta p}}-m\frac{x}{p_{0}}\approx\frac{x}{v}\left({-\frac{{\delta p}}{{p_{0}}}+\frac{{\delta{p^{2}}}}{{p_{0}^{2}}}}\right)\hfill
\end{array}
\end{equation}

To first order:
\begin{equation}\frac{{\delta p}}{{{p_0}}} \approx  - \frac{{\delta \tau }}{{\bar \tau }}\end{equation}
$\delta p$ and $\delta \tau$ have opposite signs; faster means earlier, slower means later.

If the gate is much wider than the wave function then the particle will pass through unscathed.
If the gate is narrower than the wave function then the particle will clipped to the width of the gate.

The wave function just before the gate (using $p = \frac{{mB}}{{A + \delta \tau }}$) is:
\begin{equation}
{{\bar \varphi }_{{G^{\left(  pre  \right)}}}}\left( {\delta \tau } \right) = \sqrt[4]{{\frac{1}{{\pi \sigma _G^2}}}}{e^{ - \frac{{\delta {\tau ^2}}}{{2\sigma _G^2}} - \imath \frac{1}{{2m}}{{\left( {\frac{{mB}}{{A + \delta \tau }}} \right)}^2}\left( {A + \delta \tau } \right)}}
\end{equation}
with $\sigma_{G}$ defined by:
\begin{equation}{\sigma _G} \equiv A\frac{{{\sigma _p}}}{{{p_0}}}\end{equation}

Note that the wave function is correctly normalized for integration over $\tau$.
And that to lowest order:
\begin{equation}\frac{{\delta \tau }}{{{\sigma _G}}} =  - \frac{{\delta p}}{{{\sigma _p}}}\end{equation}
$\sigma_{G}$ is the measure of how wide the beam is in clock time when it reaches the gate.  It is linear in $A$ and proportional to $\sigma_{p}$.

For the SQM case, we assume no diffraction in time: the wave function
will be clipped by the gate, not diffracted. This means that to get the wave function post-gate
we must multiply the wave function pre-gate by the gate function:
\begin{equation}
{{\bar{\varphi}}_{{G^{\left(post\right)}}}}\left({\delta\tau}\right)={e^{-\frac{{{\left({\delta\tau}\right)}^{2}}}{2{W^{2}}}}}{{\bar{\varphi}}_{{G^{\left(pre\right)}}}}\left({\delta\tau}\right)
\end{equation}

Now we convert back to $p$ space, but on the far side of the gate:
\begin{equation}
{{\hat{\bar{\varphi}}}_{{G^{\left(post\right)}}}}\left(p\right)=\sqrt[4]{{\frac{1}{{\pi\sigma_{p}^{2}}}}}{e^{-\frac{{\delta{p^{2}}}}{{2\sigma_{p}^{2}}}-\frac{{A^{2}}}{{2{W^{2}}}}\frac{{\delta{p^{2}}}}{{p_{0}^{2}}}-\imath\frac{{p^{2}}}{{2m}}{\tau_{G}}}}
\end{equation}

The evolution from gate to detector only adds a phase:
\begin{equation}{{\hat {\bar {\varphi}} }_D}\left( p \right) = \sqrt[4]{{\frac{1}{{\pi \sigma _p^2}}}}{e^{ - \frac{{\delta {p^2}}}{{2\sigma _p^2}} - \frac{{\delta {p^2}}}{{2p_0^2}}\frac{{{A^2}}}{{{W^2}}} - \imath \frac{{{p^2}}}{{2m}}{\tau _D}}}\end{equation}

We again switch  to $\tau$ space at the detector. 
The significant
change is from ${\sigma_{p}}\to{{\sigma'}_{p}}$ where the primed
dispersion in $p$ is:
\begin{equation}\frac{1}{{{{\left( {{{\sigma '}_p}} \right)}^2}}} = \frac{1}{{\sigma _p^2}} + \frac{{{A^2}}}{{{W^2}p_0^2}}\end{equation}

With this we can write out the wave function at the detector by inspection.
In $p$ space:
\begin{equation}
{{\hat{\bar{\varphi}}}_{D}}\left(p\right)=\sqrt[4]{{\frac{1}{{\pi\sigma_{p}^{2}}}}}{e^{-\frac{{\delta{p^{2}}}}{{\left({{\sigma'}_{p}}\right)}^{2}}-\imath\frac{{p^{2}}}{{2m}}{\tau_{D}}}}
\end{equation}
In $\tau$ space:
\begin{equation}
{{\bar{\varphi}}_{D}}\left({\delta\tau}\right)=\sqrt[4]{{\frac{1}{{\pi\sigma_{G}^{2}}}}}{e^{-\frac{{\delta{\tau^{2}}}}{{\left({{\sigma'}_{G}}\right)}^{2}}-\imath\frac{{m{L^{2}}}}{{2T}}\left({1-\frac{{\delta\tau}}{T}+\frac{{\delta{\tau^{2}}}}{{T^{2}}}}\right)}}
\end{equation}

The primed dispersion in clock time at the detector is a scaled
version of the dispersion in momentum post gate:
\begin{equation}
{{\sigma'}_{G}}\equiv T\frac{{{\sigma'}_{p}}}{{p_{0}}}
\end{equation}

Note the factor $\sqrt[4]{{\frac{1}{{\pi\sigma_{G}^{2}}}}}$ is unchanged.
The ratio:
\begin{equation}
{\left({\frac{{{\sigma'}_{G}}}{{\sigma_{G}}}}\right)^{2}}
\end{equation}
tells us what percentage of the particles get through the gate. But
this does not affect the calculation of the dispersion, which is normalized. 

The effective dispersion of the wave function is proportional to $T$. 
It is useful to scale that out.
The result is controlled by the ratio of the angular width of the beam in momentum space ${\theta _p} \equiv \frac{{{\sigma _p}}}{{{p_0}}}$
to the angular width of the gate in time ${\theta _G} \equiv \frac{W}{A}$:
\begin{equation}\frac{{{{\left( {{{\sigma '}_G}} \right)}^2}}}{{{T^2}}} = \frac{{\theta _p^2\theta _G^2}}{{\theta _p^2 + \theta _G^2}}\end{equation}

When the gate is wide open, the dispersion at the detector is proportional to the initial dispersion of the beam:
\begin{equation}W \to \infty  \Rightarrow \frac{{{{\sigma '}_G}}}{T} \to \frac{{{\sigma _p}}}{{{p_0}}}\end{equation}

But when the gate is much narrower than the beam, the dispersion at
the detector proportional to the  width of the gate:
\begin{equation}W \to 0 \Rightarrow \frac{{{{\sigma '}_G}}}{T} \to \frac{W}{A}\end{equation}

The narrower the gate, the narrower the beam.

While the approximations we have used have been simple, this is a
fundamental implication of the SQM view of time, of time as classical.
In SQM, wave functions are not, by assumption, diffracted by a gate,
they are clipped. And therefore their dispersion in time must be reduced
by the gate rather than increased by it. 

And we have therefore the uncertainty in time-of-arrival:
\begin{equation}\frac{{\Delta \tau }}{T} = \frac{1}{{\sqrt 2 }}\frac{{{{\sigma '}_G}}}{T}\end{equation}

\subsubsection{Single slit in TQM\label{subsec:single-tq}}

We now have a baseline from SQM; we turn to the single slit in time
in TQM using the SQM treatment as a starting point.

\begin{figure}[h]
\begin{centering}
\includegraphics[width=10cm]{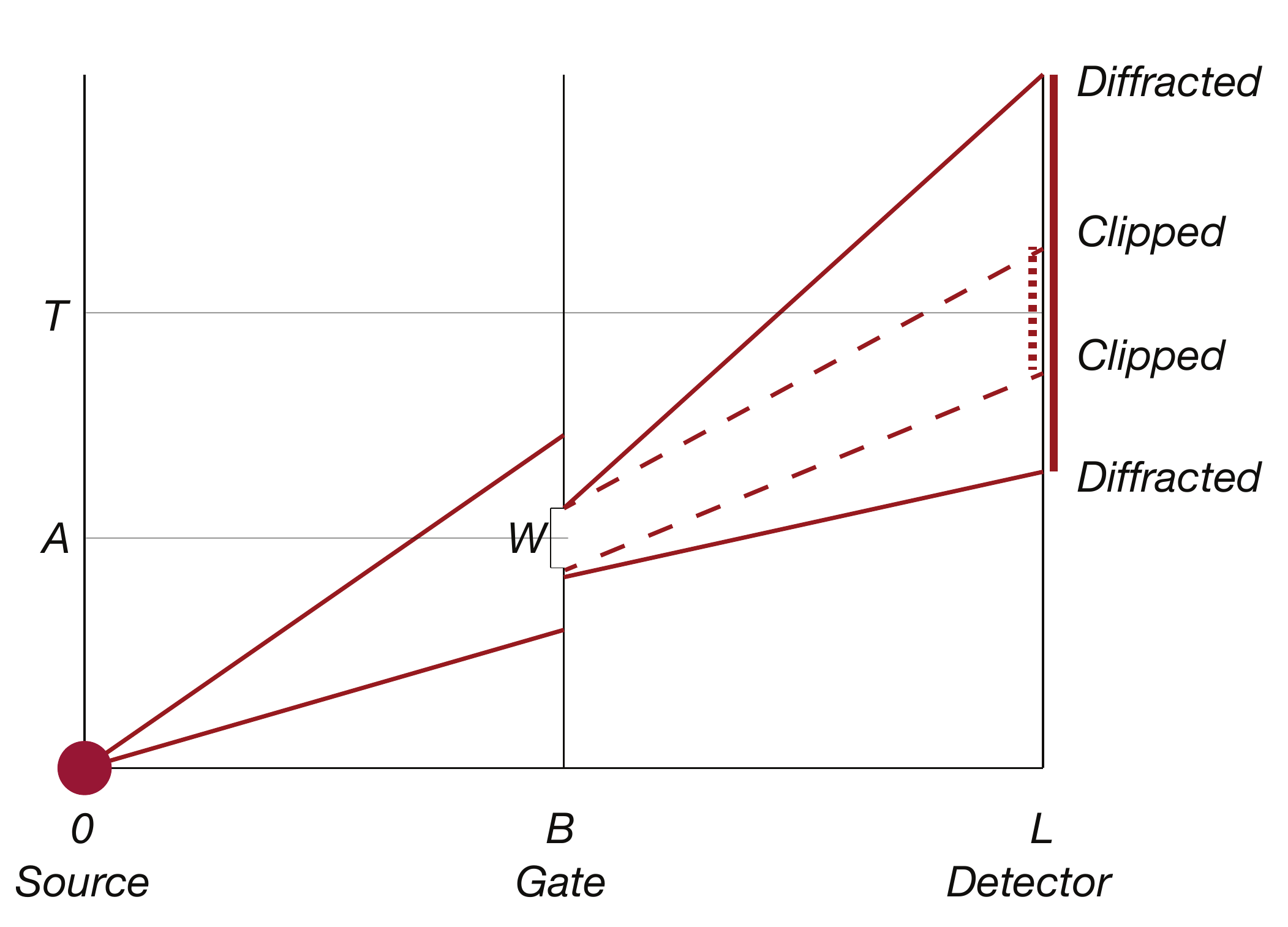} 
\par\end{centering}
\caption{Single slit in TQM\label{fig:free-toa-tq}}
\end{figure}

By assumption, the gate will act only on the time part:
\begin{equation}
\widetilde{G}_{t}=e^{-\frac{\left(t-A\right)^{2}}{2W^{2}}}
\end{equation}

For TQM, we start with a wave function factored in time and momentum:
\begin{equation}
{\psi_{\tau}}\left(t\right)={{\tilde{\varphi}}_{\tau}}\left(t\right){{\hat{\bar{\varphi}}}_{\tau}}\left(p\right)
\end{equation}

We are treating the $p$ part as a carrier, using the wave function
from above. The momentum part will be unchanged throughout.

We take the particle as starting with $t_{0}=0$. The wave function
pre-gate is:
\begin{equation}{{\tilde \varphi }_\tau }\left( t \right) = \sqrt[4]{{\frac{1}{{\pi \sigma _t^2}}}}\sqrt {\frac{1}{{f_\tau ^{\left( t \right)}}}} {e^{ - \imath {E_0}t - \frac{1}{{2\sigma _t^2f_\tau ^{\left( t \right)}}}{{\left( {t - \frac{{{E_0}}}{m}\tau } \right)}^2} + \imath \frac{{E_0^2}}{{2m}}\tau }}\end{equation}

We will again take the particle as non-relativistic: $E_{0}$ such
that $\frac{E_{0}}{m}\approx1$.

On arrival at the gate the time part of the wave function is:
\begin{equation}{{\tilde \varphi }_{{G^{\left(  pre \right)}}}}\left( {{t_G}} \right) = \sqrt[4]{{\frac{1}{{\pi \sigma _t^2}}}}\sqrt {\frac{1}{{f_G^{\left( t \right)}}}} \exp \left( { - \imath {E_0}{t_G} - \frac{{{{\left( {{t_G} - A} \right)}^2}}}{{2\sigma _t^2\left( {1 - \imath \frac{A}{{m\sigma _t^2}}} \right)}} + \imath \frac{{E_0^2}}{{2m}}A} \right)\end{equation}

Post gate:
\begin{equation}
{{\tilde{\varphi}}_{{G^{\left(post\right)}}}}\left({t_{G}}\right)={e^{-\frac{{{\left({{t_{G}}-A}\right)}^{2}}}{{2W{^{2}}}}}}{{\tilde{\varphi}}_{{G^{\left(pre\right)}}}}\left({t_{G}}\right)
\end{equation}

The effect of the gate on the wave function is to rescale it:
\begin{equation}
\frac{1}{2\sigma_{t}^{2}\left(1-\imath\frac{A}{m\sigma_{t}^{2}}\right)}\to\frac{1}{2W^{2}}+\frac{1}{2\sigma_{t}^{2}\left(1-\imath\frac{A}{m\sigma_{t}^{2}}\right)}
\end{equation}

We define rescaling constants $\sigma _t^{*2},{\tau ^*}$ by:
\begin{equation}\frac{1}{{{W^2}}} + \frac{1}{{\sigma _t^2 - \imath \frac{A}{m}}} = \frac{1}{{\sigma _t^{*2} - \imath \frac{{{\tau ^*}}}{m}}}\end{equation}

The effect of the gate is to change the shape of the wave function so that it looks \textit{as if} it had started at time $A-\tau^{*}$ with width $\sigma^{*}$.
This parameterization makes it easy to see what the gate is doing to the wave function.

We have by inspection that 
as the width of the gate goes to infinity, it lets the wave function through unchanged:
\begin{equation}
W\to\infty\Rightarrow{{\sigma^{*} _t}}\to{\sigma _t},{\tau ^*}\to A
\end{equation}

While as the gate gets narrow, the effective width of the wave function in time goes to the width of the gate:
\begin{equation}
W\to0\Rightarrow{{\sigma^{*} _t}}\to W,{\tau ^*}\to0
\end{equation}

To compute the rescaling factors between these limits we clear the denominators:
\begin{equation}\left( {\sigma _t^2 - \imath \frac{A}{m}} \right)\left( {\sigma _t^{*2} - \imath \frac{{{\tau ^*}}}{m}} \right) + {W^2}\left( {\sigma _t^{*2} - \imath \frac{{{\tau ^*}}}{m}} \right) = {W^2}\left( {\sigma _t^2 - \imath \frac{A}{m}} \right)\end{equation}

And equate the real and imaginary parts:
\begin{equation}\begin{array}{c}
  \left( {\sigma _t^2 + {W^2}} \right)\sigma _t^{*2} - \frac{A}{m}\frac{{{\tau ^*}}}{m} = {W^2}\sigma _t^2 \hfill \\
   - \frac{A}{m}\sigma _t^{*2} - \left( {\sigma _t^2 + {W^2}} \right)\frac{{{\tau ^*}}}{m} =  - {W^2}\frac{A}{m} \hfill \\ 
\end{array} \end{equation}

This gives us a two by two matrix equation for $\sigma _t^{*2},{\tau ^*}$ which we invert and apply to the right hand side:
\begin{equation}\left( {\begin{array}{*{20}{c}}
  {\sigma _t^{*2}} \\ 
  {\frac{{{\tau ^*}}}{m}} 
\end{array}} \right) = \frac{{{W^2}}}{D}\left( {\begin{array}{*{20}{c}}
  {\sigma _t^2 + {W^2}}&{ - \frac{A}{m}} \\ 
  { - \frac{A}{m}}&{ - \left( {\sigma _t^2 + {W^2}} \right)} 
\end{array}} \right)\left( {\begin{array}{*{20}{c}}
  {\sigma _t^2} \\ 
  { - \frac{A}{m}} 
\end{array}} \right)\end{equation}
with determinant:
\begin{equation}D = {\left( {\sigma _t^2 + {W^2}} \right)^2} + \frac{{{A^2}}}{{{m^2}}}\end{equation}
getting:
\begin{equation}\begin{array}{c}
  \sigma _t^{*2} = \frac{{{W^2}}}{D}\left( {\left( {\sigma _t^2 + {W^2}} \right)\sigma _t^2 + \frac{{{A^2}}}{{{m^2}}}} \right) \hfill \\
  {\tau ^*} = \frac{A}{D}{W^4} \hfill \\ 
\end{array} \end{equation}

As a double check, we see we get the expected limits for $\sigma _t^{*2}$ and ${\tau ^*} $ as $W\to\infty,W\to0$.
 
$\sigma _t^{*2}$ and ${\tau ^*} $ in turn gives us the $t$ wave function at the detector. We replace
$\sigma_{t}$ and $\tau_{G}$ in the initial wave function by $\sigma^{*}_{t}$
and ${\tau ^*}$:
\begin{equation}{{\tilde \varphi }_D}\left( {{t_D}} \right) = N\sqrt[4]{{\frac{1}{{\pi \sigma _t^{*2}}}}}\sqrt {\frac{1}{{{{f'}_G}}}} {e^{ - \imath {E_0}{t_D} - \frac{1}{{2\sigma _t^{*2}{{f'}_D}}}{{\left( {{t_D} - {{\bar t}_D}} \right)}^2} + \imath \frac{{E_0^2}}{{2m}}{\tau _D}}}\end{equation}
with:
\begin{equation}\begin{array}{c}
  {{f'}_D} \equiv 1 - \imath \frac{{{\tau ^*} + \left( {T - A} \right)}}{{m\sigma _t^{*2}}} \hfill \\
  N \equiv \sqrt[4]{{\frac{{\sigma _t^{*2}}}{{\sigma _t^2}}}}\sqrt {\frac{{{{f'}_G}}}{{{f_G}}}}  \hfill \\ 
\end{array} \end{equation}

The value of $N$ drops out in the normalization when we calculation the uncertainty.

The use of rescaling constants makes clear that the effect of the gate is to reset the wave function, changing its shape and leaving no other memory of the wave function before the gate. 
The wave function gets a fresh start with the gate.

Of particular interest here is the limit $W\to0$:  an extremely narrow gate -- an extremely fast camera shutter -- resets 
the wave function to have a width of order width of the gate.

And therefore, given the Heisenberg uncertainty principle, the uncertainty in energy post-gate goes as $\frac{1}{W}$.
And this in turn forces the wave to diverge more rapidly in time than would otherwise have been the case.

At large distances from the gate the dispersion of the wave function in time $\tilde \sigma _\tau ^2 \equiv \sigma _t^{*2}{{f'}_\tau }$ goes as:
\begin{equation}{{\tilde \sigma }_\tau }\sim\frac{\tau }{m}{\sigma _E} = \frac{\tau }{{mW}}\end{equation}

We can reapply the analysis from above (subsection \ref{subsec:free-toa-sqm})
to get:
\begin{equation}\Delta \tau  = \frac{1}{{\sqrt 2 }}\sqrt {\tilde \sigma _\tau ^2 + \bar \sigma _\tau ^2}  = \frac{1}{{\sqrt 2 }}\frac{T}{m}\sqrt {\frac{1}{{{W^2}}} + \frac{1}{{{v^2}\sigma _x^2}}} \end{equation}

So in principle by making $W$ small enough we can make the uncertainty in time of arrival as large as we wish.

Therefore: 
\begin{enumerate}
\item In SQM the uncertainty in clock time is \textit{directly} proportional to the width
of the gate. 
\item In TQM the uncertainty in clock time is \textit{inversely} proportional
to the width of the gate. 
\end{enumerate}
These effects are results of the fundamental assumptions. In SQM,
time is a parameter and the gate must \textsl{clip} the incoming wave
function. In TQM, time is an operator and the gate must \textsl{diffract}
the incoming wave function.

Therefore the observable effect may be made, in principle, \textsl{arbitrarily}
large. 

To go from observable in principle to observable in practice will require us to resolve two difficulties.

First we have to include the relativistic case:  it is clear that the $\frac{1}{v}$ factor will make it hard for the effects of uncertainty in time to compete with the effects of uncertainty in space.  This is a large part of the motivation for looking at the multiple particle case in the next section:  
relativistic velocities imply particle creation and therefore a need to understand what TQM looks like when there is more than one particle in play.

Secondly we need a way to separate measurements of uncertainty in time from measurements of uncertainty in space; 
an approach is suggested below (subsubsection \ref{paragraph:mul-abs-two-different}).

\subsection{Double slit in time\label{subsec:slits-double}}

The two principle effects we have been focused on are additional dispersion
in time and the uncertainty principle in time. We have not looked
at effects related to interference in time, mostly notably the double
slit in time experiment. 

The double slit experiment has been described by Feynman \cite{Feynman:1965ah}
as the ``only mystery'' of quantum mechanics. We therefore take
a look at it in the TQM context.

There have been many tests of the double slit experiment in space.
The double slit in time variation has been performed by Lindner et
al \cite{Lindner:2005vv}. In the Linder et al experiment argon atoms are ionized
using few-cycle electric pulses. If the electron is ionized by a single
peak, that scores as a single slit in time; if by two, that scores
as a double slit in time. The times are of order 500 attoseconds,
much greater then the times we have been concerned with. But the principles
of the experiment are of great interest here.

Horwitz \cite{Horwitz:2005ix} argues that the analysis in Lindner
et al is not properly characterized as interference in time, since
the analysis in Lindner et al is based on SQM, in which time is not
an operator. 

But Horwitz then shows you get the same diffraction spacing using
as a starting point the relativistic dynamics equation\footnote{Note the sign of the evolution parameter $s$ here is opposite from
the sign used in Land and Horwitz \cite{Land:1996aj}.}:
\begin{equation}
\imath\frac{\partial}{{\partial s}}{\psi_{s}}=-\frac{{{E^{2}}-{{\vec{p}}^{2}}}}{{2m}}{\psi_{s}}
\end{equation}

As noted above, this differs from our Schr\"{o}dinger equation by only
a gauge term. As gauge terms are not physically meaningful, Horwitz
has provided us with the diffraction pattern for TQM. 

And as Horwitz gets the same spacing as Lindner did, the double slit
in time experiment does not help us differentiate between SQM and
TQM, at least not to first order\footnote{To higher orders, a preliminary examination suggests that while the spacing peak-to-peak is unaffected,
the widths of the individual peaks are increased.}.
 And therefore -- somewhat surprisingly
-- the double slit in time experiment does not help us to falsify
TQM

And it follows in turn that it is not the double slit but the single slit experiment
which provides the decisive test of TQM.

\section{Multiple particles\label{sec:Multiple-particle-case}}
\begin{quotation}
``In what follows we assume that even though $A=\delta m$ is formally
divergent, it is still \`{ }small\'{ } in the sense that it is of
the order of 1/137 times the electron mass.'' -- J. J. Sakurai \cite{Sakurai:1967tl} 
\end{quotation}
\subsection{Why look at the multiple particle case?\label{subsec:mul-overview}}

In principle, we have enough to falsify TQM. Why then look at the
multiple particle case? 
\begin{enumerate}
\item From the above we see the effects of TQM will be largest at relativistic
velocities and short times. This implies we need to look at high energies
and, since high energies imply particle creation, at the multiple
particle case. (We thank Dr. Steve Libby for bringing this point to
our attention.) 
\item There is the further concern that TQM may not be renormalizable. If
the loop integrals in SQM are barely renormalizable, with the addition
of one more dimension the loop integrals in TQM may well become completely
intractable. 
\item Extending TQM to the multiple particle case opens up some new effects. 
\item And then there is the intrinsic interest of the question. 
\end{enumerate}

At the same time, there are formidable difficulties: the literature
on quantum field theory is vast and complex (among the references
we have found helpful are \cite{Bjorken:1965el,Bjorken:1965mo,Roman:1969lr,Rivers:1987ma,Ramond:1990ab,Kaku:1993xj,Peskin:1995rv,Weinberg:1995hu,Huang:1998bd,Weinberg:1995rl,Itzykson:2005cv,Maggiore:2005hc,Zee:2010oy,Nazarov:2013uq,Das:2014fk,Lancaster:qy,Horwitz:2015jk}).

We will focus on analyzing a toy model in a series of toy
situations -- but in a way that will make clear how to extend the
toy model to more realistic and useful cases.

Since the complications of spin are inessential here, we will look
at a simple model with three massive spinless particles $A,$ $B$,
and $C$ (correspondingly very loosely to the electron, photon, and
proton).

As in the single particle case, we will first work out the rules for
the various pieces of the path integral, then apply these rules to
a few simple cases. There is no possibility of covering
all cases, but we will do enough to make clear how to extend the approach
to an arbitrary problem. 

We will require, as before, manifest covariance and consistency with established results.
We will also require that the treatment of the single and multiple cases be consistent; the single particle case should appear 
at the end to be a specialized version of the multiple particle case.

The analysis divides into two parts:

First, we use a Lagrangian approach to derive the appropriate Feynman path integrals. 
This approach will be loosely parallel to the approach in the single particle case:
\begin{enumerate}
\item Extend Fock space from 3D to 4D. 
\item Reuse the existing field theory Lagrangian densities by interpreting
the time as the coordinate time. 
\item Compute the action by integrating over both clock and coordinate times,
so that the integrals over the Lagrangian density goes from $\int{d\tau d\vec{x}}\to\int{d\tau dtd\vec{x}}$. 
\item Verify that the free particle propagator computed with this approach
matches the single particle free particle propagator. 
\end{enumerate}

The result is that we get the TQM Feynman diagrams from their SQM equivalents by
keeping the topology of lines and vertexes the same,
but replacing the SQM parts with their TQM equivalents. 
We replace:
\begin{enumerate}
\item{SQM wave functions with TQM},
\item{SQM propagators with TQM},
\item{and SQM integrals over three space dimensions and clock time with TQM integrals over the three space dimensions, coordinate time, and clock time.}
\end{enumerate}

In principle, we could simply have postulated these substitutions; 
the derivation from first principles adds depth to our understanding, 
and also makes possible the extension of TQM to problems 
not amenable to attack via the Feynman diagram approach.

In the second part, we apply the derived Feynman rules to a few simple cases:
\begin{enumerate}
\item Free particle,
\item Emission of a particle, 
\item Absorption of a particle, 
\item Exchange of a particle, 
\item Loop correction to the mass. 
\end{enumerate}
By combining these elements we can in principle calculate any Feynman diagram.

The loop calculation provides a preliminary success for TQM. If treated
naively, the loop diagrams in TQM are divergent. But the combination
of Morlet wavelet decomposition and entanglement in time -- neither
alone sufficient -- causes the loop integrals to converge.

As we develop the multiple particle case, we will see a number of
additional effects:
\begin{enumerate}
\item anti-symmetry in time, 
\item forces of anticipation and regret,
\item interference and entanglement in time.
\end{enumerate}

With the approach taken here we have no free or adjustable parameters.
We have a simple transition from the single to the multiple particle treatment.
We do not have the familiar ultraviolet divergences. 
And we have a
large number of opportunities for experimental test.

\subsection{$ABC$ model\label{subsec:mul-abc}}

\begin{figure}[h]
\begin{centering}
\includegraphics[width=10cm]{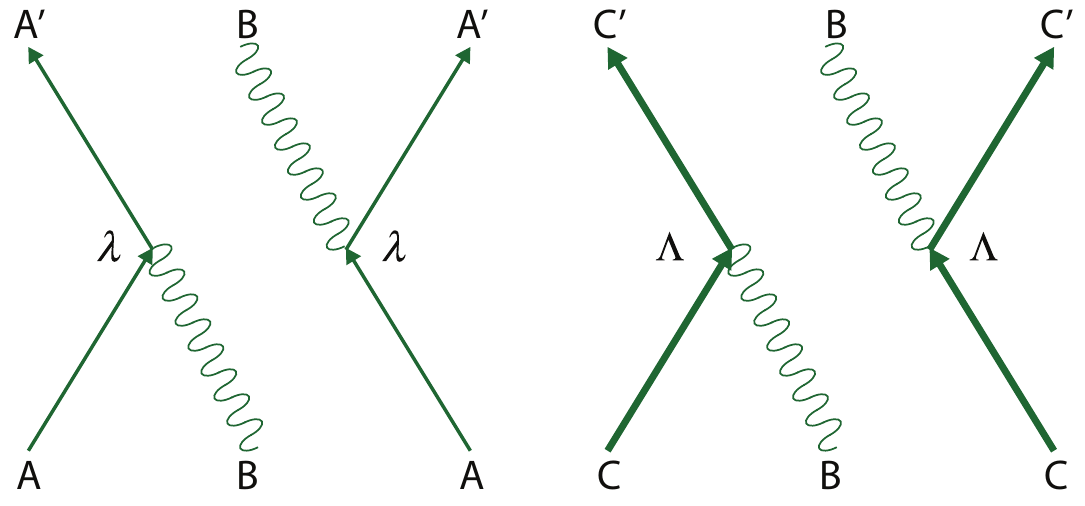} 
\par\end{centering}
\caption{Interactions in the $ABC$ model\label{fig:mul-ab-vertex}}
\end{figure}

The simplest model we can find that lets us cover the basic interactions
has three spinless, massive particle species $A,$ $B$, and $C$.
We assign them non-zero masses $m$, $\mu$, and $M$ respectively.
We will take $\mu$ as small as it needs to be. They are real fields.
$A$\textsc{'}s emit and absorb $B$\textsc{'}s with amplitude $\lambda$.
$C$\textsc{'}s emit and absorb $B$\textsc{'}s with amplitude $\Lambda$.
$A$\textsc{'}s and $C$\textsc{'}s do not talk with each other directly.
There are no other interactions.

Loosely, $A$ is a spinless model of an electron, $B$ of a photon,
and $C$ of a proton.

We will focus primarily on the $A$ and $B$ particles. We will need
the $C$ for the discussion of particle exchange.

\subsection{Fock space\label{subsec:mul-fock}}

In the single particle case we generalized the wave functions from
three dimensions to four. Here we generalize Fock space from three
dimensions to four. As paths may be seen as a series of wave functions,
one wave function per clock tick, this implicitly generalizes the
associated paths from three dimensions to four as well.

We use box normalization. The box runs from $-L\to L$ in
all four coordinates: coordinate time and the three space dimensions.
It is taken to extend well past the wavelets we are working with.

Using box normalization means that the Fourier transforms will be
discrete. Discrete Fourier transforms are convenient for discussing
various points of principle and to help in visualizing the field theory
calculations. For the actual calculations we will use continuous wave
functions.

In general the extrusion from three to four dimensions is straight
forward; a few specific points require attention.

\subsubsection{Fock space in three dimensions\label{subsec:mul-fock-3d}}

\begin{figure}[h]
\begin{centering}
\includegraphics[width=8cm]{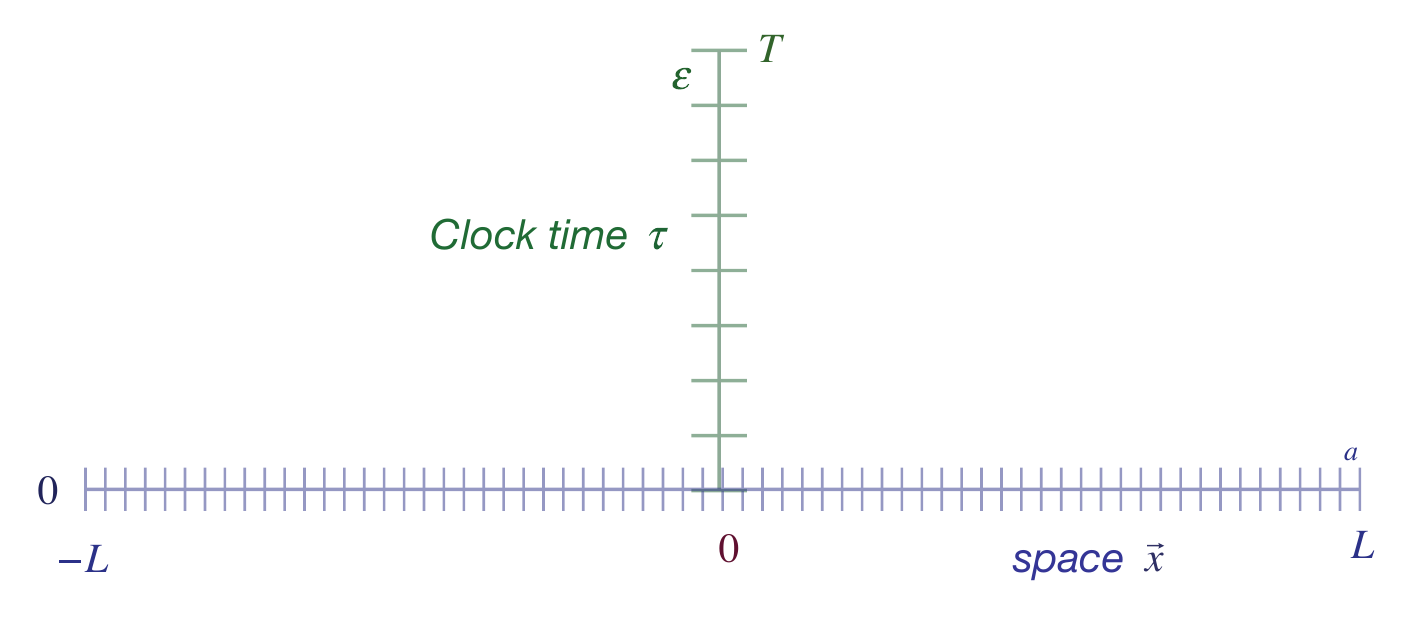} 
\par\end{centering}
\caption{Fock space in three dimensions\label{fig:mul-fock}}
\end{figure}

\paragraph{Single particle basis wave functions}

We will focus on the $x$ coordinate here; $y$ and $z$ are the same.
We break our box into 2M pieces, implying a lattice spacing $a\equiv\frac{L}{M}$.
$a$ has dimensions of length.
This lets us replace the smoothly varying $x$ with the values of
$x$ at a series of points:
\begin{equation}
x\in\left[{-L,L}\right]\to x=\left({-aM,\ldots,0\ldots,aM}\right)
\end{equation}

The continuous/discrete translation table is:
\begin{equation}
\left({x,y,z}\right)\leftrightarrow\left({ai,aj,ak}\right)
\end{equation}

%We are using $i$ with a dot for the index, $\imath$ without a dot for the square root of -1.

Integrals over space go to sums over $i,j,k$:
\begin{equation}
\int\limits _{-L}^{L}{dxdydz}\leftrightarrow{a^{3}}\sum\limits _{i,j,k=-M}^{M}{}
\end{equation}

The coordinate basis is trivial, just Kronecker $\delta$ functions:
\begin{equation}
{\phi_{\vec{x}'}}\left({\vec{x}}\right)\equiv{\delta^{3}}\left({\vec{x}-\vec{x}'}\right)\leftrightarrow{\delta_{ii'}}{\delta_{jj'}}{\delta_{kk'}}
\end{equation}

We assume the wave functions are periodic in $2L$. The periodic condition
is not important; $L$ will be chosen large enough that all interesting
wave functions are well inside of it. They will be trivially periodic
because they are zero on both sides of each dimension.

We normalize the basis wave functions to one:
\begin{equation}
\int\limits _{-L,-L,-L}^{L,L,L}{d\vec{x}\phi_{\vec{k}}^{*}\left({\vec{x}}\right){\phi_{\vec{k}'}}\left({\vec{x}}\right)}={\delta_{\vec{k}\vec{k}'}}
\end{equation}
giving:
\begin{equation}
{\phi_{\vec{k}}}\left({\vec{x}}\right)=\frac{1}{{{\sqrt{2L}}^{3}}}\exp\left({\imath\vec{k}\cdot\vec{x}}\right)
\end{equation}

Now we can expand an arbitrary wave function in terms of the basis
functions:
\begin{equation}
\phi\left({\vec{x}}\right)=\sum\limits _{\vec{k}}{c_{\vec{k}}}{\phi_{\vec{k}}}\left({\vec{x}}\right)
\end{equation}

The measure in the path integrals is in terms of the $c$\textsc{'}s:

\begin{equation}
\mathcal{D}\phi\equiv\prod\limits _{n=0}^{N}{{\mathcal{D}_{n}}\phi},{\mathcal{D}_{n}}\phi\equiv\prod\limits _{\vec{k}}{d{c_{\vec{k}}}}
\end{equation}
so there is one set of space integrals at each clock tick.

At the end of the discrete part of the calculation we will be letting
$M$, $N$, and $L$ go to infinity. As the effects of TQM are averaged
out over larger times we will \textit{not} be letting $T$ go to infinity,
even in the SQM case.

In the continuum limit we have:
\begin{equation}
{\phi_{\vec{k}}}\left({\vec{x}}\right)\to\frac{1}{{{\sqrt{2\pi}}^{3}}}\exp\left({\imath\vec{k}\cdot\vec{x}}\right)
\end{equation}

\paragraph{Multi-particle wave functions}

Things become interesting when we go to multiple particle wave functions.
For two particles:
\begin{equation}
{\phi_{\vec{k}\vec{k}'}}\left({1,2}\right)\equiv\frac{1}{{\sqrt{2}}}\left({{\phi_{\vec{k}}}\left(1\right){\phi_{\vec{k}'}}\left(2\right)+{\phi_{\vec{k}}}\left(2\right){\phi_{\vec{k}'}}\left(1\right)}\right)
\end{equation}

As we get to larger and larger numbers of particles, these wave functions
become tricky to write out and manage. To simplify, we use the familiar
annihilation and creation operators, defined by:
\begin{equation}
\begin{array}{c}
a_{\vec{k}}^{\dag}\left|{n_{\vec{k}}}\right\rangle =\sqrt{{n_{\vec{k}}}+1}\left|{{n_{\vec{k}}}+1}\right\rangle \hfill\\
{a_{\vec{k}}}\left|{n_{\vec{k}}}\right\rangle =\sqrt{{n_{\vec{k}}}}\left|{{n_{\vec{k}}}-1}\right\rangle \hfill
\end{array}
\end{equation}
with the usual commutation operators:
\begin{equation}
\left[{{a_{\vec{k}}},a_{\vec{k}'}^{\dag}}\right]={\delta_{\vec{k}\vec{k}'}}
\end{equation}

We define the single particle operator:
\begin{equation}
\bar{\phi}\left({\vec{x}}\right)=\sum\limits _{\vec{k}}{{a_{\vec{k}}}\bar{\phi}_{\vec{k}}^{\dag}\left({\vec{x}}\right)+a_{\vec{k}}^{\dag}{{\bar{\phi}}_{\vec{k}}}\left({\vec{x}}\right)}
\end{equation}

An arbitrary multiple particle basis state may be built up as products of these:
\begin{equation}
\left|{\left\{ {n_{\vec{k}}}\right\} }\right\rangle =\frac{1}{{\sqrt{\prod\limits _{\vec{k}}{{n_{\vec{k}}}!}}}}\prod\limits _{\vec{k}}{{\left({a_{\vec{k}}^{\dag}}\right)}^{{n_{\vec{k}}}}}\left|0\right\rangle 
\end{equation}
where ${\left\{ {n_{\vec{k}}}\right\} }$ is a specific set of occupation
numbers.

We define the general wave function in the occupation number basis:
\begin{equation}
\sum\limits _{\left\{ {n_{\vec{k}}}\right\} }{c_{\left\{ {n_{\vec{k}}}\right\} }}\left|{\left\{ {n_{\vec{k}}}\right\} }\right\rangle 
\end{equation}
with normalization:
\begin{equation}1 = \sum\limits_{\left\{ {{n_{\vec k}}} \right\}} {c_{\left\{ {{n_{\vec k}}} \right\}}^2} \end{equation}

Note we are \textit{not} defining the creation and annihilation operators
in terms of an infinite set of harmonic oscillators; we are defining
them by their effects on Fock space. We can think of movement in Fock
space as a giant game of snakes and ladders, with the creation operators
as the ladders, the annihilation operators as the snakes.

\subsubsection{Fock space in four dimensions\label{subsec:mul-fock-4d}}

\begin{figure}[h]
\begin{centering}
\includegraphics[width=10cm]{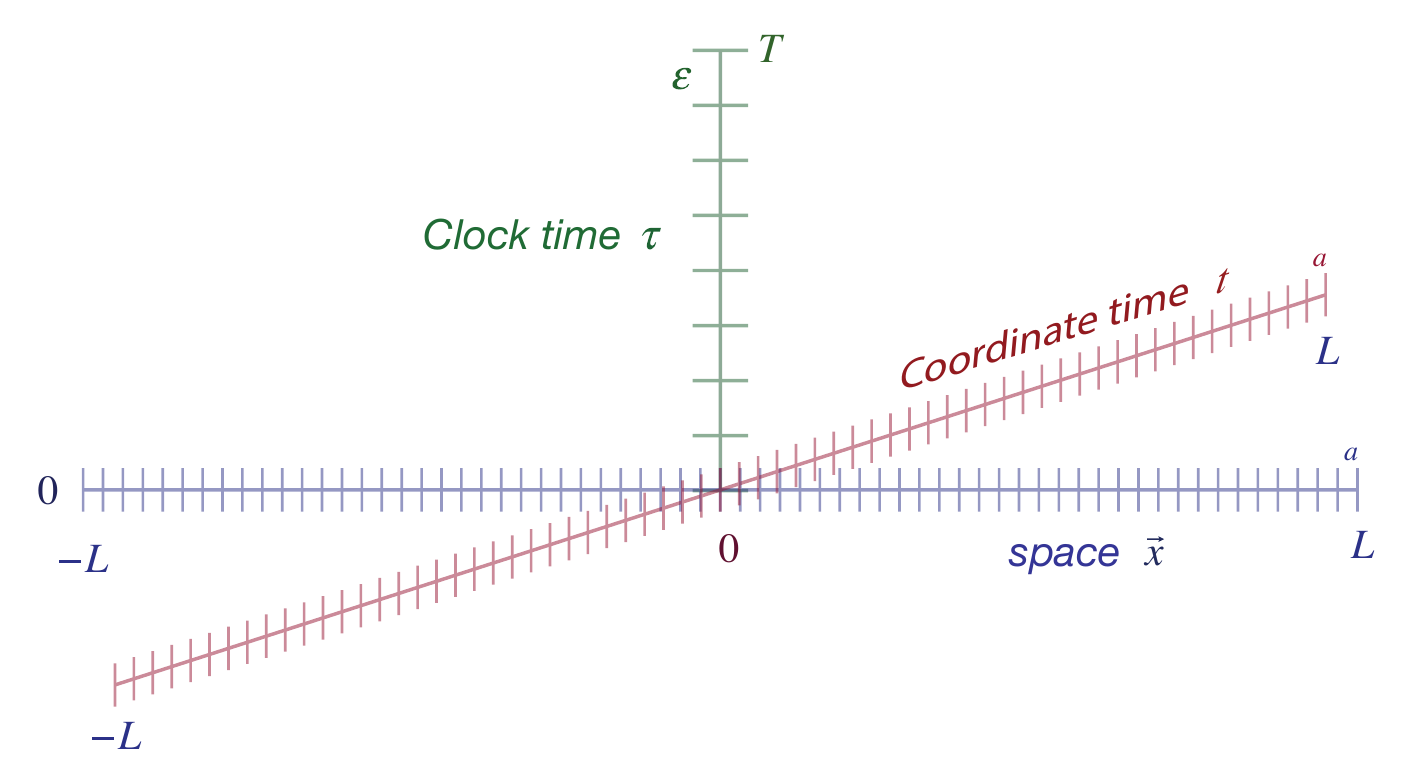} 
\par\end{centering}
\caption{Fock space in four dimensions\label{fig:mul-fock-1}}
\end{figure}

\paragraph{Single particle basis wave functions}

The development in TQM works along the same lines but with one more
dimension $t$ and a fourth index $h$. We are treating coordinate
time essentially like a fourth spatial dimension, much like the $x_{4}\equiv\imath ct$
trick of earlier works on special relativity. Paths in TQM will typically
start at $\tau=0$ and finish at some defined clock time $\tau=T$.
But their paths in coordinate time may well dive before $t=0$ and
climb past $t=T$. To make sure that all of the relevant paths are
included in our box, we require that $-L\ll0$ and $L\gg{T}$. Using
the same $L$ for time as for space, and requiring that $L\to\infty$
accomplishes this.

The continuous/discrete translation table is now: 
\begin{equation}
\left({t,x,y,z}\right)\leftrightarrow\left({ah,ai,aj,ak}\right)
\end{equation}

So we have:
\begin{equation}
t=ah
\end{equation}

Note there is no requirement that the lattice spacing $a$ in coordinate
time $t$ match the step spacing $\epsilon$ in clock time $\tau$;
In general we will have: $a\ne\varepsilon$.

We promote integrals and sums over three dimensions to integrals and
sums over four:
\begin{equation}
\int{d\vec{x}\to}\int{dtd\vec{x}},\sum\limits _{ijk}{{f_{ijk}}\to}\sum\limits _{hijk}{f_{hijk}}
\end{equation}
\begin{equation}
{\phi_{x'}}\left(x\right)\equiv{\delta^{4}}\left({x-x'}\right)\leftrightarrow{\delta_{hh'}}{\delta_{ii'}}{\delta_{jj'}}{\delta_{kk'}}
\end{equation}

The $k$\textsc{'}s are again periodic. We need to think for a moment
about how our wave functions get their start in life. At clock time
zero, the TQM wave function will look something like:
\begin{equation}
{\tilde{\varphi}_{0}}\left(t\right)\sim\exp\left({-\frac{{{\left({t-{t_{0}}}\right)}^{2}}}{{2\sigma_{t}^{2}}}}\right)
\end{equation}

One subtlety: we always develop our wave functions from $\tau=0\to\tau=T$.
Consider the wave function at $\tau=0$. How did it get there? what
of its past? Won't at some time the wave function have existed before
$-L$, before the box appeared? The working answer is that the shape
of the wave function at $\tau=0$ tells us all we need to know of
its history before that time. If the wave function \textit{at} $\tau=0$
is well within the box, if $\sigma_{t}\ll L$, we have what we need.
What happened in the past, stays in the past.

We normalize the basis wave functions to one:
\begin{equation}
\int\limits _{-L,-L,-L,-L}^{L,L,L,L}{dtd\vec{x}\phi_{k}^{*}\left({t,\vec{x}}\right){\phi_{k'}}\left({t,\vec{x}}\right)}={\delta_{kk'}}
\end{equation}
giving:
\begin{equation}
{\phi_{k}}\left(x\right)=\frac{1}{{4{L^{2}}}}\exp\left({-\imath kx}\right)
\end{equation}

Now we can expand an arbitrary wave function in terms of the basis
functions:
\begin{equation}
\phi\left({x}\right)=\sum\limits _{k}{c_{k}}{\phi_{k}}\left({x}\right)
\end{equation}

Again, the measure in the path integrals is in terms of the $c$\textsc{'}s::
\begin{equation}
\mathcal{D}\phi\equiv\prod\limits _{n=0}^{N}{{\mathcal{D}_{n}}\phi},{\mathcal{D}_{n}}\phi\equiv\prod\limits _{k}{d{c_{k}}}
\end{equation}
so there is one set of time and space integrals at each clock tick.
At the end of the discrete part of the calculation we will
be letting $M$, $N$, and $L$ go to infinity. And we will \textit{not}
be letting $T$ go to infinity.

In the continuum limit we have:
\begin{equation}
{\phi_{k}}\left(x\right)\to\frac{1}{{4{\pi^{2}}}}\exp\left({-\imath wt+\imath\vec{k}\cdot\vec{x}}\right)
\end{equation}

\paragraph{Multiple particle wave functions\label{par:Symmetrized-wave-functions}}

Symmetrization in four dimensions works exactly as in three. For two
particles we have:
\begin{equation}
{\phi_{kk'}}\left({1,2}\right)\equiv\frac{1}{{\sqrt{2}}}\left({{\phi_{k}}\left(1\right){\phi_{k'}}\left(2\right)+{\phi_{k}}\left(2\right){\phi_{k'}}\left(1\right)}\right)
\end{equation}

And the creation and annihilation operators work in the same way:
\begin{equation}
\begin{array}{c}
a_{k}^{\dag}\left|{n_{k}}\right\rangle =\sqrt{{n_{k}}+1}\left|{{n_{k}}+1}\right\rangle \hfill\\
{a_{k}}\left|{n_{k}}\right\rangle =\sqrt{{n_{k}}}\left|{{n_{k}}-1}\right\rangle \hfill
\end{array}
\end{equation}

With the commutators:
\begin{equation}
\left[{{a_{k}},a_{k'}^{\dag}}\right]={\delta_{kk'}}
\end{equation}
Single particle operator:
\begin{equation}
\phi\left(x\right)=\sum\limits _{k}{{a_{k}}\phi_{k}^{\dag}\left(x\right)+a_{k}^{\dag}{\phi_{k}}\left(x\right)}
\end{equation}

Arbitrary multiple particle:
\begin{equation}
\left|{\left\{ {n_{k}}\right\} }\right\rangle =\frac{1}{{\sqrt{\prod\limits _{k}{{n_{k}}!}}}}\prod\limits _{k}{{\left({a_{k}^{\dag}}\right)}^{{n_{k}}}}\left|0\right\rangle 
\end{equation}
where ${\left\{ {n_{k}}\right\} }$ is a specific set of occupation
numbers, now over all possibilities in four dimensions.

And we again define the general wave function as a sum over all possible
${\left\{ {n_{k}}\right\} }$:
\begin{equation}
\sum\limits _{\left\{ {n_{k}}\right\} }{c_{\left\{ {n_{k}}\right\} }}\left|{\left\{ {n_{k}}\right\} }\right\rangle 
\end{equation}
with normalization:
\begin{equation}1 = \sum\limits_{\left\{ {{n_k}} \right\}} {c_{\left\{ {{n_k}} \right\}}^2} \end{equation}

This defines the Fock space in four dimensions, along the same lines
as the one in three. We are again playing snakes and ladders, but
with four dimensional snakes and ladders rather than three.

\subsubsection{Anti-symmetry in time\label{subsec:mul-fock-symmetry}}

We assume the same overall symmetry properties are required in four
dimensions as in three. This implies that wave functions can use the
coordinate time to help meet their symmetry responsibilities, with
potentially amusing implications. In particular if the wave function
is anti-symmetric in time, it will have the ``wrong'' symmetry properties
in space.

This is testable, at least in principle.

Say we have wide wave functions in time and space $A\left(t\right)$
and $B\left(x\right)$, and narrow wave functions in time and space
$a\left(t\right)$ and $b\left(x\right)$. The particles are identified
as 1 and 2.

An acceptable initial wave function is:
\begin{equation}
{\varphi_{sym}}\left({1,2}\right)=\frac{1}{{\sqrt{2}}}\left({{A\left({t_{1}}\right)}{B\left({x_{1}}\right)}{a\left({t_{2}}\right)}{b\left({x_{2}}\right)}+{A\left({t_{2}}\right)}{B\left({x_{2}}\right)}{a\left({t_{1}}\right)}{b\left({x_{1}}\right)}}\right)
\end{equation}

This clearly has the right symmetry between particles 1 and 2.
We wish to break this down into sums over products of wave functions
in time and space.

The symmetrical basis functions in time and space are:
\begin{equation}
\begin{array}{c}
{{\tilde{\varphi}}_{sym}}\left({1,2}\right)=\frac{1}{{\sqrt{2}}}\left({{A\left({t_{1}}\right)}{a\left({t_{2}}\right)}+{A\left({t_{2}}\right)}{a\left({t_{1}}\right)}}\right)\hfill\\
{{\bar{\varphi}}_{sym}}\left({1,2}\right)=\frac{1}{{\sqrt{2}}}\left({{B\left({x_{1}}\right)}{b\left({x_{2}}\right)}+{B\left({x_{2}}\right)}{b\left({x_{1}}\right)}}\right)\hfill
\end{array}
\end{equation}

If we use these as a product we get:
\begin{equation}
{\tilde \varphi _{sym}}\left( {1,2} \right){\bar \varphi _{sym}}\left( {1,2} \right) = \frac{1}{2}\left( \begin{array}{c}
  A\left( {{t_1}} \right)B\left( {{x_1}} \right)a\left( {{t_2}} \right)b\left( {{x_2}} \right) \hfill \\
   + A\left( {{t_1}} \right)B\left( {{x_2}} \right)a\left( {{t_2}} \right)b\left( {{x_1}} \right) \hfill \\
   + A\left( {{t_2}} \right)B\left( {{x_1}} \right)a\left( {{t_1}} \right)b\left( {{x_2}} \right) \hfill \\
   + A\left( {{t_2}} \right)B\left( {{x_2}} \right)a\left( {{t_1}} \right)b\left( {{x_1}} \right) \hfill \\ 
\end{array}  \right)
\end{equation}
where the two middle terms do not belong.

The anti-symmetric basis functions in time and space are:
\begin{equation}
\begin{array}{c}
{{\tilde{\varphi}}_{anti}}\left({1,2}\right)=\frac{1}{{\sqrt{2}}}\left({{A\left({t_{1}}\right)}{a\left({t_{2}}\right)}-{A\left({t_{2}}\right)}{a\left({t_{1}}\right)}}\right)\hfill\\
{{\bar{\varphi}}_{anti}}\left({1,2}\right)=\frac{1}{{\sqrt{2}}}\left({{B\left({x_{1}}\right)}{b\left({x_{2}}\right)}-{B\left({x_{2}}\right)}{B\left({x_{1}}\right)}}\right)\hfill
\end{array}
\end{equation}
and their product is:
\begin{equation}
{{\tilde \varphi }_{anti}}\left( {1,2} \right){{\bar \varphi }_{anti}}\left( {1,2} \right) = \frac{1}{2}\left( \begin{array}{c}
  A\left( {{t_1}} \right)B\left( {{x_1}} \right)a\left( {{t_2}} \right)b\left( {{x_2}} \right) \hfill \\
   - A\left( {{t_1}} \right)B\left( {{x_2}} \right)a\left( {{t_2}} \right)b\left( {{x_1}} \right) \hfill \\
   - A\left( {{t_2}} \right)B\left( {{x_1}} \right)a\left( {{t_1}} \right)b\left( {{x_2}} \right) \hfill \\
   + A\left( {{t_2}} \right)B\left( {{x_2}} \right)a\left( {{t_1}} \right)b\left( {{x_1}} \right) \hfill \\ 
\end{array}  \right)
\end{equation}

Therefore the sum of the completely symmetric and the completely anti-symmetric
gives the target wave function:
\begin{equation}
{\varphi_{sym}}\left({1,2}\right)=\frac{1}{2}\left({{\tilde{\varphi}}_{sym}}\left({1,2}\right){{\bar{\varphi}}_{sym}}\left({1,2}\right)+{{\tilde{\varphi}}_{anti}}\left({1,2}\right){{\bar{\varphi}}_{anti}}\left({1,2}\right)\right)
\end{equation}

To get a wave function which is completely symmetric in time and space
together we need to use both the symmetric and the anti-symmetric
basis functions.

\subsection{Lagrangian\label{subsec:mul-lag}}

What should we use as a Lagrangian? We will look at this first from
a classical perspective.

In SQM the Lagrangian for a massive spin 0 free particle is given
by:
\begin{equation}
\bar{\mathcal{L}}^{free}\left[{\phi,\dot{\phi}}\right]=\frac{1}{2}\frac{{\partial{\phi}}}{{\partial\tau}}\frac{{\partial\phi}}{{\partial\tau}}-\frac{1}{2}\nabla{\phi}\nabla\phi-\frac{{m^{2}}}{2}\phi^{2}
\end{equation}

In classical mechanics, the wave functions may be written as sums
over the basis plane waves:
\begin{equation}
{\phi_{\tau}}\left({\vec{x}}\right)\sim\sum\limits _{\vec{k}}{{c_{\tau,\vec{k}}}{\phi_{\tau,\vec{k}}}\left({\vec{x}}\right)}
\end{equation}

The action is the integral of this over space and clock time:
\begin{equation}
\int\limits _{0}^{T}{d\tau d\vec{x}}\bar{\mathcal{L}}^{free}\left[{\phi,\dot{\phi}}\right]
\end{equation}

Typically we let the limits in clock time go to $\pm\infty$, usually
somewhere near the end of the analysis. Here that would average out
the effects of any dispersion in time. So just as in the definition
of Fock space, we keep the total clock time finite.

How to extend this Lagrangian to include coordinate time?

By our first requirement, $x$ and $t$ have to rotate into each other
under a Lorentz transformation. The only way to do this is to change
clock time to coordinate time: $\tau\to t$. So we start with:
\begin{equation}
\mathcal{L}^{free}\left[{\phi}\right]=\frac{1}{2}{\partial_{t}}\phi{\partial_{t}}\phi-\frac{1}{2}\nabla\phi\nabla\phi-{\frac{m^{2}}{2}}{\phi^{2}}
\end{equation}

The wave functions may be written as sums over the basis plane waves,
with all the clock time dependence in the coefficients $c$.
\begin{equation}
{\phi _\tau }\left( {t,\vec x} \right)~\sum\limits_k {c_\tau ^{\left( k \right)}{\phi _k}\left( {t,\vec x} \right)} 
\end{equation}

Neither basis functions nor operators are functions of clock time.
Therefore the Lagrangian is not.
To include the dependence on clock time we will need to include the
integral over clock time from 0 to $T$
\begin{equation}
{S_{0}}\sim\int\limits _{0}^{T}{d\tau}\int{dtd\vec{x}{\mathcal{L}^{free}}\left[{\varphi}\right]}
\end{equation}

We can write the Lagrangian in momentum space as:
\begin{equation}
{\mathcal{\hat{L}}^{free}}\sim\frac{{{w^{2}}-{{\vec{k}}^{2}}-{m^{2}}}}{2}
\end{equation}

The integral over clock time gives:
\begin{equation}
\imath\int{d\tau}\to\imath\frac{{{w^{2}}-{{\vec{k}}^{2}}-{m^{2}}}}{2}
\end{equation}

This has two problems: it is not dimensionless and it does not match
the results for the single particle propagator. We can fix both by
adding a factor of $\frac{1}{m}$:
\begin{equation}{\mathcal{L}^{\left( {free} \right)}} \to \frac{1}{{2m}}\left( {{\partial _t}\phi {\partial _t}\phi  - \nabla \phi \nabla \phi  - {m^2}{\phi ^2}} \right)\end{equation}

With that noted, we will get the single particle propagator:
\begin{equation}
\hat{K}\sim\exp\left({\imath\frac{{{w^{2}}-{{\vec{k}}^{2}}-{m^{2}}}}{{2m}}\tau}\right)
\end{equation}

Our recipe for going from an existing SQM Lagrangian and action to the equivalent TQM Lagrangian and action is therefore:
\begin{enumerate}
\item{replace the time in the SQM Lagrangian with coordinate time},
\item{replace the integral over three space dimensions with an integral over coordinate time and the three space dimensions},
\item{add an overall integral over clock time},
\item{divide the Lagrangian by the mass of the particle,}
\item{and in the case of a massless particle -- not covered here -- use the familiar trick of taking the limit as the mass goes to zero}
\end{enumerate}

Spelling this out for the free Lagrangian we get:
\begin{equation}S = \int\limits_0^T {d\tau dtd\vec x} \frac{1}{{2m}}\left( {{\partial _t}\phi {\partial _t}\phi  - \nabla \phi \nabla \phi  - {m^2}{\phi ^2}} \right)\end{equation}
The extension to include $B,C$ particles is:
\begin{equation}{\mathcal{L}_{AB}}\left[ {A,B,C} \right] = {\mathcal{L}^{free}}\left[ A \right] + {\mathcal{L}^{free}}\left[ B \right] + {\mathcal{L}^{free}}\left[ C \right] - \frac{\lambda }{2}ABA - \frac{\Lambda }{2}CBC\end{equation}

Since there is no longer any explicit dependence on clock time in
the TQM Lagrangian, the corresponding Hamiltonian is merely $-\mathcal{L}$,
with the slightly disconcerting result that there are no non-trivial
canonical momenta.

In SQM the next step is to promote the classical fields to operators:
\begin{equation}
\bar{\phi}\left({\vec{x}}\right)\to\sum\limits _{\vec{k}}{{a_{\vec{k}}}\bar{\phi}_{\vec{k}}^{\dag}\left({\vec{x}}\right)+a_{\vec{k}}^{\dag}{{\bar{\phi}}_{\vec{k}}}\left({\vec{x}}\right)}
\end{equation}
so in TQM we do the same:
\begin{equation}
\phi\left(x\right)\to\sum\limits _{k}{{a_{k}}\phi_{k}^{\dag}\left(x\right)+a_{k}^{\dag}{\phi_{k}}\left(x\right)}
\end{equation}

\subsection{Path integrals\label{subsec:mul-paths}}

With the Lagrangian defined, we can write the full kernel as:
\begin{equation}
{K_{T}}\equiv\int{\mathcal{D}\phi\exp\left({\imath\int\limits _{0}^{T}{d\tau\int{{d^{4}}x\mathcal{L}\left[\phi\right]}}}\right)}
\end{equation}

This notation conceals much complexity. We start with the zero dimensional
free case.

\subsubsection{Zero dimensional free case \label{subsubsec:mul-free}}

\begin{figure}[h]
\begin{centering}
\includegraphics[width=6cm]{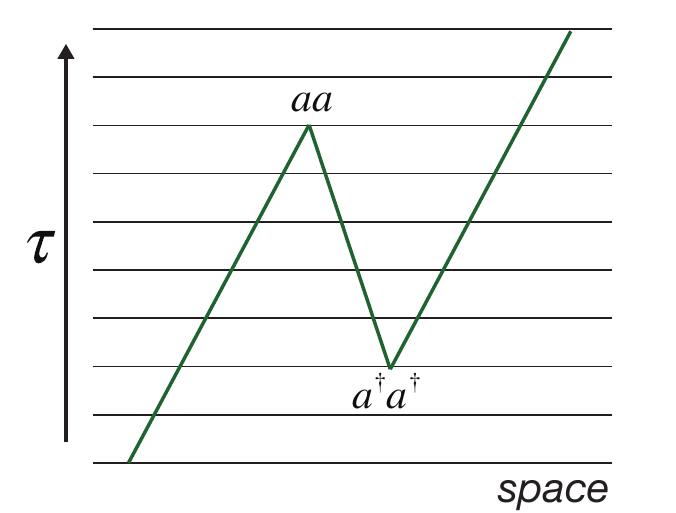} 
\par\end{centering}
\caption{Zigzag paths in clock time\label{fig:mul-abs-1}}
\end{figure}

The basis of wave functions for the zero dimensional case is the set
of possible occupation numbers from 0 to infinity. Any wave function
may be written as a sum over these:
\begin{equation}
\psi=\sum\limits _{l=0}^{\infty}{{c_{l}}\left|l\right\rangle }
\end{equation}
with the normalization condition that:
\begin{equation}
1=\sum\limits _{l=0}^{\infty}{c_{l}^{2}}
\end{equation}

We take the zero single particle wave function as simply the number one.
For multiple particles we have:
\begin{equation}
\begin{array}{c}
  {\phi _2} = \frac{1}{{\sqrt {2!} }}\left( {{\phi _1}\left( 1 \right){\phi _1}\left( 2 \right) + {\phi _1}\left( 2 \right){\phi _1}\left( 1 \right)} \right) \hfill \\
  {\phi _3} = \frac{1}{{\sqrt {3!} }}\left( {{\phi _1}\left( 1 \right){\phi _1}\left( 2 \right){\phi _1}\left( 3 \right) +  \ldots } \right) \hfill \\
   \ldots  \hfill
\end{array} 
\end{equation}

The number states are orthogonal:
\begin{equation}
\left\langle {l}\mathrel{\left|{\vphantom{l{l'}}}\right.\kern-\nulldelimiterspace}{l'}\right\rangle ={\delta_{ll'}}
\end{equation}

The amplitude to go from one wave function to another is given by
a Fock space sandwich:

\begin{equation}
\langle\psi'|{K_{T}}\left|\psi\right\rangle =\left\langle {\sum\limits _{l'=0}^{\infty}{c_{l'}}{\phi_{l'}}}\right|\int{\mathcal{D}\phi\exp\left({\imath\int\limits _{0}^{T}{d\tau\mathcal{L}}}\right)}\left|{\sum\limits _{l=0}^{\infty}{c_{l}}{\phi_{l}}}\right\rangle 
\end{equation}
with measure:
\begin{equation}
\mathcal{D}\phi=\prod\limits _{n=0}^{n=N}{\mathcal{D}{\phi^{\left(n\right)}}},\mathcal{D}{\phi^{\left(n\right)}}=\prod\limits _{l=0}^{\infty}{dc_{l}^{\left(n\right)}}
\end{equation}
and Lagrangian:
\begin{equation}
\mathcal{L}=-\frac{m}{2}\left({aa+a{a^{\dag}}+{a^{\dag}}a+{a^{\dag}}{a^{\dag}}}\right)
\end{equation}

Each term is a pair of operators. Two terms change the particle numbers; two
do not. The term with two annihilation operators will reduce the number
of particles by two, the term with two creation operators will increase
the number of particles by two.

This means that even the simple free case has us moving up and down
in Fock space. We have to deal with populations of particles.

We proceed discretely:
\begin{equation}
\exp\left({\imath\int\limits _{0}^{T}{d\tau\mathcal{L}}}\right)\to\exp\left({\imath\varepsilon\sum\limits _{n=1}^{N}\mathcal{L}}\right)
\end{equation}
and one step at a time:
\begin{equation}
\exp\left({\imath\varepsilon\mathcal{L}}\right)\approx1+\imath\varepsilon\mathcal{L}
\end{equation}

We will start with a common tactic: we will throw out all disconnected
diagrams. For instance at one clock tick an $a^{\dag}a^{\dag}$ term
could create a virtual particle/anti-particle pair which a few clock
ticks later an ${a}{a}$ then deletes. These represent self-interactions
of the vacuum and are mere background noise, in common across all
diagrams.

We will also use normal ordering (``always annihilate before you
create''):
\begin{equation}
\frac{1}{2}\left({{a}a^{\dag}+a^{\dag}{a}}\right)\to a^{\dag}{a}+\frac{1}{2}=\hat{n}+\frac{1}{2}
\end{equation}

Here $\hat{n}$ is the number operator. The $\frac{1}{2}$ term gives
us an overall constant, which we can also ignore, since it also is present for all diagrams.

There is a third problem: sometimes a particle can interact with a
virtual pair. The particle is evolving in clock time and encounters
a term with two annihilation operators. By chance, a particle in the
vacuum encounters the same term at the same clock tick. To an outside
observer, it looks as if our particle has reversed direction in clock
time and is now headed backwards. This is now a connected diagram,
however, so the previous rules do not exclude it.

This is however implicitly included in our analysis of particle exchange
below -- the middle part of the diagram is the exchanged particle.
We will drop these terms unless we explicitly need them, e.g. if we
are looking at pair creation or annihilation.

With this the Lagrangian reduces to a sum over number operators:
\begin{equation}
\mathcal{L}\to-\imath m\frac{\hat{n}}{2}=-\imath\frac{m}{2}\sum\limits _{l=0}^{\infty}{{l}{\delta_{ll'}}}
\end{equation}

We return the infinitesimal Lagrangian to the exponential:
\begin{equation}
\exp\left({-\imath\varepsilon\sum\limits _{n=1}^{N}{\frac{m}{2}\sum\limits _{l=0}^{\infty}{{l}{\delta_{ll'}}}}}\right)
\end{equation}
and get:
\begin{equation}
{K_{\tau}}\left({l;l'}\right)=\exp\left({-\imath\frac{m}{2}l\tau}\right){\delta_{ll'}}
\end{equation}
which makes sense as the zero dimensional matrix element: if there
are $l$ particles present, they oscillate $l$ times as quickly as
one.

\subsubsection{Four dimensional free case}

We now extend the zero dimensional treatment to four dimensions. Free
Lagrangian:
\begin{equation}
{\mathcal{L}^{free}}\left[\phi\right]=\frac{1}{2m}\frac{{\partial\phi}}{{\partial t}}\frac{{\partial\phi}}{{\partial t}}-\frac{1}{2m}\nabla\phi\nabla\phi-\frac{{m^{2}}}{2m}{\phi^{2}}
\end{equation}
where $\phi$ is an operator:
\begin{equation}
\phi\left(x\right)=\int{{d^{4}}k{{\hat{a}}_{k}}\phi_{k}^{\dag}\left(x\right)+\hat{a}_{k}^{\dag}{\phi_{k}}\left(x\right)}
\end{equation}

Again the amplitude to go from one state to another is computed by
constructing a Fock space sandwich:
\begin{equation}
{A_{\varepsilon}}=\langle\left\{ {n_{k'}}\right\} |\exp\left({\imath\varepsilon\int{{d^{4}}x\mathcal{L}\left[{\phi,\partial\phi}\right]}}\right)\left|{\left\{ {n_{k}}\right\} }\right\rangle 
\end{equation}

We will use as the single particle wave functions the exponentials
${\phi_{k}}\left(x\right)\equiv\frac{1}{{4{\pi^{2}}}}\exp\left({-\imath kx}\right)$.
In momentum space the partial derivatives $\partial_{x}$ turn into
powers of $k$. We have the overall integral over $x$ and (from the
transition to momentum space) integrals over $k$ and $k'$ for each
of the two operators in each term. The basis functions integrated
over $x$ give $\delta$ functions in $k,k'$. The integral over k'
gives an integral over $k$. We are left with:
\begin{equation}
\int{d^{4}x}\phi\phi\to\int{{d^{4}}k\left({{a_{k}}{a_{-k}}+{a_{k}}a_{k}^{\dag}+a_{k}^{\dag}{a_{k}}+a_{k}^{\dag}a_{-k}^{\dag}}\right)}
\end{equation}

This is  the zero dimensional case  with an index $k$.
All four terms conserve momentum, but as before two terms change the
particle numbers, two do not.

The momentum space integral of the Lagrangian is:
\begin{equation}
-\imath\int\limits _{0}^{\tau}{d\tau'}{f_{k}}{n_{k}}=-\imath{f_{k}}\tau
\end{equation}
giving the kernel for one frequency as:
\begin{equation}
{K_{T}}\sim\exp\left({-\imath{f_{k}}{n_{k}}\tau}\right)
\end{equation}
If ${n_{k}}=1$ we have the TQM single particle propagator:
\begin{equation}
{{\hat{K}}_{\tau}}\left({k;k'}\right)=\exp\left({-\imath{f_{k}}\tau}\right)\delta\left({k-k'}\right)
\end{equation}

\subsubsection{Measure}

Fock space consists of products of basis wave functions:
\begin{equation}
\left|{\left\{ {n_{k}}\right\} }\right\rangle \to\left|{{n_{0}}{k_{0}}}\right\rangle \left|{{n_{1}}{k_{1}}}\right\rangle \left|{{n_{2}}{k_{2}}}\right\rangle \ldots
\end{equation}

A path is a series of positions in this Fock space. The measure weights
each possible position equally, so the measure is:
\begin{equation}
\mathcal{D}\phi\equiv\prod\limits _{n=0}^{N}{{\mathcal{D}_{n}}\phi,}{\mathcal{D}_{n}}\phi\equiv\prod\limits _{k,{n_{k}}}{dc_{k,{n_{k}}}^{\left(n\right)}}
\end{equation}

\subsubsection{Interaction terms\label{subsubsec:mul-potential}}

\begin{figure}[h]
\begin{centering}
\includegraphics[width=8cm]{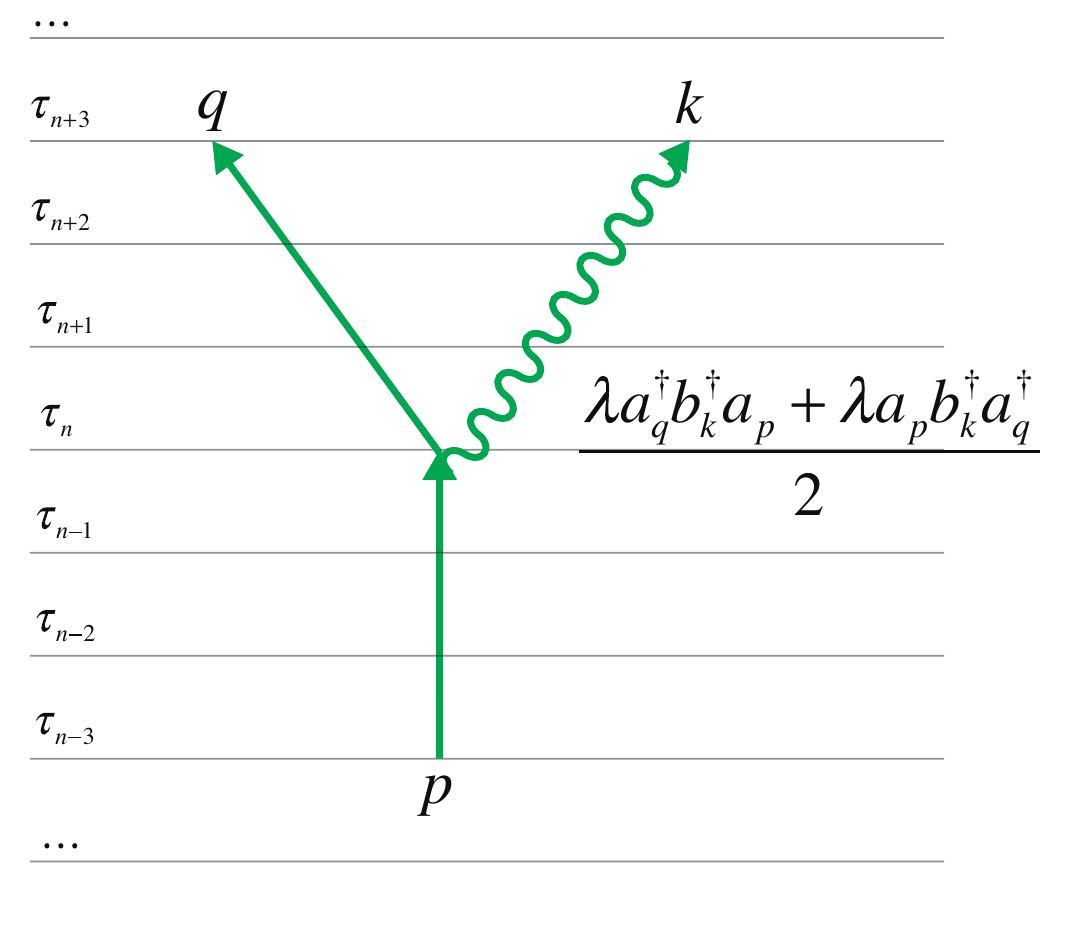} 
\par\end{centering}
\caption{Interaction term in $ABC$ model \label{fig:mul-paths-interaction}}
\end{figure}

The coupling
term is represented by a factor of the form:
\begin{equation}
\exp\left({-\imath\varepsilon\lambda\int{{d^{4}}x\frac{{ABA}}{2}}}\right)
\end{equation}

Assume we
are looking at the case where this happens at clock tick number $n$.
Note -- in striking contrast to the SQM case -- the interaction
term has no dependence on $n$, on the clock time.

As  $N\to\infty$ $\epsilon\to0$ so we can approximate the exponential by:
\begin{equation}
1-\imath\varepsilon\lambda\int{{d^{4}}x\frac{{ABA}}{2}}
\end{equation}

The path integral is formed by doing the integrals from $0\to n-1$,
then the integral over the interaction term at $n$, then the integrals
from $n+1\to N$. The terms before $n$ are included in the free
propagator(s) from $0$ to $n$; the terms after $n$ are included
in the free propagators from $n$ to $N$, we have the interaction
term to consider here.

The integral over space at step $n$ will give us a $\delta$ function
in momentum at step $n$: $\delta\left(k+q-p\right)$. Notice that
four momentum is conserved at the vertex. This is another point of
difference with SQM. In SQM only the three momentum is conserved at
a vertex, the conservation of energy comes from the integral over
the clock time.

For the $nth$ time in our path integral from 0 to $T$, we role the
dice in our game of snakes and ladders. Spelled out in terms of $a$
and $b$ operators we have:
\begin{equation}
\begin{array}{c}
\\
{A_{p}}={a_{p}}\phi_{p}^{\dag}+a_{p}^{\dag}{\phi_{p}}\hfill\\
{B_{k}}={b_{k}}\phi_{k}^{\dag}+b_{k}^{\dag}{\phi_{k}}\hfill\\
{A_{q}}={a_{q}}\phi_{q}^{\dag}+a_{q}^{\dag}{\phi_{q}}\hfill
\end{array}
\end{equation}

We might drop down one step in terms of $A$ particles with momentum
$p$ while going up one step for $A$ particles with momentum $q$
and one step for $B$ particles with momentum $k$. This would be
accomplished by a term of the form $\lambda a_{q}^{\dag}b_{k}^{\dag}{a_{p}}$.
There are two such terms in the interaction term, neatly canceling
out the factor of $\frac{1}{2}$. To contribute to the first order
perturbation diagram the interaction must hit exactly once on the
way from 0 to N. Result:
\begin{equation}
{\hat{\psi}_{T}}\left({q,k}\right)=-\imath\lambda\int\limits _{0}^{T}{d\tau}\int{dp\hat{K}_{T\tau}^{\left(m\right)}\left(q\right)\hat{K}_{T\tau}^{\left(\mu\right)}\left(k\right)\delta\left({q+k-p}\right)\hat{K}_{\tau}^{\left(m\right)}\left(p\right)\hat{\varphi}\left(p\right)}
\end{equation}

\subsubsection{Full propagator}

We now have what is required to compute an arbitrary propagator. 

The
topology of the diagrams is unchanged from SQM: we get exactly the
same set of diagrams, but with the intermediate integrals and initial
wave function(s) over four dimensions rather than three. 

The general
propagator is given by:
\begin{equation}
\langle\left\{ {n_{k'}}\right\} |\int{\mathcal{D}\phi\exp\left({\imath\int\limits _{0}^{T}{d\tau{\mathcal{L}^{free}}\left[\phi\right]-V\left[\phi\right]}}\right)}\left|{\left\{ {n_{k}}\right\} }\right\rangle 
\end{equation}

The Feynman diagrams are generated
by expanding this expression in powers of the coupling constant.

\subsection{Free particles\label{subsec:mul-prop}}

How does the free propagator in TQM compare to the free propagator
in SQM?

\begin{figure}[h]
\begin{centering}
\includegraphics[width=9cm]{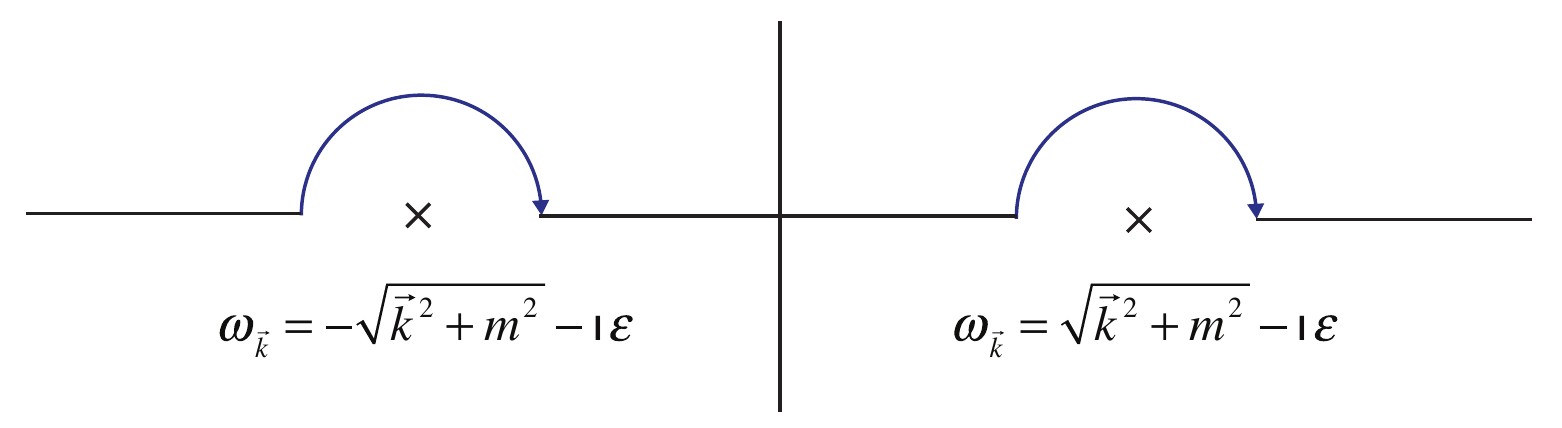} 
\par\end{centering}
\caption{Contour for retarded propagator\label{fig:mul-free-contour}}
\end{figure}

We compute the free propagators for SQM and then TQM. We work these
out for an $A$ particle; the $B$ and $C$ are the same.

Our goal here to establish clearly the relationship between the SQM
and TQM propagators; to make an apples-to-apples comparison between
the two. The best way to do this is to look not just at the propagators
but -- as usual -- at their effects on Gaussian test functions.

We will take as the starting point the respective differential equations
for the SQM and TQM propagators. These may be derived from the path
integral approach using the powerful generating approaches described
in for instance Kashiwa \cite{Kashiwa:1997xt} and Zee \cite{Zee:2010oy}.
Here we take them as a given of the analysis.

\subsubsection{Free particle in SQM}

We start with the Klein-Gordon equation. We define the propagator
by:
\begin{equation}
\left({-\frac{{\partial^{2}}}{{\partial{\tau^{2}}}}+{\nabla^{2}}-{m^{2}}}\right){{\bar{K}}_{\tau}}\left({\vec{x};\vec{x}'}\right)=\imath\delta\left({\tau}\right){\delta^{3}}\left({\vec{x}-\vec{x}'}\right)
\end{equation}

In momentum space we have:
\begin{equation}
\frac{-\imath}{{{k_{0}^{2}}-{{\vec{k}}^{2}}-{m^{2}}}}=\frac{1}{{{\left({2\pi}\right)}^{4}}}
\end{equation}

We choose retarded boundary conditions. This implies that both poles
have a small negative imaginary part:
\begin{equation}
{k_{0}}=\pm{\omega_{\vec{k}}}-\imath\varepsilon,{\omega_{\vec{k}}}\equiv\sqrt{{m^{2}}+{{\vec{k}}^{2}}}
\end{equation}
and the inverse Fourier transform is:
\begin{equation}
{{\bar{K}}_{\tau}}\left({\vec{x},\vec{x}'}\right)=-\imath\mathop{\lim}\limits _{\varepsilon\to0}\frac{1}{{{\left({2\pi}\right)}^{4}}}\int{d{k_{0}}d\vec{k}\frac{{\exp\left({-\imath{k_{0}}\tau+\imath\vec{k}\cdot\left(\vec{x}-\vec{x}'\right)}\right)}}{{{{\left({{k_{0}}+\imath\varepsilon}\right)}^{2}}-{{\vec{k}}^{2}}-{m^{2}}}}}
\end{equation}

Doing the $k_{0}$ integral explicitly we get:
\begin{equation}
{{\bar{K}}_{\tau}}\left({\vec{x},\vec{x}'}\right)=\frac{1}{{{\left({2\pi}\right)}^{3}}}\int{d\vec{k}\frac{{\exp\left({-\imath{\omega_{\vec{k}}}\tau}\right)-\exp\left({\imath{\omega_{\vec{k}}}\tau}\right)}}{{2{\omega_{\vec{k}}}}}\exp\left({\left({\imath\vec{k}\cdot\left(\vec{x}-\vec{x}'\right)}\right)}\right)}
\end{equation}

This corresponds to the basis (see for instance section 2.3 of Peskin \cite{Peskin:1995rv}):
\begin{equation}\frac{{\sqrt {2{\omega _{\vec k}}} }}{{{{\sqrt {2\pi } }^3}}}\exp \left( {\imath\vec k \cdot \vec x} \right)\end{equation}

To make a closer comparison to TQM we shift to the basis functions:
\begin{equation}\frac{1}{{{{\sqrt {2\pi } }^3}}}\exp \left( {i\vec k \cdot \vec x} \right)\end{equation}
This gives:
\begin{equation}{{\bar K}_\tau }\left( {\vec x,\vec x'} \right) = \int {d\vec k\frac{{\exp \left( { - \imath {\omega _{\vec k}}\tau } \right) - \exp \left( {\imath {\omega _{\vec k}}\tau } \right)}}{{{{\left( {2\pi } \right)}^3}}}\exp \left( {\left( {\imath \vec k \cdot \left( {\vec x - \vec x'} \right)} \right)} \right)} \end{equation}

We apply this to a Gaussian test function. We choose one centered
on $\vec{k}_{0}$,
initial position $\vec{x}_{0}$. 
We  choose one which is separable
in the three space directions.
We use the obvious definition $\delta\vec{k}\equiv\vec{k}-\vec{k}_{0}$:
\begin{equation}
{{\hat{\bar{\varphi}}}_{0}}\left({\vec{k}}\right)=\hat{\bar{\varphi}}_{0}^{\left(y\right)}\left({k_{y}}\right)\hat{\bar{\varphi}}_{0}^{y}\left({k_{y}}\right)\hat{\bar{\varphi}}_{0}^{\left(z\right)}\left({k_{z}}\right)
\end{equation}
where the $x$ Gaussian test function is:
\begin{equation}
\hat{\bar{\varphi}}_{0}^{\left(x\right)}\left({k_{x}}\right)=\sqrt[4]{{\frac{1}{{\pi\hat{\sigma}_{x}^{2}}}}}{e^{-\imath{k_{x}}{x_{0}}-\frac{{{\left({{k_{x}}-k_{x}^{\left(0\right)}}\right)}^{2}}}{{2\hat{\sigma}_{x}^{2}}}}}=\sqrt[4]{{\frac{1}{{\pi\hat{\sigma}_{x}^{2}}}}}{e^{-\imath{k_{x}}{x_{0}}-\frac{{\delta k_{x}^{2}}}{{2\hat{\sigma}_{x}^{2}}}}}
\end{equation}
and $y,z$ the same.

We include both positive and negative
frequencies for each wave vector.

We now take as a first working assumption that our incoming wave function
is dominated by the positive frequency part. (This is the same trick
we used in the analysis of the time-of-arrival measurements in subsection
\ref{subsec:free-toa}). If we were going to examine phenomena like
Zitterbewegung we would need to relax this assumption.

As a second assumption, we will assume our incoming Gaussian test function is reasonably well-focused, $\hat{\sigma_{k}}$ small,
so that we can replace the factor overall factor by its expectation:
\begin{equation}\frac{1}{{\sqrt {2{\omega _{\vec k}}} }} \to \frac{1}{{\sqrt {2{\omega _{{{\vec k}_0}}}} }}\end{equation}

We therefore simplify our kernel to:
\begin{equation}{{\hat{\bar{K}}}_\tau }\left( {\vec k;\vec k'} \right) = \frac{1}{{2{\omega _{{{\vec k}_0}}}}}\exp \left( { - \imath {\omega _{\vec k}}\tau } \right){\delta ^3}\left( {\vec k - \vec k'} \right)\end{equation}

We next expand ${\omega_{\vec{k}}}$ in powers of the kinetic energy:
\begin{equation}{\omega _{\vec k}} \approx m + \frac{{{{\vec k}^2}}}{{2m}} - \frac{{{{\vec k}^4}}}{{8{m^3}}} + O\left( {{{\vec k}^6}} \right)\end{equation}

%Alternatively we could take advantage of the fact
%that all our calculations start with Gaussian test functions and expand,
%not around $\vec{k}$ = 0, but around $\vec{k}={\vec{k}}_{0}$:

%\begin{equation}
%{\omega_{\vec{k}}}=\sqrt{{m^{2}}+{{\left({{{\vec{k}}_{0}}+\delta\vec{k}}\right)}^{2}}}\approx{\omega_{0}}+\frac{{\delta\vec{k}\cdot{{\vec{k}}_{0}}}}{{\omega_{0}}}+\frac{{\omega_{0}}}{2}\left({\frac{{\delta\vec{k}\cdot\delta\vec{k}}}{{\omega_{0}^{2}}}-\frac{{{\left({\delta\vec{k}\cdot{{\vec{k}}_{0}}}\right)}^{2}}}{{\omega_{0}^{4}}}}\right)+\ldots
%\end{equation}

Applied to the Gaussian test function we get:
\begin{equation}
{{\hat{\bar{\varphi}}}_{\tau}}\left({\vec{k}}\right)=\int{d\vec{k}'{{\hat{\bar{K}}}_{\tau}}\left({\vec{k};\vec{k}'}\right){{\hat{\bar{\varphi}}}_{0}}\left({\vec{k}}\right)}
\end{equation}
or:
\begin{equation}{{\hat{\bar{\varphi}}}_\tau }\left( {\vec k} \right) = \exp \left( { - \imath m\tau  - \imath \frac{{{{\vec k}^2}}}{{2m}}\tau  + \imath\frac{{{{\vec k}^4}}}{{8{m^3}}}\tau } \right){{\hat{\bar{\varphi}}}_0}\left( {\vec k} \right)\end{equation}

We see on the SQM side the relationship between the various levels
of analysis: the particle at rest ($m$ term), the particle moving
slowly (${\vec k}^{2}$) term, and the particle moving relativistically
(${\vec k}^{4}$ and higher corrections).

We summarize the SQM propagator (as applied to a Gaussian test function)
as:
\begin{equation}{\hat {\bar K}_\tau }\left( {\vec k} \right) = \exp \left( { - \imath m\tau  - \imath \frac{{{{\vec k}^2}}}{{2m}}\tau  + \imath \frac{{{{\vec k}^4}}}{{8{m^3}}}\tau } \right)\end{equation}

\subsubsection{Free particle in TQM}

We have the free propagator from above:
\begin{equation}
{{\hat{K}}_{\tau}}\left({\vec{k}}\right)=\exp\left({-\imath\frac{m}{2}\tau+\imath\frac{{w^{2}}}{{2m}}\tau-\imath\frac{{{\vec{k}}^{2}}}{{2m}}\tau}\right)
\end{equation}

There are three differences between the TQM and SQM propagators. The
first two are not that interesting for our purposes; the third is critical.

First we have an overall factor of: 
\begin{equation}
\exp\left(-{\imath\frac{m}{2}\tau}\right)
\end{equation}
as compared to the factor of: 
\begin{equation}
\exp\left({-\imath m\tau}\right)
\end{equation}
in the SQM propagator.

The frequency associated with, say, the mass of an electron is: 
\begin{equation}
\frac{{m_{e}}}{\hbar}\approx\frac{{.51\cdot{{10}^{6}}eV}}{{6.6\cdot{{10}^{-16}}eV\sec}}=7.7\cdot{10^{-24}}{\sec^{-1}}
\end{equation}

This is of order the frequencies associated with Zitterbewegung, about
one million times the frequencies we are dealing with here. 
As Zitterbewegung has not itself been measured, we would have no comparison
point on the SQM side. We will therefore ignore this factor.

The relativistic correction factor: 
\begin{equation}\exp \left( { + \imath \frac{{{{\vec k}^4}}}{{8{m^3}}}\tau } \right)\end{equation}
is present in SQM but not in TQM.

These represent higher order corrections. As TQM is predicting significant
differences from SQM in even the non-relativistic case, relativistic
corrections to SQM are not needed (and clutter up the analysis).

The key difference is the factor of:
\begin{equation}
\exp\left({+\imath\frac{{w^{2}}}{{2m}}\tau}\right)
\end{equation}

This represents the extension of the function in time/energy. Basically
the TQM propagator is the SQM propagator with additional fuzziness
in time/energy. This difference is the focus of attention in this
work.

So we will take the SQM propagator as: 
\begin{equation}
{{\hat{\bar{K}}}_{\tau}}\left({\vec{k}}\right)=\exp\left(-{\imath\frac{{{\vec{k}}^{2}}}{{2m}}\tau}\right)
\end{equation}
the TQM propagator as: 
\begin{equation}
{{\hat{K}}_{\tau}}\left({\vec{k}}\right)=\exp\left({\imath\frac{{w^{2}}}{{2m}}\tau-\imath\frac{{{\vec{k}}^{2}}}{{2m}}\tau}\right)=\exp\left({\imath\frac{{w^{2}}}{{2m}}\tau}\right){{\hat{\bar{K}}}_{\tau}}\left({\vec{k}}\right)
\end{equation}

We apply this to a Gaussian test function. We use a Gaussian test
function in energy times the previous Gaussian test function in three
momentum.
\begin{equation}
{{\hat{\varphi}}_{0}}\left({w,\vec{k}}\right)={{\hat{\tilde{\varphi}}}_{0}}\left(w\right){{\hat{\bar{\varphi}}}_{0}}\left({\vec{k}}\right)
\end{equation}
with energy part:
\begin{equation}
{{\hat{\tilde{\varphi}}}_{0}}\left(w\right)\equiv\sqrt[4]{{\frac{1}{{\pi\hat{\sigma}_{t}^{2}}}}}{e^{\imath w{t_{0}}-\frac{{{\left({w-{w_{0}}}\right)}^{2}}}{{2\hat{\sigma}_{t}^{2}}}}}
\end{equation}

From the entropic analysis above (subsection \ref{subsec:free-initial}):
\begin{equation}
\begin{array}{c}
{w_{0}}\sim{\omega_{\vec{k}}}\hfill\\
\hat{\sigma}_{t}^{2}\sim\hat{\sigma}_{x}^{2}+\hat{\sigma}_{y}^{2}+\hat{\sigma}_{t}^{2}\hfill
\end{array}
\end{equation}

Application of the TQM kernel to the Gaussian test function is trivial:
\begin{equation}
{{\hat{\varphi}}_{\tau}}\left(k\right)=\exp\left({\imath\frac{{{k^{2}}-{m^{2}}}}{{2m}}\tau}\right){{\hat{\varphi}}_{0}}\left(k\right)
\end{equation}

We have therefore recovered, at the cost of a few approximations, the single particle propagator for TQM.

We turn now to specific applications.

\subsection{Emission of a particle\label{subsec:mul-emis}}

What does the emission of a particle look like in TQM?

\begin{figure}[h]
\begin{centering}
\includegraphics[width=8cm]{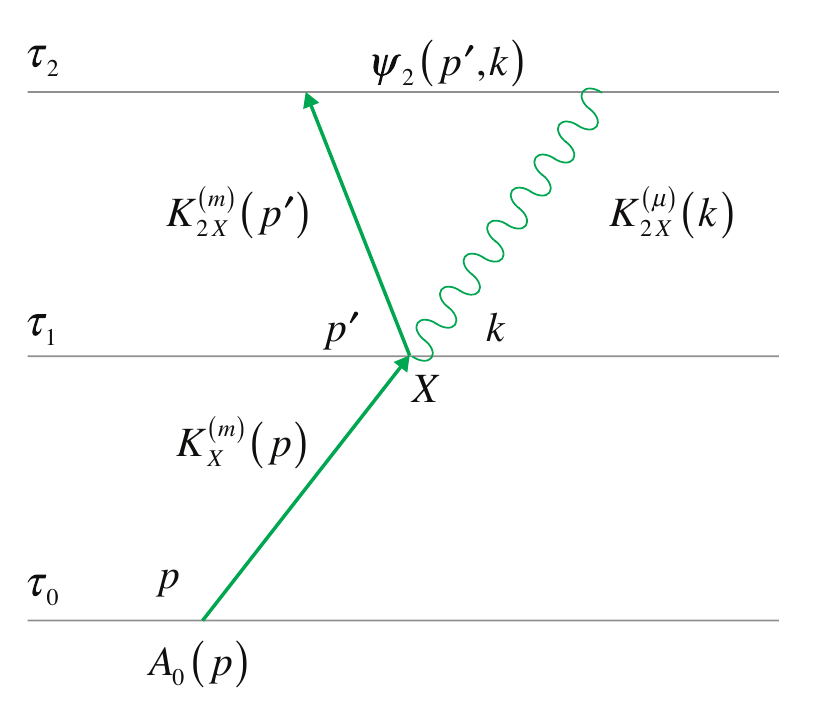} 
\par\end{centering}
\caption{An $A$ particle emits a $B$ particle}
\end{figure}

\subsubsection{Overview}

We look at the case where an $A$ particle emits a $B$. The initial
particle expectation and dispersion are given; we wish to compute
the outgoing particle expectations and dispersions.

We expect that this emission diagram will be part of a larger calculation.
We will assume for convenience here that the momentum space expectations of the incoming particle are on-shell, but that need not be true in general.

We start with a Gaussian test function $A$, extended in space for
the SQM case, in time and space for TQM.

The initial expectation and dispersion are given at clock time $\tau_{0}$.
We wish to compute the final expectations and dispersions at clock
time $\tau_{2}$. The $B$ has an amplitude $\lambda$ to be emitted
at each intermediate clock time $\tau_{1}$.

In a first order perturbation expansion we would integrate over the intermediate
clock time $\tau_{1}$. But here we will focus on a smaller piece
of the puzzle, looking at the contribution to the final wave function
from a single point in clock time. We look at:
\begin{equation}
{\psi_{2}}\left({p,k}\right)=-\imath\lambda\int{dpK_{21}^{\left(m\right)}\left({p'}\right)K_{21}^{\left(\mu\right)}\left(k\right)\delta\left({p'+k-p}\right)K_{10}^{\left(m\right)}\left(p\right)A_{0}\left(p\right)}
\end{equation}

We look at this first in SQM then in TQM.

\subsubsection{Emission of a particle in SQM}

We start our $A$ particle as a Gaussian test function in momentum
space:

\begin{equation}
{{\hat{\bar{A}}}_{0}}\left({\vec{p}}\right)=\hat{\bar{\varphi}}_{a}^{\left(x\right)}\left({p_{y}}\right)\hat{\bar{\varphi}}_{a}^{\left(y\right)}\left({p_{x}}\right)\hat{\bar{\varphi}}_{a}^{\left(z\right)}\left({p_{x}}\right)
\end{equation}

For the $x$ component we have:
\begin{equation}
\hat{\bar{\varphi}}_{a}^{\left(x\right)}\left({p_{x}}\right)\equiv\sqrt[4]{{\frac{1}{{\pi\hat{\sigma}_{x}^{2}}}}}{e^{-\imath{p_{x}}{x_{a}}-\frac{{{\left({{p_{x}}-p_{x}^{\left(a\right)}}\right)}^{2}}}{{2\hat{\sigma}_{x}^{2}}}}}
\end{equation}
with $y,z$ in parallel. The full wave function at $\tau_{0}$ is:
\begin{equation}
{\hat{\bar{A}}}_{0}\left({\vec{p}}\right)=\sqrt[4]{{\frac{1}{{{\pi^{3}}det\left({\hat{\bar{\Sigma}}}\right)}}}}{e^{-\imath\vec{p}\cdot{{\vec{x}}_{a}}-\frac{1}{2}\delta\vec{p}\cdot{{\hat{\bar{\Sigma}}}^{-1}}\cdot\delta\vec{p}}}
\end{equation}
with ancillary definitions:
\begin{equation}
\begin{array}{c}
\delta\vec{p}\equiv\vec{p}-{{\vec{p}}_{a}}\hfill\\
\hat{\bar{\Sigma}}\equiv\left({\begin{array}{ccc}
{\hat{\sigma}_{x}^{2}} & 0 & 0\\
0 & {\hat{\sigma}_{x}^{2}} & 0\\
0 & 0 & {\hat{\sigma}_{x}^{2}}
\end{array}}\right)\hfill
\end{array}
\end{equation}

We are dropping the parts of the kernel that depend on the rest mass.
For the $A$ particle these are independent of the interaction and
therefore irrelevant. For the $B$ particle the $exp\left(-\imath\mu\tau\right)$
is not independent of the interaction, but in the limit as $\mu\to0$
this factor is constant and therefore also irrelevant.

The kernel that carries A from $\tau_{0}\to\tau_{X}$ is:
\begin{equation}
\hat{\bar{K}}_{X}^{\left(m\right)}\left({\vec{p}}\right)=\exp\left(-{\imath\frac{{{\vec{p}}^{2}}}{{2m}}{\tau_{X}}}\right)
\end{equation}

So the $A$ wave function at $X$ is:
\begin{equation}
{\hat{\bar{A}}_{X}}\left({\vec{p}}\right)=\sqrt[4]{{\frac{1}{{{\pi^{3}}det\left({\hat{\bar{\Sigma}}}\right)}}}}{e^{-\imath\vec{p}\cdot{{\vec{x}}_{a}}-\frac{1}{2}\delta\vec{p}\cdot{{\hat{\bar{\Sigma}}}^{-1}}\cdot\delta\vec{p}-\imath\frac{{{\vec{p}}^{2}}}{{2m}}{\tau_{X}}}}
\end{equation}

The integral over the $\delta$ function at $\tau_{X}$ gives:
\begin{equation}
{{\hat{\bar{\psi}}}_{2}}\left({\vec{p}',\vec{k}}\right)=-\imath\lambda\hat{\bar{K}}_{2X}^{\left(m\right)}\left({\vec{p}'}\right)\hat{\bar{K}}_{2X}^{\left(\mu\right)}\left({\vec{k}}\right){\hat{\bar{A}}}_{X}\left({\vec{p}'+\vec{k}}\right)
\end{equation}

We have replaced the initial momentum with the sum of the final momenta.

The post-vertex kernels are:
\begin{equation}
\hat{\bar{K}}_{2X}^{\left(m\right)}\left({\vec{p}'}\right)=\exp\left({-\imath\frac{{\left({\vec{p}'}\right)^{2}}}{{2m}}{\tau_{2X}}}\right),\hat{\bar{K}}_{2X}^{\left(\mu\right)}\left({\vec{k}}\right)=\exp\left({-\imath\frac{{{\vec{k}}^{2}}}{{2\mu}}{\tau_{2X}}}\right)
\end{equation}
with ${\tau_{2X}}\equiv{\tau_{2}}-{\tau_{X}}$.

The joint wave function at $\tau_{2}$ is therefore:
\begin{equation}
{{\hat{\bar{\psi}}}_{2}}\left({\vec{p}',\vec{k}}\right)=\sqrt[4]{{\frac{1}{{{\pi^{3}}det\left({\hat{\bar{\Sigma}}}\right)}}}}e^{\left({-\imath\left(\vec{p}'+\vec{k}\right)\cdot{{\vec{x}}_{a}}-\frac{1}{2}\delta\vec{p'}\cdot{{\hat{\bar{\Sigma}}}^{-1}}\cdot\delta\vec{p'}}{-\imath{\Omega_{X}}{\tau_{2X}}-\imath{\Omega_{0}}{\tau_{X}}}\right)}
\end{equation}
with change in momentum:
\begin{equation}
\delta\vec{p'}\equiv\vec{p}'+\vec{k}-{{\vec{p}}_{a}}
\end{equation}
and initial and final energies:
\begin{equation}
{\Omega_{0}}\equiv\frac{{{\left({\vec{p}'+\vec{k}}\right)}^{2}}}{{2m}},{\Omega_{X}}\equiv\frac{{{\left({\vec{p}'}\right)}^{2}}}{{2m}}+\frac{{{\vec{k}}^{2}}}{{2\mu}}
\end{equation}

We can see that an integral over $\tau_{X}$ would tend to subtract
out components where $\Omega_{0}\ne\Omega_{X}$, giving us an effective
$\delta$ function in the SQM energy.

The final wave function is strongly correlated between left and right.
The conservation condition at the vertex means $A',B$ are each sharing
part of the same initial momentum. They are like Siamese twins --
separated at birth but still connected. This is the source of the
mysterious spooky action at a distance complained of in the initial
EPR paper \cite{Einstein:1935er}.

\subsubsection{Emission of a particle in TQM}

We take the same basic approach, but now with the coordinate energy/coordinate
time included, and with the conservation condition at the vertex being
for four momentum rather than three momentum.

We write the initial wave function as direct product of time and space
parts: $A=\tilde{A}\bar{A}$.
\begin{equation}
{\hat{A}_{0}}\left(p\right)={{\hat{\tilde{A}}}_{a}}\left(E\right){{\hat{\bar{A}}}_{a}}\left({\vec{p}}\right)
\end{equation}

The three momentum part is the same as in SQM. For the energy part
we have:
\begin{equation}
{{\hat{\tilde{A}}}_{a}}\left(E\right)\equiv\sqrt[4]{{\frac{1}{{\pi\hat{\sigma}_{t}^{2}}}}}{e^{\imath E{t_{a}}-\frac{{{\left({E-{E_{a}}}\right)}^{2}}}{{2\hat{\sigma}_{t}^{2}}}}}
\end{equation}

We  take:
\begin{equation}
\begin{array}{c}
{t_{a}}\approx{\tau_{0}}=0\hfill\\
{E_{a}}\approx\sqrt{{m^{2}}+\vec{p}_{a}^{2}}\hfill\\
\hat{\sigma}_{E}^{2}\approx\hat{\sigma}_{x}^{2}+\hat{\sigma}_{y}^{2}+\hat{\sigma}_{z}^{2}\hfill
\end{array}
\end{equation}

We write the entire wave function function as:
\begin{equation}
\hat{A}_{0}\left(p\right)=\sqrt[4]{{\frac{1}{{\pi^{4}{\det}\left({\hat{\Sigma}}\right)}}}}{e^{-\imath p{x_{a}}-\frac{1}{2}\delta p{\hat{\Sigma}}^{{-1}}\delta p}}
\end{equation}
with ancillary definitions:
\begin{equation}
\begin{array}{c}
\delta p\equiv p-{p_{a}}\hfill\\
\hat{\Sigma}\equiv\left({\begin{array}{cccc}
{\hat{\sigma}_{t}^{2}} & 0 & 0 & 0\\
0 & {\hat{\sigma}_{x}^{2}} & 0 & 0\\
0 & 0 & {\hat{\sigma}_{y}^{2}} & 0\\
0 & 0 & 0 & {\hat{\sigma}_{z}^{2}}
\end{array}}\right)\hfill
\end{array}
\end{equation}

The kernel that carries A from $\tau_{0}\to\tau_{X}$ is:
\begin{equation}
K_{X}^{\left(m\right)}\left(p\right)=\exp\left({-\imath f_{p}{\tau_{X}}}\right)=\hat{\tilde{K}}_{X}^{\left(m\right)}\left(E\right)\hat{\bar{K}}_{X}^{\left(m\right)}\left({\vec{p}}\right)\exp\left({-\imath\frac{m}{2}{\tau_{X}}}\right)
\end{equation}
with ${f_{p}}\equiv-\frac{{{E^{2}}-{{\vec{p}}^{2}}+{m^{2}}}}{{2m}}$
and with energy part:
\begin{equation}
\hat{\tilde{K}}_{X}^{\left(m\right)}\left(E\right)\equiv\exp\left({\imath\frac{{E^{2}}}{{2m}}{\tau_{X}}}\right)
\end{equation}

So the $A$ wave function at $X$ is:
\begin{equation}{{\hat A}_X}\left( p \right) = \sqrt[4]{{\frac{1}{{{\pi ^4}det\left( {\hat \Sigma } \right)}}}}\exp \left( { - \imath p \cdot {x_a} - \frac{1}{2}\delta p{{\hat \Sigma }^{ - 1}}\delta p - \imath {f_p}{\tau _X}} \right)\end{equation}

The energy part at $A$ is now:
\begin{equation}{{\hat {\tilde {A}}}_X}\left( E \right) = \sqrt[4]{{\frac{1}{{\pi \hat \sigma _t^2}}}}\exp \left( {  \imath E{t_a} - \frac{{{{\left( {E - {E_a}} \right)}^2}}}{{2\hat \sigma _t^2}} + \imath \frac{{E_a^2}}{{2m}}{\tau _X}} \right)\end{equation}

The integral over the $\delta$ function at $X$ gives:
\begin{equation}
{{\hat{\psi}}_{2}}\left({p',k}\right)=-\imath\lambda\hat{K}_{2X}^{\left(m\right)}\left(p'\right)\hat{K}_{2X}^{\left(\mu\right)}\left(k\right)\hat{A}_{X}\left({p'+k}\right)
\end{equation}

Again we are pushing the sum of the final momenta back into the initial
wave function.

Post vertex kernels:
\begin{equation}
\hat{K}_{2X}^{\left(m\right)}\left(p'\right)=\exp\left({-\imath{f_{p'}}{\tau_{2X}}}\right),\hat{K}_{2X}^{\left(\mu\right)}\left(k\right)=\exp\left({-\imath{f_{k}}{\tau_{2X}}}\right)
\end{equation}

So the wave function at $\tau_{2}$ is:
\begin{equation}{\hat \psi _2}\left( {p',k} \right) = \sqrt[4]{{\frac{1}{{{\pi ^4}\det \left( {\hat \Sigma } \right)}}}}\exp \left( {\imath \left( {p' + k} \right){x_a} - \frac{1}{2}\delta p'{{\hat \Sigma }^{ - 1}}\delta p' - \imath {F_X}{\tau _{2X}} - \imath {F_0}{\tau _X}} \right)\end{equation}
with change in four momentum:
\begin{equation}
\delta p'\equiv p'+k-{p_{a}}
\end{equation}
and initial and final clock frequencies:
\begin{equation}
\begin{array}{c}
{F_{0}}\equiv-\frac{{{{\left({E'+w}\right)}^{2}}-{{\left({\vec{p}'+\vec{k}}\right)}^{2}}-{m^{2}}}}{{2m}},\hfill\\
{F_{X}}\equiv{f_{p'}}+{f_{k}}=-\frac{{{{\left({E'}\right)}^{2}}-{{\left({\vec{p}'}\right)}^{2}}-{m^{2}}}}{{2m}}-\frac{{{w^{2}}-{{\vec{k}}^{2}}-{\mu^{2}}}}{{2\mu}}\hfill
\end{array}
\end{equation}

Per the long, slow approximation, we expect that both $F_{0}$ and
$F_{X}$ will be small. 
And an integral over $\tau_{X}$ would make their difference still smaller.
As noted in the free particle section most
of the dependence on clock time will be carried by the coordinate
time part of the wave function.

The parts dependent on the rest masses do not play a critical role,
for the same reasons as in the SQM case.

The energy part of the joint wave function is:
\begin{equation}
{{\hat{\tilde{\psi}}}_{2}}\left({E',w}\right)=\sqrt[4]{{\frac{1}{{\pi\hat{\sigma}_{t}^{2}}}}}{e^{\imath\left({E'+w}\right){t_{a}}-\frac{{{\left({E'+w-{E_{a}}}\right)}^{2}}}{{2\hat{\sigma}_{t}^{2}}}+\imath\left({\frac{{{\left({E'}\right)}^{2}}}{{2m}}+\frac{{w^{2}}}{{2m}}}\right){\tau_{2X}}+\imath\frac{{{\left({E'+w}\right)}^{2}}}{{2m}}{\tau_{X}}}}
\end{equation}

The left and right halves are again strongly correlated -- now in energy/time as well as in three-momentum/space -- even though with increasing clock time
they are separated by greater and greater distances. Again, they are
like Siamese twins separated at birth but still connected, now across
time as well as space.

\subsubsection{Discussion of particle emission}

With TQM, to the correlations in three-momentum complained of in the
initial EPR paper we add correlations in energy. These provide raw
material for a Bell\textsc{'}s theorem ``in time''.

Presumably Einstein would still be unhappy about the ``spooky action
at a distance'', but perhaps he would be partly consoled by the inclusion
of time on the same basis as space.

\subsection{Absorption of a particle\label{subsec:mul-abs}}

What does the absorption of a particle look like in TQM?

\begin{figure}[h]
\begin{centering}
\includegraphics[width=8cm]{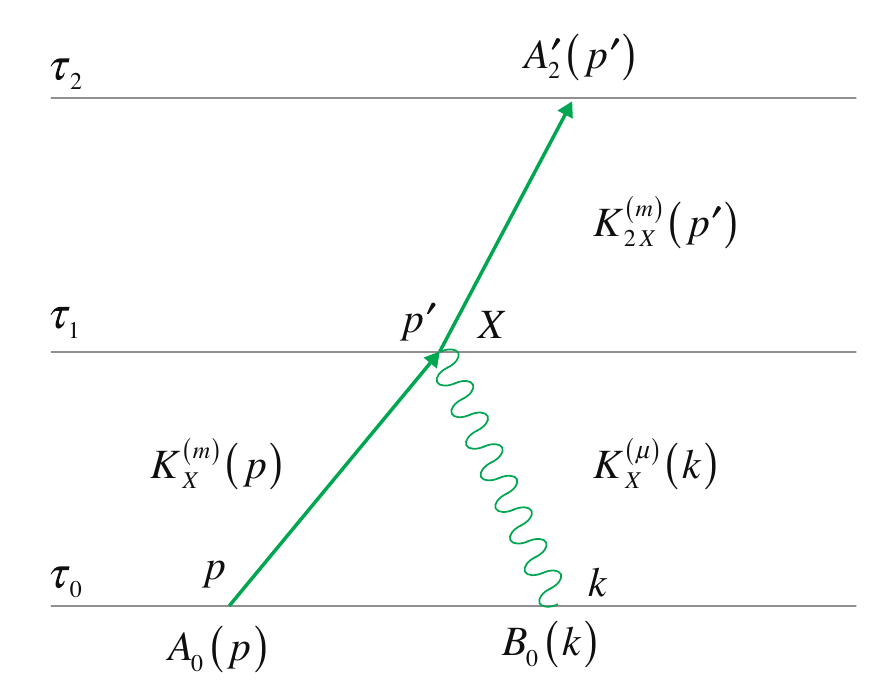} 
\par\end{centering}
\caption{An $A$ particle absorbs a $B$ particle\label{fig:mul-abs}}
\end{figure}

\subsubsection{Overview}

We look at the case where an $A$ absorbs a $B$. The initial particle
expectations and dispersions are given; we wish to compute the outgoing
particle\textsc{'}s expectation and dispersion.

As with emission,
we  expect that this absorption diagram will be part of a larger calculation.
We will assume for convenience here that the momentum space expectations of the incoming particle are on-shell, but that need not be true in general.

We start with two Gaussian test functions $A,B$. These are centered
on momenta $p_{a},k_{b}$ with initial expectations for position $x_{a},x_{b}$.
We define $p'_{a}\equiv p_{a}+k_{b}$.

For simplicity we take the $A$ particle as coming in from the left
and the $B$ as coming in from the right. Without loss of generality
we can assume both are coming in along the x-axis with relative offset
$b$ along the $y$ axis. With a slight loss of generality we will
assume $b\to0$.

We have starting velocities, $v>0,u>0$:
\begin{equation}
{{\vec{p}}_{a}}=mv\hat{x},{{\vec{k}}_{b}}=-\mu u\hat{x}
\end{equation}
and starting points on left and right:
\begin{equation}
{\vec{x}_{a}}=-l\hat{x},{\vec{x}_{b}}=d\hat{x}
\end{equation}

In first order perturbation theory we would compute the final amplitude
by integrating over all intermediate clock times $\tau_{1}$. But
as with emission we focus on the
interaction at a specific clock time $\tau_{X}$. Since we have a
natural clock time to work with -- the time defined by the intersection
of the classical paths of $A,B$ -- we will use that.

This is defined by:
\begin{equation}
x=-l+v{\tau_{X}}=d-u\tau_{X}
\end{equation}
giving crossing time ${\tau_{X}}$: 
\begin{equation}
{\tau_{X}}=\frac{{d+l}}{{v+u}}
\end{equation}
and crossing position $x_{X}$:
\begin{equation}
x_{X}=\frac{{vd-ul}}{{v+u}}
\end{equation}
Trivially $y_{X}=z_{X}=0$.

We are left with one integral to do,
a convolution of the initial momenta:
\begin{equation}
{{\hat{A}}_{2}}\left({p'}\right)=-\imath\lambda\int{dk\hat{K}_{2X}^{\left(m\right)}\left({p'}\right){{\hat{A}}_{X}}\left({p'-k}\right){{\hat{\bar{B}}}_{X}}\left(k\right)}
\end{equation}

We will first treat the SQM case, then TQM.

\subsubsection{Absorption of a particle in SQM}

\paragraph{Initial wave functions}

The initial particles are given by:
\begin{equation}
\begin{array}{c}
{{\hat{\bar{A}}}_{0}}\left({\vec{p}}\right)=\sqrt[4]{{\frac{1}{{{\pi^{3}}det\left({\hat{\bar{\Sigma}}}\right)}}}}{e^{-\imath\vec{p}\cdot{{\vec{x}}_{a}}-\frac{1}{2}\vec{p}\cdot{{\hat{\bar{\Sigma}}}^{-1}}\cdot\vec{p}}}\hfill\\
{{\hat{\bar{B}}}_{0}}\left({\vec{k}}\right)=\sqrt[4]{{\frac{1}{{{\pi^{3}}\det\left({\hat{\bar{S}}}\right)}}}}{e^{-\imath\vec{k}\cdot{{\vec{x}}_{b}}-\frac{1}{2}\vec{k}\cdot{{\hat{\bar{S}}}^{-1}}\cdot\vec{k}}}\hfill
\end{array}
\end{equation}
with expectations and dispersions: 
\begin{equation}
\begin{array}{c}
\vec{p}\equiv\vec{p}-{{\vec{p}}_{a}},\vec{k}\equiv\vec{k}-{{\vec{k}}_{b}}\hfill\\
\hat{\bar{\Sigma}}\equiv\left({\begin{array}{ccc}
{\hat{\sigma}_{x}^{2}} & 0 & 0\\
0 & {\hat{\sigma}_{x}^{2}} & 0\\
0 & 0 & {\hat{\sigma}_{x}^{2}}
\end{array}}\right),\hat{\bar{S}}\equiv\left({\begin{array}{ccc}
{\hat{s}_{x}^{2}} & 0 & 0\\
0 & {\hat{s}_{y}^{2}} & 0\\
0 & 0 & {\hat{s}_{z}^{2}}
\end{array}}\right)\hfill
\end{array}
\end{equation}

In momentum space the kernels from start to $X$ are:
\begin{equation}
\hat{\bar{K}}_{X}^{\left(m\right)}\left({\vec{p}}\right)=\exp\left(-{\imath\frac{{{\vec{p}}^{2}}}{{2m}}{\tau_{X}}}\right),\hat{\bar{K}}_{X}^{\left(\mu\right)}=\exp\left({-\imath\frac{{{\vec{k}}^{2}}}{{2\mu}}{\tau_{X}}}\right)
\end{equation}

The wave functions at $X$ are therefore:
\begin{equation}
\begin{array}{c}
{{\hat{\bar{A}}}_{X}}\left({\vec{p}}\right)=\sqrt[4]{{\frac{1}{{{\pi^{3}}det\left({\hat{\bar{\Sigma}}}\right)}}}}{e^{-\imath\vec{p}\cdot{{\vec{x}}_{a}}-\frac{1}{2}\vec{p}\cdot{{\hat{\bar{\Sigma}}}^{-1}}\cdot\vec{p}-\imath\frac{{{\vec{p}}^{2}}}{{2m}}{\tau_{X}}}}\hfill\\
{{\hat{\bar{B}}}_{X}}\left({\vec{k}}\right)=\sqrt[4]{{\frac{1}{{{\pi^{3}}\det\left({\hat{\bar{S}}}\right)}}}}{e^{-\imath\vec{k}\cdot{{\vec{x}}_{b}}-\frac{1}{2}\vec{k}\cdot{{\hat{\bar{S}}}^{-1}}\cdot\vec{k}-\imath\frac{{{\vec{k}}^{2}}}{{2\mu}}{\tau_{X}}}}\hfill
\end{array}
\end{equation}

\paragraph{Interaction}

The final wave function at $\tau_{2}$ will be given by a convolution
of all possible incoming momenta:
\begin{equation}
{\hat{\bar{A}}}_{2}\left({\vec{p}'}\right)=-\imath\lambda\hat{K}_{2X}^{\left(m\right)}\left({\vec{p}'}\right)\int{d\vec{k}{{\hat{A}}_{X}}\left({\vec{p}'-\vec{k}}\right){{\hat{B}}_{X}}\left({\vec{k}}\right)}
\end{equation}

We  focus on the convolution integral over $x$:
\begin{equation}
\hat{I}_{X}^{\left(x\right)}\left({{p'}_{x}}\right)\equiv\int{d{k_{x}}{\hat{A}_{X}}\left({{{p'}_{x}}-{k_{x}}}\right){\hat{B}_{X}}\left({k_{x}}\right)}
\end{equation}

Since this is the integral of a Gaussian it can be solved exactly.
However we will get more insight by shifting to the position basis.
The convolution in momentum space becomes a multiplication in position
space:
\begin{equation}
{I_{X}^{\left(x\right)}}\left(x\right)=\sqrt{2\pi}{A_{X}}\left(x\right){B_{X}}\left(x\right)
\end{equation}

The Gaussian test functions for $A$ and $B$ are centered on their
corresponding classical paths.

The close relationship of classical and quantum trajectories is
an attractive feature of the approach here; we can think of a particle
as traveling along a classical line with quantum fuzz around it.
In SQM the fuzz extends out in the three space dimensions;
in TQM in the time dimension as well.

At the crossing time, the coordinate forms for $A$ and $B$ at $X$
are therefore:
\begin{equation}
\begin{array}{c}
{A_{X}}\left(x\right)=F_{X}^{\left(x\right)}{e^{\imath p_{x}^{\left(a\right)}\left({x-{x_{a}}}\right)-\frac{1}{{2\sigma_{x}^{2}f_{X}^{\left(x\right)}}}{{\left({x-{x_{X}}}\right)}^{2}}-\imath\frac{{p_{x}^{\left(a\right)2}}}{{2m}}{\tau_{X}}}}\hfill\\
{B_{X}}\left(x\right)=G_{X}^{\left(x\right)}{e^{\imath k_{x}^{\left(b\right)}\left({x-{x_{b}}}\right)-\frac{1}{{2s_{x}^{2}g_{X}^{\left(x\right)}}}{{\left({x-{x_{X}}}\right)}^{2}}-\imath\frac{{k_{x}^{\left(b\right)2}}}{{2\mu}}{\tau_{X}}}}\hfill
\end{array}
\end{equation}

The dispersion factors $f,g$ and normalization factors $F,G$ are
spelled out in \ref{subsec:app-conv-gtfs}.

We work along the same lines as for the single slit above (subsection \ref{subsec:single-tq}).

We define effective crossing times $\tau^{*}$ and dispersions
${\sigma}_{x}^{*2}$ via:
\begin{equation}\frac{1}{{\sigma _x^2 + \imath \frac{{{\tau _X}}}{m}}} + \frac{1}{{s_x^2 + \imath \frac{{{\tau _X}}}{\mu }}} = \frac{1}{{\sigma _x^{*2} + \imath \frac{{{\tau ^*}}}{m}}}\end{equation}

The effect of the interaction is to change the shape of the $A$ wave function so that it looks 
as if it had started at time $\tau_{X}-{\tau^{*}}$ with dispersion ${\sigma}_{x}^{*2}$. 

We clear the denominators:
\begin{equation}\left( {\sigma _x^2 + s_x^2} \right)\sigma _x^{*2} + \imath \left( {\frac{{{\tau _X}}}{m} + \frac{{{\tau _X}}}{\mu }} \right)\sigma _x^{*2} - \left( {\frac{{{\tau _X}}}{m} + \frac{{{\tau _X}}}{\mu }} \right)\frac{{{\tau ^*}}}{m} + \imath \left( {\sigma _x^2 + s_x^2} \right)\frac{{{\tau ^*}}}{m} = \left( {\sigma _x^2s_x^2 - \frac{{\tau _X^2}}{{m\mu }}} \right) + \imath \left( {\sigma _x^2\frac{{{\tau _X}}}{\mu } + s_x^2\frac{{{\tau _X}}}{m}} \right)\end{equation}
We equate the real and imaginary parts:
\begin{equation}\begin{array}{c}
  \left( {s_x^2 + \sigma _x^2} \right)\sigma _t^{*2} - \left( {\frac{{{\tau _X}}}{m} + \frac{{{\tau _X}}}{\mu }} \right)\frac{{{\tau ^*}}}{m} = \sigma _x^2s_x^2 - \frac{{\tau _X^2}}{{m\mu }} \hfill \\
  \left( {\frac{{{\tau _X}}}{m} + \frac{{{\tau _X}}}{\mu }} \right)\sigma _x^{*2} + \left( {\sigma _x^2 + s_x^2} \right)\frac{{{\tau ^*}}}{m} = \sigma _x^2\frac{{{\tau _X}}}{\mu } + s_x^2\frac{{{\tau _X}}}{m} \hfill \\ 
\end{array} \end{equation}

We have a two by two matrix equation for $\sigma _x^{*2},{\tau ^*}$, which we invert and apply to the right hand side:
\begin{equation}\left( {\begin{array}{*{20}{c}}
  {\sigma _x^{*2}} \\ 
  {\frac{{{\tau ^*}}}{m}} 
\end{array}} \right) = \frac{1}{D}\left( {\begin{array}{*{20}{c}}
  {s_x^2 + \sigma _x^2}&{\frac{{{\tau _X}}}{m} + \frac{{{\tau _X}}}{\mu }} \\ 
  { - \left( {\frac{{{\tau _X}}}{m} + \frac{{{\tau _X}}}{\mu }} \right)}&{s_x^2 + \sigma _x^2} 
\end{array}} \right)\left( {\begin{array}{*{20}{c}}
  {\sigma _x^2s_x^2 - \frac{{\tau _X^2}}{{m\mu }}} \\ 
  {\sigma _x^2\frac{{{\tau _X}}}{\mu } + s_x^2\frac{{{\tau _X}}}{m}} 
\end{array}} \right)\end{equation}
with determinant $D$:
\begin{equation}D = {\left( {\sigma _x^2 + s_x^2} \right)^2} + {\left( {\frac{{{\tau _X}}}{m} + \frac{{{\tau _X}}}{\mu }} \right)^2}\end{equation}

The resulting expressions for the rescaling constants are relatively complex; 
the shape of $A'$ is a marriage of the shapes of $A,B$:
\begin{equation}\begin{array}{c}
  \sigma _x^{*2} = \frac{1}{D}\left( {\left( {s_x^2 + \sigma _x^2} \right)\left( {\sigma _x^2s_x^2 - \frac{{\tau _X^2}}{{m\mu }}} \right) + \left( {\frac{{{\tau _X}}}{m} + \frac{{{\tau _X}}}{\mu }} \right)\left( {\sigma _x^2\frac{{{\tau _X}}}{\mu } + s_x^2\frac{{{\tau _X}}}{m}} \right)} \right) \hfill \\
  \frac{{{\tau ^*}}}{m} = \frac{1}{D}\left( { - \left( {\frac{{{\tau _X}}}{m} + \frac{{{\tau _X}}}{\mu }} \right)\left( {\sigma _x^2s_x^2 - \frac{{\tau _X^2}}{{m\mu }}} \right) + \left( {s_x^2 + \sigma _x^2} \right)\left( {\sigma _x^2\frac{{{\tau _X}}}{\mu } + s_x^2\frac{{{\tau _X}}}{m}} \right)} \right) \hfill \\ 
\end{array} \end{equation}

The resulting wave function is:
\begin{equation}
{I_{X}^{\left(x\right)}}\left(x\right)=N_{X}^{\left(x\right)}{\varphi}_{X}^{\left({*}\right)}\left(x\right)
\end{equation}
with:
\begin{equation}\varphi _X^{\left( * \right)}\left( x \right) = F_\tau ^{\left( x \right)*}{e^{\imath p_x^{\left( {a'} \right)}\left( {x - {x_X}} \right) - \frac{1}{{2\sigma _x^{*2}f_X^{*\left( x \right)}}}{{\left( {x - {x_X}} \right)}^2} - \imath \frac{{p_x^{\left( {a'} \right)2}}}{{2m}}{\tau _X}}}\end{equation}
and an overall constant $N_{X}^{\left(x\right)}$ independent of $x$.
The overall constant will drop out when we calculate the final expectation
and dispersion, so the effective dispersion and crossing time carry
all the physically significant information.

\paragraph{Final wave function}

Since all $x$ dependence is carried by $\varphi_{X}^{\left(*\right)}\left(x\right)$,
we have the Fourier transform by inspection:
\begin{equation}
\hat{I}_{X}^{\left(x\right)}\left({{p'}_{x}}\right)=N_{X}^{\left(x\right)}\hat{\varphi}_{X}^{\left({*}\right)}\left({{p'}_{x}}\right)
\end{equation}
with the momentum space form of the starred wave function:
\begin{equation}{{\hat{\bar{\varphi}} }^*}_X\left( {{{p'}_x}} \right) = \sqrt[4]{{\frac{1}{{\pi \hat \sigma _X^{*2}}}}}{e^{ - \imath p_x^\prime {x_X} - \frac{{{{\left( {{{p'}_x} - p_x^{\left( {a'} \right)}} \right)}^2}}}{{2\hat \sigma _X^{*2}}} - \imath \frac{{{{\vec p}^{\prime 2}}}}{{2m}}{\tau _X}}}\end{equation}

The other two dimensions work in parallel. Now that we are back in
momentum space we have the resulting wave function at $\tau_{2}$:
\begin{equation}
{\hat{\bar{A}}}_{2}\left({\vec{p}'}\right)=-\imath\lambda\exp\left({-\imath\frac{\vec{p}^{\prime2}}{{2m}}{\tau_{2X}}}\right)N_{X}^{\left(x\right)}N_{X}^{\left(y\right)}N_{X}^{\left(z\right)}{{\hat{\bar{\varphi}}^{*}}_{X}}\left({\vec{p}'}\right)
\end{equation}

For our purposes the most interesting aspect is the associated uncertainty
in momentum. The overall normalization drops out:
\begin{equation}
\left\langle {{\left({\vec{p}'-{{\vec{p}}^{\left({a'}\right)}}}\right)}^{2}}\right\rangle =\frac{{\int{d\vec{p}'{{\left({\vec{p}'-{{\vec{p}}^{\left({a'}\right)}}}\right)}^{2}}{{\left|{{{\hat{\bar{\varphi}}}_{X}}\left({\vec{p}'}\right)}\right|}^{2}}}}}{{\int{d\vec{p}'{{\left|{{{\hat{\bar{\varphi}}}_{X}}\left({\vec{p}'}\right)}\right|}^{2}}}}}
\end{equation}

The uncertainty in $p$ is defined by the post-interaction wave function.
For the $x$ direction this is:
\begin{equation}
\hat{\sigma}_{X}^{*2}=\frac{1}{{\sigma_{X}^{*2}}\ensuremath{}}
\end{equation}

We can read off the physically important part of the resulting wave
function from the starred dispersion. 

The simplest case is when $B$ is both narrow in space $s_{x}\ll \sigma_{x}$ and heavy $\mu \gg m$.
In this case the effective dispersions and start time take the simple form:
\begin{equation}\begin{array}{c}
  \sigma _x^{*2} \to s_x^2 \hfill \\
  \frac{{{\tau ^*}}}{m} \to 0 \hfill \\ 
\end{array} \end{equation}
Operationally,
if we detect $A'$ at all, we know to within $s_{x}\ll\sigma_{x}$
where $A$ was at $\tau_{X}$. 
But this $A'$ has momentum $p'_{x}$
with a large effective dispersion $\hat{\sigma}_{x}^{*}\sim{1\mathord{\left/{\vphantom{1{s_{x}}}}\right.\kern-\nulldelimiterspace}{s_{x}}}$.
The momentum of $A'$ has become much more uncertain.

So the $B$ functions as a classical gate,
with the familiar reciprocal relationship between the position and momentum.

\subsubsection{Absorption of a particle in TQM}

We work along the same lines to extend the analysis to TQM. The classical
trajectories now include a time component:
\begin{equation}
t={t_{0}}+\gamma\tau
\end{equation}

We will treat the time parts of the wave functions as non-relativistic,
$\gamma\approx1$. This is consistent with our use of the non-relativistic
approximation for the space parts.

We will assume our initial wave functions are centered on $\tau_{0}$:
\begin{equation}
{t_{0}}={\tau_{0}}\Rightarrow{t_{0}}=0
\end{equation}

As a result we have the same intersection point in coordinate time
that we have in clock time:
\begin{equation}
{t_{X}}={\tau_{X}}
\end{equation}

\paragraph{Initial wave functions}

For the initial wave functions we have $A$ and $B$ as products of
their time and space parts:

\begin{equation}
\begin{array}{c}
{{\hat{A}}_{0}}\left(p\right)=\sqrt[4]{{\frac{1}{{\pi det\left({\hat{\Sigma}}\right)}}}}{e^{-\imath p{x_{a}}-\frac{1}{2} p\hat{\Sigma} p}}={{\hat{\tilde{A}}}_{0}}\left(E\right){{\hat{\bar{A}}}_{0}}\left({\vec{p}}\right)\hfill\\
{{\hat{B}}_{0}}\left(k\right)=\sqrt[4]{{\frac{1}{{{\pi^{4}}\det\left({\hat{S}}\right)}}}}{e^{\imath k{x_{b}}-\frac{1}{2} k{{\hat{S}}^{-1}} k}}={{\hat{\tilde{B}}}_{0}}\left(w\right){{\hat{\bar{B}}}_{0}}\left({\vec{k}}\right)\hfill
\end{array}
\end{equation}
with time parts:
\begin{equation}
\begin{array}{c}
{{\hat{\tilde{A}}}_{0}}\left(E\right)=\sqrt[4]{{\frac{1}{{\pi\hat{\sigma}_{t}^{2}}}}}{e^{\imath E{t_{a}}-\frac{{{\left({E-{E_{a}}}\right)}^{2}}}{{2\hat{\sigma}_{t}^{2}}}}}\hfill\\
{{\hat{\tilde{B}}}_{0}}\left(w\right)=\sqrt[4]{{\frac{1}{{\pi\hat{s}_{t}^{2}}}}}{e^{\imath w{t_{b}}-\frac{1}{{2\hat{s}_{t}^{2}}}{{\left({w-{w_{b}}}\right)}^{2}}}}\hfill
\end{array}
\end{equation}
and:
\begin{equation}
\begin{array}{c}
 p\equiv p-{p_{a}}, k\equiv k-{k_{b}}\hfill\\
\hat{\Sigma}\equiv\left({\begin{array}{cccc}
{\hat{\sigma}_{t}^{2}} & 0 & 0 & 0\\
0 & {\hat{\sigma}_{x}^{2}} & 0 & 0\\
0 & 0 & {\hat{\sigma}_{y}^{2}} & 0\\
0 & 0 & 0 & {\hat{\sigma}_{z}^{2}}
\end{array}}\right),\hat{S}\equiv\left({\begin{array}{cccc}
{\hat{s}_{t}^{2}} & 0 & 0 & 0\\
0 & {\hat{s}_{x}^{2}} & 0 & 0\\
0 & 0 & {\hat{s}_{y}^{2}} & 0\\
0 & 0 & 0 & {\hat{s}_{z}^{2}}
\end{array}}\right)\hfill
\end{array}
\end{equation}
and expectations and dispersions in time/energy:
\begin{equation}
\begin{array}{c}
{t_{a}}\approx{\tau_{0}}=0,{t_{b}}\approx{\tau_{0}}=0\hfill\\
{E_{a}}\approx\sqrt{{m^{2}}+\vec{p}_{a}^{2}},{w_{b}}\approx\sqrt{{\mu^{2}}+\vec{k}_{b}^{2}}\hfill\\
\hat{\sigma}_{E}^{2}\approx\hat{\sigma}_{x}^{2}+\hat{\sigma}_{y}^{2}+\hat{\sigma}_{z}^{2},\hat{s}_{t}^{2}\approx\hat{s}_{x}^{2}+\hat{s}_{y}^{2}+\hat{s}_{z}^{2}\hfill
\end{array}
\end{equation}

The wave functions at $X$ are:
\begin{equation}
\begin{array}{c}
{{\hat{A}}_{X}}\left(p\right)=\sqrt[4]{{\frac{1}{{{\pi^{4}}det\left({\hat{\Sigma}}\right)}}}}{e^{\imath p{x_{a}}-\frac{1}{2} p{{\hat{\Sigma}}^{-1}} p}}{e^{-\imath{f_{p}}{\tau_{X}}}}\hfill\\
{{\hat{B}}_{X}}\left(k\right)=\sqrt[4]{{\frac{1}{{{\pi^{4}}\det\left({\hat{S}}\right)}}}}{e^{\imath k{x_{b}}-\frac{1}{2} k{{\hat{S}}^{-1}} k-\imath{f_{k}}{\tau_{X}}}}\hfill
\end{array}
\end{equation}

\paragraph{Interaction}

To compute the wave function at $\tau_{2}$ we again convolute the
incoming wave functions:
\begin{equation}
{{\hat{A}}_{2}}\left({p'}\right)=-\imath\lambda\hat{K}_{2X}^{\left(m\right)}\left({p'}\right)\int{dk{{\hat{A}}_{X}}\left({p'-k}\right){{\hat{B}}_{X}}\left({k}\right)}
\end{equation}

The coordinate energy part of the integral in momentum space is:
\begin{equation}
\hat{\tilde{I}}_{X}\left({E'}\right)\equiv\int{d{w}{\hat{\tilde{A}}}\left({{E'}-{w}}\right){\hat{\tilde{B}}}\left({w}\right)}
\end{equation}

In coordinate time:
\begin{equation}
\tilde{I}_{X}\left(t\right)=\sqrt{2\pi}\tilde{A}_{X}\left(t\right)\tilde{B}_{X}\left(t\right)
\end{equation}

The wave functions in time are:
\begin{equation}
\begin{array}{c}
{{\tilde{A}}_{X}}\left(t\right)=F_{X}^{\left(t\right)}{e^{-\imath{E_{a}}t-\frac{1}{{2\sigma_{t}^{2}f_{X}^{\left(t\right)}}}{{\left({t-{t_{a}}-\frac{{E_{a}}}{m}{\tau_{X}}}\right)}^{2}}+\imath\frac{{E_{a}^{2}}}{{2m}}{\tau_{X}}}}\hfill\\
{{\tilde{B}}_{X}}\left(t\right)=G_{X}^{\left(t\right)}{e^{-\imath{w_{b}}t-\frac{1}{{2s_{t}^{2}g_{X}^{\left(t\right)}}}{{\left({t-{t_{b}}-\frac{{w_{b}}}{\mu}{\tau_{X}}}\right)}^{2}}+\imath\frac{{w_{b}^{2}}}{{2\mu}}{\tau_{X}}}}\hfill
\end{array}
\end{equation}

The quadratic arguments of the Gaussians both reduce to ${\left({t-{\tau_{X}}}\right)^{2}}$
in our non-relativistic approximation.

As before, we define effective crossing times $\tau^{*}$ and dispersions
${\sigma}_{t}^{*2}$ via:
\begin{equation}\frac{1}{{\sigma _t^2 - \imath \frac{{{\tau _X}}}{m}}} + \frac{1}{{s_t^2 - \imath \frac{{{\tau _X}}}{\mu }}} = \frac{1}{{\sigma _t^{*2} - \imath \frac{{{\tau ^*}}}{m}}}\end{equation}

Again, effect of the interaction is to change the shape of the $A$ wave function so that it looks 
as if it had started at time $\tau_{X}-{\tau^{*}}$ with dispersion ${\sigma}_{t}^{*2}$. 

As before, we clear the denominators, equate the real and imaginary parts, 
and invert the two by two matrix equation for $\sigma _t^{*2},{\tau ^*}$ to get:
\begin{equation}\begin{array}{c}
  \sigma _x^{*2} = \frac{1}{D}\left( {\left( {\sigma _t^2 + s_t^2} \right)\left( {\sigma _t^2s_t^2 - \frac{{\tau _X^2}}{{m\mu }}} \right) + \left( {\frac{{{\tau _X}}}{m} + \frac{{{\tau _X}}}{\mu }} \right)\left( {\sigma _t^2\frac{{{\tau _X}}}{\mu } + s_t^2\frac{{{\tau _X}}}{m}} \right)} \right) \hfill \\
  \frac{{{\tau ^*}}}{m} = \frac{1}{D}\left( { - \left( {\frac{{{\tau _X}}}{m} + \frac{{{\tau _X}}}{\mu }} \right)\left( {\sigma _t^2s_t^2 - \frac{{\tau _X^2}}{{m\mu }}} \right) + \left( {\sigma _t^2 + s_t^2} \right)\left( {\sigma _t^2\frac{{{\tau _X}}}{\mu } + s_t^2\frac{{{\tau _X}}}{m}} \right)} \right) \hfill \\ 
\end{array} \end{equation}
with determinant $D$:
\begin{equation}D = {\left( {\sigma _t^2 + s_t^2} \right)^2} + {\left( {\frac{{{\tau _X}}}{m} + \frac{{{\tau _X}}}{\mu }} \right)^2}\end{equation}

As before we have:
\begin{equation}
{{\tilde{I}}_{X}}\left(t\right)={{\tilde{N}}_{X}}\tilde{\varphi}_{X}^{\left(*\right)}\left(t\right)
\end{equation}
with starred wave function:
\begin{equation}
\tilde{\varphi}_{X}^{\left(*\right)}\left(t\right)\equiv F_{X}^{\left(t\right)*}{e^{\imath{{E'}_{a}}\left({t-{t_{X}}}\right)-\frac{1}{{2\sigma_{t}^{*2}f_{X}^{*\left(t\right)}}}{{\left({t-{t_{X}}}\right)}^{2}}-\imath\frac{{{{E'}_{a}}^{2}}}{{2m}}\tau_{X}}}
\end{equation}

Since all $t$ dependence is carried by $\tilde{\varphi}_{X}^{\left(*\right)}\left(t\right)$,
we have the Fourier transform by inspection:
\begin{equation}
\hat{I}_{X}^{\left(x\right)}\left({{p'}_{x}}\right)=\tilde{N}_{X}\hat{\tilde{\varphi}}_{X}^{\left({*}\right)}\left({E'}\right)
\end{equation}
with the momentum space form of the starred wave function:
\begin{equation}
\hat{\tilde{\varphi}}_{X}^{\left(*\right)}\left({E'}\right)=\sqrt[4]{{\frac{1}{{\pi\hat{\sigma}_{X}^{\left(t\right)*2}}}}}{e^{\imath E'{x_{X}}-\frac{{{\left({E'-{{E'}_{a}}}\right)}^{2}}}{{2\hat{\sigma}_{X}^{\left(t\right)*2}}}+\imath\frac{{{\left({E'}\right)}^{2}}}{{2m}}{\tau_{X}}}}
\end{equation}
and therefore the final wave function:
\begin{equation}
\begin{array}{c}
{{\hat{A}}_{2}}\left({p'}\right)={{\hat{\tilde{A}}}_{2}}\left({E'}\right){{\hat{\bar{A}}}_{2}}\left({\vec{p}'}\right)\hfill\\
{{\hat{\tilde{A}}}_{2}}\left({E'}\right)=\exp\left({\imath\frac{{{E'}^{2}}}{{2m}}{\tau_{2X}}}\right)N_{X}^{\left(t\right)}{{\hat{\tilde{\varphi}}}_{X}}\left({E'}\right)\hfill
\end{array}
\end{equation}

\paragraph{Uncertainty in time}

For our purposes the most interesting aspect is the associated uncertainty
in energy. The overall normalization drops out:
\begin{equation}
\left\langle {{\left({E'-{E^{\left({a'}\right)}}}\right)}^{2}}\right\rangle =\frac{{\int{dE'{{\left({E-{E^{\left({a'}\right)}}}\right)}^{2}}{{\left|{{{\hat{\tilde{\varphi}}}_{X}}\left(E\right)}\right|}^{2}}}}}{{\int{dE'{{\left|{{{\hat{\tilde{\varphi}}}_{X}}\left(E\right)}\right|}^{2}}}}}
\end{equation}

The post-interaction dispersion in $E$ is given by the post-interaction
dispersion in $t$:
\begin{equation}
\sigma_{E}^{*2}\equiv\hat{\sigma}_{t}^{*2}=\frac{1}{{\sigma_{t}^{*2}}}
\end{equation}
and the uncertainty in $E$ by the usual formula:
\begin{equation}
\Delta E=\sqrt{\frac{{\sigma_{E}^{*2}}}{2}}
\end{equation}

With the dispersion in coordinate time we can calculate the dispersion
in clock time at a detector using the formulas above in subsection
\ref{subsec:free-toa}.

The increase in complexity from using a quantum particle as the gate,
rather than a classical camera shutter, is significant.

The simplest case is when $B$ is both narrow in time $s_{x}\ll \sigma_{x}$ and  heavy $\mu \gg m$.
In this case the effective dispersions and start time take the form:
\begin{equation}\begin{array}{c}
  \sigma _t^{*2} \to s_t^2 \hfill \\
  \frac{{{\tau ^*}}}{m} \to 0 \hfill \\ 
\end{array} \end{equation}

In this case we can apply the same formulae as with the single slit in time;
getting the same arbitrarily large increase in the dispersion in time-of-arrival.

In the more general case, a certain amount of ingenuity may be required to tease out the
effects of the increased dispersion in time.

\subsubsection{Distinguishing between uncertainty in space and in time}\label{paragraph:mul-abs-two-different}

And this point takes us to a general
problem in looking for uncertainty in time with time-of-arrival measurements.
If we use time-of-arrival measurements we are measuring dispersion
in time, yes, but we are also measuring dispersion in momentum along
the axis of flight. Since we expect the initial dispersions in time
to be of order of those in space, it may be difficult to \textit{prove}
there was no dispersion in time. Perhaps it was lost in the error
bars?

The basic problem is the time-of-arrival measurement is being used
to measure two different things: one measurement cannot serve two
masters.

One way to separate the two measurements would be to run the post-interaction
particle through a magnetic field. Let\textsc{'}s say the particle
will be bent to the right by the magnetic field. The $y$ position
will serve as usual as a measurement of velocity. But if at each $y$
position we also record the time-of-arrival, the time-of-arrival should
now serve as a measurement of dispersion in coordinate time.

If we graph the clicks on a $y,\tau$ grid, the faster particles will
hit earlier in time and more to the left. In SQM we would expect to
see a relatively narrow trace from small $y,\tau$ to large $y,\tau$.
In TQM we would expect to see the same trace on average, but significantly
broader in $\tau$ at each $y$.   

The SQM trace will look more like the thin scar left by a rapier; 
the TQM trace more like the thicker scar left by a saber.

\subsection{Exchange of a particle\label{subsec:mul-exch}}

What does the exchange of a particle look like in TQM?

\begin{figure}[h]
\begin{centering}
\includegraphics[width=11cm]{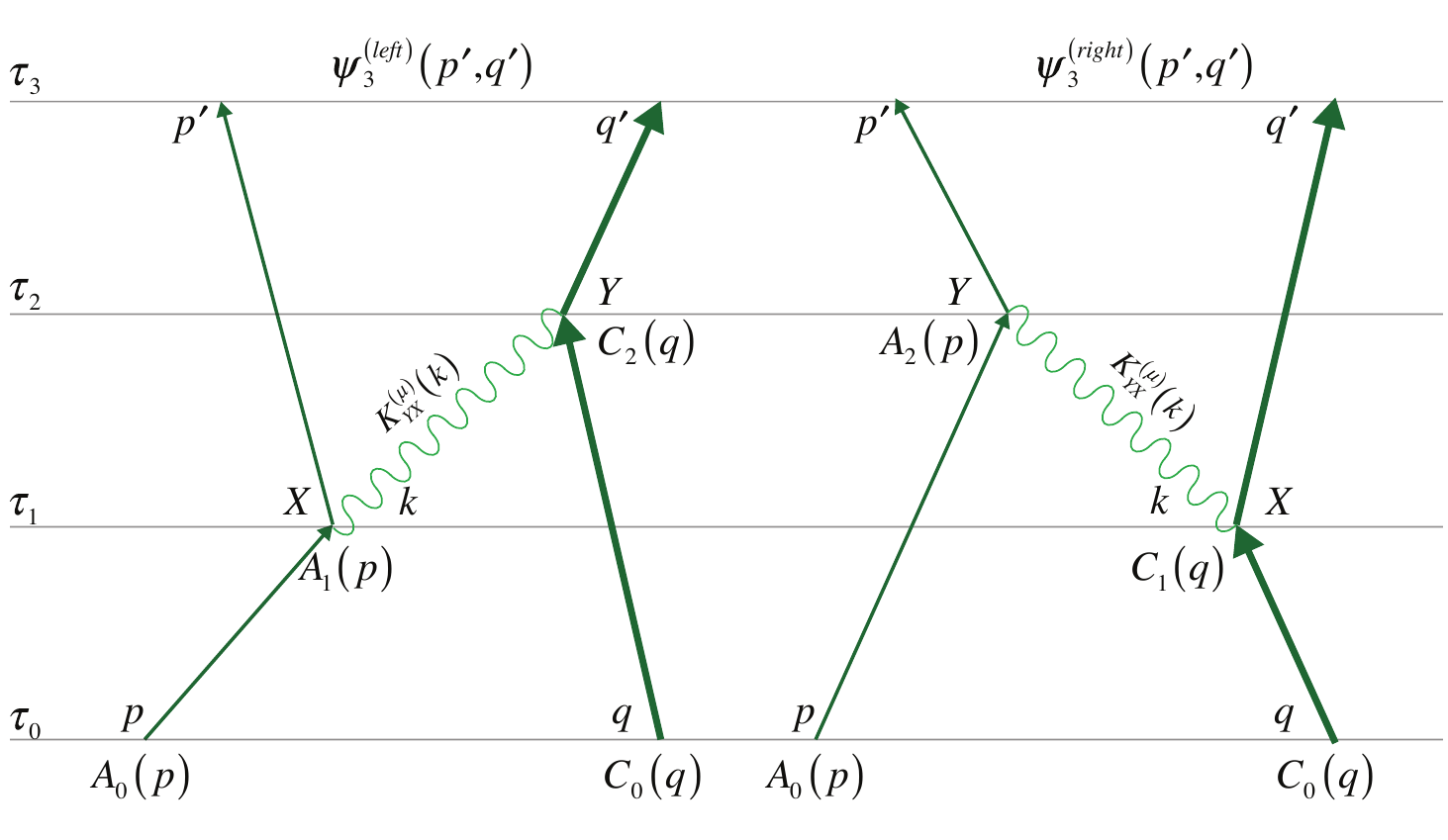} 
\par\end{centering}
\caption{An $A$ particle exchanges a $B$ particle with a $C$ particle}
\end{figure}

\subsubsection{Overview}

We look at the case where an $A$ and a $C$ exchange a $B$. The
initial particle expectations and dispersions are given; we wish to
compute the outgoing particle expectation and dispersions. We have
two cases: the $A$ emits a $B$ which is then absorbed by $C$ and
the $C$ emits a $B$ which is then absorbed by the $A$. We will
call these the left and right cases.

As with emission and absorption,
we expect that this exchange diagram will be part of a larger calculation.
We will assume for convenience here that the momentum space expectations of the incoming particles are on-shell, but that need not be true in general.
Note that this is the first time it is possible for both incoming and outgoing particles to have their 
momentum space expectations on-shell;
the exchanged particle will still remain in general off-shell.

To lowest order in perturbation expansion we have convolutions over
the two intermediate clock times $\tau_{1},\tau_{2}$:
\begin{equation}
\begin{array}{c}
\hat{\psi}_{3}^{\left({left}\right)}\left({p',q'}\right)=-\lambda\Lambda\int\limits _{0}^{{\tau_{3}}}{d{\tau_{2}}\hat{K}_{32}^{\left(M\right)}\left({q'}\right)\int{{d^{4}}q{d^{4}}k{{\hat{C}}_{2}}\left(q\right)\int\limits _{0}^{{\tau_{2}}}{d{\tau_{1}}\int{{d^{4}}p\hat{K}_{21}^{\left(\mu\right)}\left(k\right){{\hat{A}}_{1}}\left(p\right)}}}}\hfill\\
\hat{\psi}_{3}^{\left({right}\right)}\left({p',q'}\right)=-\lambda\Lambda\int\limits _{0}^{{\tau_{3}}}{d{\tau_{2}}\hat{K}_{32}^{\left(m\right)}\left({p'}\right)\int{{d^{4}}p{d^{4}}k{{\hat{A}}_{2}}\left(p\right)\int\limits _{0}^{{\tau_{2}}}{d{\tau_{1}}\int{{d^{4}}q\hat{K}_{21}^{\left(\mu\right)}\left(k\right){{\hat{C}}_{1}}\left(q\right)}}}}\hfill
\end{array}
\end{equation}

In TQM, the two initial wave functions are each defined by four expectations
in position, four in momentum, and four dispersions in either momentum
or position space -- twenty-four variables total. We have twelve integrals
in momentum (or position), over the two initial particles $A,C$ and
the exchange particle $B$. And the two convolutions in clock time.
With appropriately programmed mathematics software this is hardly
a problem. But it is easy to lose sight of the physics in the course
of doing the calculations.

To keep focus we will do as we have in the two previous subsections
and fix the clock times of the vertexes as $\tau_{X},\tau_{Y}$. Different
values of $\tau_{X},\tau_{Y}$ will let us look at specific cases.
The properties of the exchanged particle are key; we will focus on
these.

\paragraph{Classical trajectories}

We will take the same starting wave functions as with absorption,
but with $B\to C$. $A$ has expectations ${x_{a}},{p^{\left(a\right)}}$,
dispersions ${{\hat{\sigma}}^{\left(a\right)}}$; $C$ expectations
${x_{c}},{q^{\left(c\right)}}$, dispersions ${{\hat{s}}^{\left(a\right)}}$.

We assume we have $A$ coming in from the left; $C$ from the right;
both along the $x$ axis. Again we assume the collision is head-on
with no offset along the $\hat{y}$ or $\hat{z}$ axes.

We start by fixing $\tau_{X},\tau_{Y}$. Taking the left case first,
we have for $A$ and $C$:
\begin{equation}
\begin{array}{c}
{x_{X}^{\left(left\right)}}={x_{a}}+\frac{{p_{x}^{\left(a\right)}}}{m}{\tau_{X}}\hfill\\
{x_{Y}^{\left(left\right)}}={x_{c}}+\frac{{q_{x}^{\left(c\right)}}}{M}{\tau_{Y}}\hfill
\end{array}
\end{equation}

And for the exchange particle:
\begin{equation}
{x_{Y}^{\left(left\right)}}={x_{X}^{\left(left\right)}}+\frac{{k_{x}^{\left(b\right)}}}{\mu}\left({{\tau_{Y}}-{\tau_{X}}}\right)
\end{equation}

With ${\delta x^{\left(left\right)}}\equiv{x_{Y}^{\left(left\right)}}-{x_{X}^{\left(left\right)}},\delta\tau\equiv{\tau_{Y}}-{\tau_{X}}$
we have the expectation of the momentum of the exchange particle:
\begin{equation}
k_{x}^{\left(left\right)}=\mu\frac{{\delta x^{\left(left\right)}}}{{\delta\tau}}
\end{equation}

The momentum of $B$ is exactly what it needs to get from $X\to Y$
in time for its rendezvous with $C$. From conservation of momentum
at each vertex we have for the expectations of the final momenta:
\begin{equation}
\left\langle {{p'}_{x}}\right\rangle =p_{x}^{\left(a\right)}-k_{x}^{\left(left\right)},\left\langle {{q'}_{x}}\right\rangle =q_{x}^{\left(c\right)}+k_{x}^{\left(left\right)}
\end{equation}

With the initial conditions specified and $\tau_{X},\tau_{Y}$ as
well, the final expectations are immediate, with the final dispersions
to be computed.

To get the right hand case we interchange the roles $A\leftrightarrow C$:
\begin{equation}
\begin{array}{c}
{x_{X}^{\left(right\right)}}={x_{c}}+\frac{{q_{x}^{\left(c\right)}}}{M}{\tau_{Y}}\hfill\\
{x_{Y}^{\left(right\right)}}={x_{a}}+\frac{{p_{x}^{\left(a\right)}}}{m}{\tau_{X}}\hfill
\end{array}
\end{equation}
and:
\begin{equation}
k_{x}^{\left(right\right)}=\mu\frac{{\delta x^{\left(right\right)}}}{{\delta\tau}}
\end{equation}
giving final expectations:
\begin{equation}
\left\langle {{p'}_{x}}\right\rangle =p_{x}^{\left(a\right)}+k_{x}^{\left(right\right)},\left\langle {{q'}_{x}}\right\rangle =q_{x}^{\left(c\right)}-k_{x}^{\left(right\right)}
\end{equation}

\subsubsection{Exchange particle}

We can write the wave function for the exchange particle in SQM as:
\begin{equation}
{{\hat{\bar{B}}}_{YX}}\left({\vec{k}}\right)=\sqrt[4]{{\frac{1}{{\pi{{\hat{\bar{\Sigma}}}^{\left(X\right)}}}}}}{e^{-\imath\vec{k}\cdot{{\vec{x}}_{X}}-\frac{1}{2}\left({\vec{k}-{{\vec{k}}_{X}}}\right)\cdot\frac{1}{{{\hat{\bar{\Sigma}}}^{\left(X\right)}}}\cdot\left({\vec{k}-{{\vec{k}}_{X}}}\right)-\imath\frac{{{\vec{k}}^{2}}}{{2\mu}}\delta\tau}}
\end{equation}
and in TQM in covariant notation:

\begin{equation}
\begin{array}{c}
{{\hat{B}}_{YX}}\left(k\right)=\sqrt[4]{{\frac{1}{{\pi{{\hat{\Sigma}}^{\left(X\right)}}}}}}{e^{\imath k{x_{X}}-\frac{1}{2}\left({k-{k_{X}}}\right)\frac{1}{{\Sigma^{\left(X\right)}}}\left({k-{k_{X}}}\right)-\imath{f_{k}}\delta\tau}}\hfill\\
{f_{k}}\equiv-\frac{{{k^{2}}-{\mu^{2}}}}{{2\mu}}\hfill
\end{array}
\end{equation}
and as a product of time and space parts:
\begin{equation}
{{\hat{B}}_{YX}}\left(k\right)={{\hat{\tilde{B}}}_{YX}}\left(w\right){{\hat{\bar{B}}}_{YX}}\left({\vec{k}}\right)\exp\left({\imath\frac{\mu}{2}\delta\tau}\right)
\end{equation}
\begin{equation}
{{\hat{\tilde{B}}}_{YX}}\left(w\right)=\sqrt[4]{{\frac{1}{{\pi{{\hat{\tilde{\sigma}}}^{\left(X\right)}}}}}}{e^{\imath w{t_{X}}-\frac{1}{2}\frac{{{\left({w-{w_{X}}}\right)}^{2}}}{{2{{\hat{\tilde{\sigma}}}^{\left(X\right)}}}}+\imath\frac{{w^{2}}}{{2\mu}}\delta\tau}}
\end{equation}

Taking the left side for definiteness, the wave function of the intermediate
state is properly the direct product of the highly correlated wave
functions of $A,B$ with the (as yet) uncorrelated wave function of
the $C$ particle. After $B$ encounters $C$ at $Y$, $B$ is gone
and now $A$ and $C$ are highly correlated.

We have proceeded a bit formally, specifying $\tau_{X},\tau_{Y}$
and then deriving the properties of the exchange particle from these.
A more physical approach might be to fix the momentum of the exchange
particle and then integrate over all values of $\tau_{X}$ and $\tau_{Y}$
consistent with that.

\subsubsection{Discussion}

How we develop the exchange depends on what we are interested in:
\begin{enumerate}
\item If we are interested in the use of one particle as a
measurement of the other, then we already have what we need: $B$
inherits its dispersions from its parent. If its parent was narrow
in time/space, then it will be as well and act as a de facto gate
with respect to the other particle.
\item If we are interested in subtler effects of dispersion in time, we can look for forces of anticipation and regret:  
if wave functions are extended in time as they are in space, then
in a collision they will start to interact earlier, cease interacting
later than would otherwise be the case.
\item If we are interested in doing a Bell\textsc{'}s theorem correlation
in time, then we need to track the correlations through. Because momentum
is conserved at each vertex, the outgoing particles will be highly
correlated in momentum space.
\item If we are interested in symmetry properties in the time direction,
then we can: 
\begin{enumerate}
\item Take our $C$ as an $A$, 
\item Start with wave functions that are symmetric under particle exchange
but which nevertheless have a component which is anti-symmetric in
time (as the wide and narrow wave function in section \ref{subsec:mul-fock}), 
\item Look for unexpected anti-symmetries in space.
\end{enumerate}
\item If we are interested in the bound case, we can look at the exchange
particle as creating a Yukawa force in time. 
%In this case we will
%need to track the factors of $\exp\left(-\imath\mu\tau\right)$. 
The $\exp\left({-\imath{f_{k}}\tau}\right)$ factor will tend
to keep the exchange particle on-shell for larger values of $\delta\tau$.
In coordinate space this will tend to make the effective potential
look like a Li\'{e}nard--Wiechert potential. 
\end{enumerate}

\subsection{Loop correction to the mass\label{subsec:mul-loop}}

How do we calculate loop diagrams in TQM?

\begin{figure}[h]
\begin{centering}
\includegraphics[width=7cm]{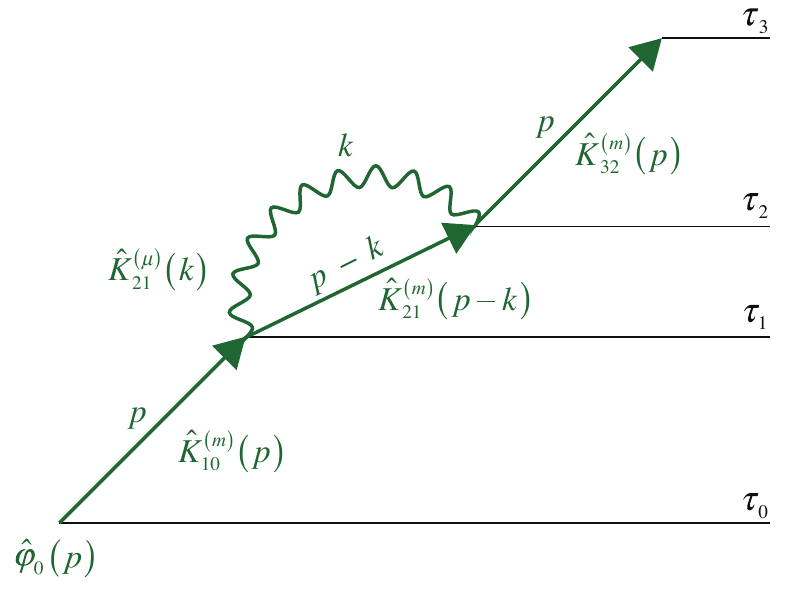} 
\par\end{centering}
\caption{Loop correction to the mass\label{fig:mul-loop-correction}}
\end{figure}

\subsubsection{Loop correction in SQM}

In the $ABC$ model there is an amplitude $\lambda$ for an $A$ to
emit a $B$ then absorb it. As is well known, in quantum field theory
this can be made to look like a correction to the mass, with the effect
of taking a bare mass to a corrected mass: $m_{0}^{2}\to m^{2}=m_{0}^{2}+\delta m^{2}$.
Unfortunately the integral for this $\delta m^{2}$ correction is
divergent. If the $A$ particle has four-momentum $p$ and emits a
$B$ with four-momentum $k$, to compute the amplitude associated
with the loop we will need to integrate over all possible values of
the intermediate $k$:
\begin{equation}
\delta{m^{2}}\sim\int{{d^{4}}k\frac{\imath}{{{{\left({p-k}\right)}^{2}}-{m^{2}}}}\frac{\imath}{{{k^{2}}-{\mu^{2}}}}}
\end{equation}

This is logarithmically divergent at large $k$:
\begin{equation}
\delta{m^{2}}\sim\int{\frac{{{d^{4}}k}}{{k^{4}}}}
\end{equation}

As a result all such integrals have to be regularized: a convergence
factor has to be inserted which acts as an effective cutoff, throwing
out the high energy part of the loop in a way that does not distort
results at lower energies.

To do this we take advantage of the general principle that all physical
measurements involve an implicit comparison between two measurements.
If we are going to use the mass of a particle as a value in one calculation
we must first have found that mass in another. An absolute, standalone
measurement is not possible even in principle. All measurements have
to be renormalized-- normalized by comparison to another -- to get
physically meaningful numbers.

In quantum field theory renormalization -- needed physically --
is also used to regularize, to contain and control the infinities.
If we make the necessary comparison in the right way, we can use it
to subtract off the infinities.

It\textsc{'}s like weighing a mouse by first weighing an aircraft
carrier without the mouse, weighing the aircraft carrier with the
mouse on board, and then subtracting out the weight of the aircraft
carrier to get the weight of the mouse.

There is no guarantee that this will work. What if the cutoff function/procedure
being used on both sides of the comparison has unintended side-effects
at lower energies? But in spite of the obvious risk the procedure
works --- and brilliantly -- producing some of the most accurate
predictions in the whole of physics.

\subsubsection{Loop correction in TQM}

With an extra dimension to integrate over we might expect
that the corresponding loop integrals in TQM would not only be divergent,
but perhaps even be divergent in a way which cannot be contained by
renormalization.

We will show that in fact the loop integrals in TQM are not divergent.  
The combination of finite initial dispersion in time with entanglement in time keep the integrals convergent.
We will work through a simple example, computing the familiar mass correction loop, first as a function of clock time, 
then as a function of clock frequency. 

\paragraph{Loop integral for fixed clock time}

We apply the loop integral to a initial Gaussian test function. In
momentum space:
\begin{equation}
{L_{\tau}}\left(p\right)=\int{{d^{4}}k\hat{K}_{\tau}^{\left(m\right)}\left({p-k}\right)\hat{K}_{\tau}^{\left(\mu\right)}\left(k\right){{\hat{\varphi}}_{0}}\left(p\right)}
\end{equation}
with initial Gaussian test function:
\begin{equation}
{{\hat{\varphi}}_{0}}\left(p\right)=\sqrt[4]{{\frac{1}{{\pi^{4}{\det}\left({\hat{\Sigma}}\right)}}}}\exp\left({-\frac{{{\left({p-{p_{0}}}\right)}^{2}}}{{2\Sigma}}}\right)
\end{equation}

The corresponding loop integral in coordinate space is:
\begin{equation}
{L_{\tau}}\left({x_{1}}\right)=4{\pi^{2}}\int{{d^{4}}{x_{0}}K_{\tau}^{\left(m\right)}\left({{x_{1}};{x_{0}}}\right)K_{\tau}^{\left(\mu\right)}\left({{x_{1}};{x_{0}}}\right){\varphi_{0}}\left({x_{0}}\right)}
\end{equation}

The kernels in coordinate space are:
\begin{equation}
\begin{array}{c}
K_{\tau}^{\left(m\right)}\left({{x_{1}};{x_{0}}}\right)=-\imath\frac{{m^{2}}}{{4{\pi^{2}}{\tau^{2}}}}\exp\left({-\frac{{\imath m}}{{2\tau}}{{\left({{x_{1}}-{x_{0}}}\right)}^{2}}-\imath\frac{m}{2}\tau}\right)\hfill\\
K_{\tau}^{\left(\mu\right)}\left({{x_{1}};{x_{0}}}\right)=-\imath\frac{{\mu^{2}}}{{4{\pi^{2}}{\tau^{2}}}}\exp\left({-\frac{{\imath\mu}}{{2\tau}}{{\left({{x_{1}}-{x_{0}}}\right)}^{2}}-\imath\frac{\mu}{2}\tau}\right)\hfill
\end{array}
\end{equation}

The product equals a single coordinate space kernel:
\begin{equation}
K_{\tau}^{\left(M\right)}\left({{x_{1}};{x_{0}}}\right)=-\imath\frac{{M^{2}}}{{4{\pi^{2}}{\tau^{2}}}}\exp\left({-\imath\frac{{M}}{{2\tau}}{{\left({{x_{1}}-{x_{0}}}\right)}^{2}}-\imath\frac{M}{2}\tau}\right)
\end{equation}
with a modified mass $M\equiv m+\mu$ and a prefactor.

So the loop integral in coordinate space is now:
\begin{equation}
{L_{\tau}}\left({x_{1}}\right)=-\imath\frac{{{m^{2}}{\mu^{2}}}}{{{\tau^{2}}{M^{2}}}}\int{{d^{4}}{x_{0}}K_{\tau}^{\left(M\right)}\left({{x_{1}};{x_{0}}}\right){\varphi_{0}}\left({x_{0}}\right)}
\end{equation}

The presence of the Gaussian test function on the right means we get
by inspection:
\begin{equation}
{L_{}{\tau}}\left({x_{1}}\right)=-\imath\frac{{{m^{2}}{\mu^{2}}}}{{{\tau^{2}}{M^{2}}}}\varphi_{\tau}^{\left(M\right)}\left({x_{1}}\right)
\end{equation}

We have a correction that shows a spread in time, but at the slightly
slower rate associated with the slightly larger mass $M$. Further
the correction is much greater at shorter clock times:

\begin{equation}
\begin{array}{c}
{{L}_{\tau}}\left({p_{1}}\right)=-\imath\frac{{{m^{2}}{\mu^{2}}}}{{{\tau^{2}}{M^{2}}}}\int{{d^{4}}{p_{0}}\hat{K}_{\tau}^{\left(M\right)}\left({{p_{1}};{p_{0}}}\right){{\hat{\varphi}}_{0}}\left({p_{0}}\right)}\hfill\\
\hat{K}_{\tau}^{\left(M\right)}\left({{p_{1}};{p_{0}}}\right)=\exp\left({\imath\frac{{p_{0}^{2}-{m^{2}}}}{{2m}}\tau}\right)\delta^{4}\left({{p_{1}}-{p_{0}}}\right)\hfill
\end{array}
\end{equation}

So the loop correction for fixed clock time is:
\begin{equation}
{L_{\tau}}\left({p}\right)=-\imath\frac{{{m^{2}}{\mu^{2}}}}{{{M^{2}}{\tau^{2}}}}\exp\left({\imath\frac{{{p^{2}}-{M^{2}}}}{{2M}}{\tau}}\right){{\hat{\varphi}}_{0}}\left({p}\right)
\end{equation}

At this point the value of the loop correction at a particular value
of $p$ is independent of the specific shape of the incoming wave
function. We are therefore free to drop the initial wave function
from the analysis:
\begin{equation}
{L_{\tau}}\left(p\right)=-\imath\frac{{{m^{2}}{\mu^{2}}}}{{{M^{2}}{\tau^{2}}}}\exp\left({\imath\frac{{{p^{2}}-{M^{2}}}}{{2M}}\tau}\right)
\end{equation}

\paragraph{Fourier transform of the loop integral over clock time}

Now that we have the loop integral for a specific value of the clock
time we can take the Fourier transform with respect to $\tau$:
\begin{equation}
{{\hat{L}}_{\omega}}\left(p\right)=-\imath\frac{1}{{\sqrt{2\pi}}}\int\limits _{-\infty}^{\infty}d\tau\frac{{{m^{2}}{\mu^{2}}}}{{{M^{2}}{\tau^{2}}}}{e^{\imath\left({\omega-{F_{p}}}\right)\tau}}
\end{equation}

We define $F_{p}\equiv-\frac{{{p^{2}}-{M^{2}}}}{{2M}}$. For small
$\mu$, $F_{p}\approx f_{p}$.
We have the value of the Fourier transform (with conventions per Mathematica):
\begin{equation}\mathcal{F}\mathcal{T}\left[ {\frac{{\exp \left( { - \imath {F_p}\tau } \right)}}{{{\tau ^2}}}} \right] =  - \sqrt {\frac{\pi }{2}} \left| {\omega  - {F_p}} \right|\end{equation}
or in our case:

\begin{equation}
{{\hat L}_\omega }\left( p \right) = \imath \frac{{{m^2}{\mu ^2}}}{{{M^2}}}\sqrt {\frac{\pi }{2}} \left| {\omega  - {F_p}} \right|
\end{equation}

The value of the loop integral is therefore finite, without need for regularization or the introduction of cutoffs or other artificial approaches.

We still need to renormalize.
But we have separated the problems associated with renormalization
from those associated with regularization.

As we have seen throughout, TQM implies the initial wave functions: 
\begin{enumerate}
\item have finite dispersion in time (as well as space) 
\item are entangled with the loop integration variables 
\end{enumerate}
This in turn means that each loop integration picks up a Gaussian
factor that guarantees its convergence -- a Gaussian easily dominates merely polynomial divergences.

The combination of finite initial dispersion and entanglement therefore forces convergence.

Note this approach will not work in SQM. In SQM there is by assumption
no finite initial dispersion in time. And even if there were, each
step is cut off from the previous since there is no entanglement in
time.

We have only looked at a toy case. But the principles established
here apply generally. As noted, the combination of Morlet wavelet
analysis and entanglement in time mean that the integrals encountered
in a diagram are self-regularized: the Gaussian functions which are
passed through a series of integrals easily dominate any polynomial
divergence.

An implication is that it is the assumption that quantum mechanics
does not apply along the time dimension that is responsible for the
ultraviolet divergences. The familiar divergences are a side-effect
of not pushing the ideas of quantum mechanics and special relativity
hard enough, of our failure to treat time and space symmetrically
in quantum mechanics.

\subsection{Discussion of the multiple particle case\label{subsec:mul-discussion}}

We have established that we can extend TQM to include the multiple
particle case. 

TQM is conceptually simpler than SQM: time and space are treated on
an equal footing and there are no ultraviolet divergences. But it
is calculationally more complex: we have coordinate time to consider
and we are required to use Gaussian test functions rather than plane
waves as the fundamental unit of analysis.

We have examined the basic parts of a Feynman diagram: free propagators;
the emission, absorption, and exchange of a particle; and simple loop
diagrams. We are therefore able to work out -- in principle at least
-- the results for any diagram in a perturbation expansion. And therefore
to compare TQM to SQM in any experiment which can be described by
such expansions.

To falsify TQM we need experiments that work at short times and with
individual wave packets. Long times and averages over wave packets
kill the effects associated with dispersion in time.

The most decisive such experiments are likely to be ones that emphasize
the effects of the time/energy uncertainty principle.

Additional effects include: 
\begin{enumerate}
\item anti-symmetry in time, 
\item forces of anticipation and regret,
\item interference and entanglement in time,
\item adjustments to the usual loop corrections.
\end{enumerate}

To be sure, these additional effects likely to be both subtle and small.
Therefore they may not be that useful for falsifiability, our primary
target in this work. But they are potentially interesting in their own right. 

We observe that the results here should be in common across the relativistic dynamics program;
these are all consequences of dispersion in time, which is found across the relativistic dynamics program.

\section{Discussion\label{sec:disc}}
\begin{quotation}
``It is difficult to see what one does not expect to see.'' --
William Feller \cite{Feller:1968he} 
\end{quotation}

We have explored the possibility that the quantum wave function should be extended in time:
\begin{equation}
\psi \left( {\vec x} \right) \to \psi \left( {t,\vec x} \right)
\end{equation}
We have argued that this idea can be developed in a way that is self-consistent, consistent with existing experimental and observational work,
and -- most importantly -- falsifiable.

\subsection{Falsifiability\label{subsec:disc-falsifiability}}

With the single slit in time we have a decisive test of temporal quantum
mechanics. In SQM, the narrower the slit, the \textit{less} the dispersion
in subsequent time-of-arrival measurements. In TQM, the narrower the
slit, the \textit{greater} the subsequent dispersion in subsequent
time-of-arrival measurements. \textsl{In principle, the difference
may be made arbitrarily great.}

To get to this point we had to develop the rules for calculation in a way that is unambiguously falsifiable.
To do this: 
\begin{enumerate}
\item We took path integrals as the defining representation. This made the
extension from three to four dimensions unambiguous. 
\item We used Morlet wavelet analysis rather than Fourier analysis to define
the initial wave functions. This let us avoid the use of unphysical
wave functions and made achieving convergence and normalization of
the path integrals possible. 
\item We distinguished carefully between ``clock time'' and ``coordinate
time''. 
\end{enumerate}
As a result, once we  applied the requirement that TQM match SQM
in the appropriate limit, we were left with no free parameters.

There are other ways to extend quantum mechanics to include time.
We have started with path integrals; one could
start with a Hamiltonian approach, see for instance Yau \cite{Yau:2015fk}.
We have focused on the extension of the wave function along coordinate time; 
at longer clock times the effects of differences in the handling of the clock time/evolution parameter may become significant.

We have therefore been careful to focus on dimensional and symmetry
arguments, which  give first order predictions which are
independent of the specifics of whatever method we might use: 
\begin{itemize}
\item The initial dispersion in time is fixed by symmetry between time and
space and the principle of maximum entropy. 
\item The evolution of the wave functions is fixed by the long, slow approximation.
This allows for give, but only over times of picoseconds, glacial
by the standards of TQM. 
\item We could choose Alice\textsc{'}s frame or Bob\textsc{'}s to do the
analysis, but the corrections due to this are of second order. 
Further, the
corrections can be eliminated entirely by selecting ``the rest frame
of the vacuum'' as the defining frame for TQM. 
\end{itemize}
The predictions of TQM are therefore falsifiable to first order.

\subsection{Experimental effects\label{subsec:disc-experimental-effects}}

We have discussed two primary effects: 
\begin{enumerate}
\item generally increased dispersion in time
\item and the time/energy uncertainty principle. 
\end{enumerate}
These effects should be present in any experimental setup in which
the sources vary in time and the detectors are time-sensitive.

Additionally we can look for: 
\begin{itemize}
\item Shadowing in time -- self-interference by detectors and sources. 
\item Interference, correlations, and entanglement in time.  
Consider for instance a Bell\textsc{'}s theorem in time: particles
that have interacted in the past, as in EPR experiments, will be entangled
in time as well as space. 
\item Forces of anticipation and regret. As the paths in TQM advance into
the future they will encounter potentials earlier (anticipation) than
in SQM. And as they dive back into the past they will continue to
interact with potentials later (regret) than is the case in SQM. 
\item Anti-symmetry in time. Wave functions are free to satisfy their symmetry
requirements using the time dimension as well as the three space dimensions. 
\item Small -- probably quite small -- corrections to the usual loop integral corrections.
\end{itemize}
In general, any quantum effect seen in space is likely have an ``in
time'' variation. TQM is to SQM with respect to time as SQM is to
classical mechanics with respect to space.

Reviews of foundational experiments in quantum mechanics (for example
Lamoreaux \cite{Lamoreaux:1992nh}, Ghose \cite{Ghose:1999zy},
and Auletta \cite{Auletta:2000vj}) provide a rich source of candidate
experiments: the single and double slit as well as many other foundational
experiments have an ``in time'' variant, typically with time and
a space dimension flipped.

The experiments are likely to be difficult. 
The attosecond times here are at the edge of the detectible. 
The
investigation here was partly inspired by Lindner\textsc{'}s ``Attosecond
Double-Slit Experiment'' \cite{Lindner:2005vv}. But the times there,
$500as$, are far too long for us. 
More recent work has reached shorter
and shorter times: 12 attoseconds in Koke, Sebastian, and Grebing
\cite{Koke:2010fk} and as noted the extraordinary sub-attosecond
times in Ossiander \cite{Ossiander:2016fp}. Further while the fundamental
scale is defined by the time taken by light to cross an atom, in the
case of Rydberg atom the width of an atom in space may be made almost
arbitrarily big. (We are indebted to Matt Riesen for this suggestion.)
Therefore the effects of dispersion in time should now be (barely)
within experimental range.

\subsection{Further extensions\label{subsec:disc-further-extensions}}

We have provided only a basic toolkit for TQM. 

As TQM falls within the relativistic dynamics program (per subsection
\ref{subsec:tqm-and-rd}), the existing literature in this area is
in general applicable. 
We get TQM from approaches to relativistic
dynamics by assuming that the dependence on clock time is negligible.
When this is reasonable, we can use the existing relativistic dynamics
literature to get TQM specific calculations by taking the sub-picosecond
limit (per subsection \ref{subsec:lsa-decoherence}). And 
we can go the other way: use TQM to get short clock time approximations
to other work in relativistic dynamics.

Areas for further investigation include: 
\begin{enumerate}
\item Generalizing the treatment of spinless massive bosons to include photons
and fermions. Extension to the Standard Model. 
\item Derivation of the bound state wave functions \cite{Arshansky:1989aa,Arshansky:1989ab,Fanchi:1993ab}.
\item More detailed treatment of scattering experiments \cite{Horwitz:1982aa,Land:1996aj}.
\item Exact treatment of the single slit in time, including paths that wander
back and forth through the slit \cite{Sawant:2013yu}. 
\item More detailed treatment of the double slit in time and of diffraction
in time \cite{Horwitz:1976aa,Horwitz:2005ix}.
\item Entanglement in time \cite{Horwitz:2005ix}.
\item Careful treatment of measurements, including paths that overshoot,
undershoot, and loop around the detector. 
\item Decoherence in time. 
\item Infrared divergences. From the point of view of TQM, these may be
the flip side of the ultraviolet divergences, suppressed in a similar
way. 
\item More detailed treatment of the ultraviolet divergences. 
\item Examination of the spin-statistics connection: the $T$ in the $CPT$
theorem must come under grave suspicion. 
\item Statistical mechanics. Any statistical ensemble should include fluctuations
in time \cite{Fanchi:1993ab,Horwitz:2015jk,HORWITZ1981306}. 
%Implicit in TQM is the possibility that the initial smooth wave function of
%wave mechanics should itself be replaced by a statistical ensemble of fluctuations in time. 
\item More detailed treatment of the choice of frame. 
\item Quantum gravity. Since TQM is by construction highly symmetric between
time and space and free of the ultraviolet divergences, it may be
a useful starting point for attacks on the problem of quantum gravity \cite{Fanchi:1993ab,Horwitz:2018aa}.
\end{enumerate}

\subsection{Five requirements\label{subsec:disc-five-requirements}}

In their delightfully titled \textit{How to Think about Weird Things} \cite{Schick-Jr.:1995je}
the philosophers Schick and Vaughn lay out five requirements that
a hypothesis such as TQM should satisfy: 
\begin{enumerate}
\item \textbf{Testability} -- \textsl{are there experimental tests?}\textit{
ideally: is the hypothesis falsifiable?} TQM has no free parameters;
it can therefore be falsified by any experiment at appropriate scale
looking at time varying quantum phenomena. 
\item \textbf{Fruitfulness} -- \textit{does the hypothesis suggest new
lines of research, new phenomena to explore?} All time-varying quantum
phenomena offers targets for investigation. The list of experimental
effects given above is doubtless far from exhaustive. 
\item \textbf{Scope} -- \textsl{how widespread are the phenomena?} TQM
applies to all time-varying quantum phenomena. 
\item \textbf{Simplicity} -- \textit{does it make the fewest possible assumptions?}
TQM eliminates the assumption that time and space should be treated
differently in quantum mechanics. It also eliminates the ultraviolet
divergences and the consequent need to regularize the loop integrals
in field theories (as QED). 
\item \textbf{Conservatism} -- \textit{is it consistent with what is known?}
TQM matches SQM in the long time (picosecond) limit. 
\end{enumerate}

\subsection{No null experiments\label{subsec:disc-absence-of-null}}

Meeting the Schick and Vaughn requirements is not enough to establish that TQM is true, 
only that it may be worth investigating experimentally.

Now suppose that one or more of the proposed experiments is done and
conclusively demonstrates that TQM is false. That would in turn raise
some interesting questions: 
\begin{enumerate}
\item Is there a frame in which TQM is maximally (or minimally) falsified?
That would be a preferred frame, anathema to relativity. 
\item Is TQM equally false in all frames? if it is \textit{equally} false
in all frames, how do we reconcile the disparate wave functions of
Alice and Bob? 
\end{enumerate}
As TQM is a straight-forward extrapolation of quantum mechanics and
special relativity, experiments that falsify TQM are likely to require
modification of our understanding of either quantum mechanics or special
relativity or both. 

Therefore it would appear that there are no null experiments. 

This concludes the argument for making an experimental investigation
of TQM.

\ack

I thank my long time friend Jonathan Smith for invaluable encouragement,
guidance, and practical assistance.

I thank Ferne Cohen Welch for extraordinary moral and practical support.

I thank Martin Land, Larry Horwitz, and the organizers of the IARD
2018 Conference for encouragement, useful discussions, and hosting
a talk on this paper at IARD 2018.

I thank Y. S. Kim for organizing the invaluable Feynman Festivals,
for several conversations, and for general encouragement and good
advice.

I thank the anonymous reviewers at the online journal Quanta, especially
for their comments on the need for an emphasis on testable and on
clarity.

I thank Catherine Asaro, Julian Barbour, Gary Bowson, Howard Brandt,
Daniel Brown, Ron Bushyager, John G. Cramer, J. Ferret, Robert Forward,
N. Gisin, Fred Herz, J. Pe\v{r}ina, Linda Kalb, A. Khrennikov, David
Kratz, Steve Libby, Andy Love, Walt Mankowski, O. Maroney, John Myers, Paul Nahin,
Marilyn Noz, R. Penrose, Stewart Personick, V. Petkov, H. Price, Matt
Riesen, Terry Roberts, J. H. Samson, Lee Smolin, L. Sk\'{a}la, Arthur
Tansky, R. Tumulka, Joan Vaccaro, L. Vaidman, A. Vourdas, H. Yadsan-Appleby,
S. Weinstein, Asher Yahalom, and Anton Zeilinger for helpful conversations.

I thank the organizers of several QUIST, DARPA, Perimeter Institute
conferences I've attended and the very much on topic conferences \textit{Quantum
Time} in Pittsburgh in 2014 and \textsl{Time and Quantum Gravity}
in San Diego in 2015.

I thank a host of friends, casual acquaintances, and passers-by for
many excellent conversations on the subjects here -- ranging from
a group of graduate students over good Czech beer at the Cafe Destiny
in Olomouc to a pair of Canadian customs agents encountered seriatim
on the way to the Perimeter institute.

And I would particularly like to thank my Olomouc experimentalist
-- I'm sorry I did not get your name -- who defined the right objective
for this work. I hope I have given you something you can prove wrong!

And none of the above are in any way responsible for any
errors of commission or omission in this work.

\appendix

\setcounter{section}{0}

\section{Conventions\label{sec:app-conv}}

We use natural units $\hbar=c=1$. 

When summing over three dimensions
we use $i,j,k$. We use $\imath$ sans dot for the square root of
-1; $i$ with dot for the index variable. When summing over four dimensions
we use $h$ for the coordinate time index so such sums run over $h,i,j,k$.

We define the relativistic time dilation factor $\gamma\equiv\frac{E}{m}$.

We use $m$ as the rest mass.  
We make no use of a relativistic mass, i.e. $\gamma m$.  Here the mass $m$ is always the rest mass.

\subsection{Clock time}

We use $\tau$ for clock time. The use of the Greek letter $\tau$
for clock time is meant to suggest that this is a ``classical''
time. We use $f$ for its complementary variable, clock frequency:
\begin{equation}
{f^{\left(op\right)}}\equiv\imath\frac{\partial}{{\partial\tau}}
\end{equation}

The clock time $\tau$ will usually be found at the bottom right of
any symbol it is indexing:
\begin{equation}
\varphi_{\tau},K_{\tau}
\end{equation}

The clock time is in its turn frequently indexed: ${\tau_{0}},{\tau_{1}},{\tau_{2}},\ldots$

As a result, deeply nested subscripts are an occasional hazard of
this analysis. To reduce the nesting level we use obvious shortenings,
i.e.:
\begin{equation}
{\varphi_{{\tau_{1}}}}\left({x_{1}}\right)\to{\varphi_{1}}\left({x_{1}}\right)\to{\varphi_{1}}
\end{equation}
\begin{equation}
{K_{{\tau_{1}}{\tau_{0}}}}\left({{x_{1}};{x_{0}}}\right)\to{K_{10}}\left({{x_{1}};{x_{0}}}\right)\to{K_{1}}\left({{x_{1}};{x_{0}}}\right)\to{K_{1}}
\end{equation}

And we represent differences in clock time by combining indexes:
\begin{equation}
{\tau_{21}}\equiv{\tau_{2}}-{\tau_{1}}
\end{equation}

\subsection{Coordinate time and space}

We use $E,\vec{p}$ for the momentum variables complementary to coordinate
time $t$ and space $\vec{x}$:
\begin{equation}
\begin{array}{c}
{E^{\left(op\right)}}\equiv\imath\frac{\partial}{{\partial t}}\hfill\\
{{\vec{p}}^{\left({op}\right)}}\equiv-\imath\nabla\hfill
\end{array}
\end{equation}

When there is a natural split into coordinate time and space parts
we use a tilde to mark the time part, an overbar to mark the space
part. For example:
\begin{equation}
\psi\left(t,\vec{x}\right)=\tilde{\psi}_{\tau}\left(t\right)\bar{\psi}_{\tau}\left(\vec{x}\right)
\end{equation}

This is to reinforce the idea that in this analysis the three dimensional
part is the average (hence overbar), while the coordinate time part
contributes a bit of quantum fuzziness (hence tilde) on top of that.
With that said, we will sometimes omit the overbar and the tilde when
they are obvious from context:
\begin{equation}
{{\tilde{\varphi}}_{\tau}}\left(t\right){{\bar{\varphi}}_{\tau}}\left({\vec{x}}\right)\to{\varphi_{\tau}}\left(t\right){\varphi_{\tau}}\left({\vec{x}}\right)
\end{equation}

We use an overdot to indicate the partial derivative with respect
to laboratory time:
\begin{equation}
\dot{g}_{\tau}\left(t,\vec{x}\right)\equiv\frac{\partial g_{\tau}\left(t,\vec{x}\right)}{\partial\tau}
\end{equation}

\subsection{Fourier transforms}

We use a caret to indicate that a function or variable is being taken
in momentum space. To keep the Fourier transform itself covariant
we use opposite signs for the coordinate time and space parts:
\begin{equation}
\begin{array}{c}
\hat{g}\left(E,\vec{p}\right)=\frac{1}{\sqrt{2\pi}^{4}}\int\limits _{-\infty}^{\infty}dtd\vec{x}e^{\imath Et-\imath\vec{p}\cdot\vec{x}}g\left(t,\vec{x}\right)\\
g\left(t,\vec{x}\right)=\frac{1}{\sqrt{2\pi}^{4}}\int\limits _{-\infty}^{\infty}dEd\vec{p}e^{-\imath Et+\imath\vec{p}\cdot\vec{x}}\hat{g}\left(E,\vec{p}\right)
\end{array}
\end{equation}

For plane waves:
\begin{equation}
\begin{array}{l}
{\phi_{p}}\left(x\right)=\tilde{\phi}\left(t\right)\bar{\phi}\left({\vec{x}}\right)=\frac{1}{{\sqrt{2\pi}}}\exp\left({-\imath Et}\right)\frac{1}{{{\sqrt{2\pi}}^{3}}}\exp\left({\imath\vec{p}\cdot\vec{x}}\right)\\
{{\hat{\phi}}_{x}}\left(p\right)=\hat{\tilde{\phi}}\left(E\right)\hat{\bar{\phi}}\left({\vec{p}}\right)=\frac{1}{{\sqrt{2\pi}}}\exp\left({\imath Et}\right)\frac{1}{{{\sqrt{2\pi}}^{3}}}\exp\left({-\imath\vec{p}\cdot\vec{x}}\right)
\end{array}
\end{equation}

To shorten the expressions we use:
\begin{equation}
x\equiv\left({t,\vec{x}}\right)=\left({t,x,y,z}\right)
\end{equation}

The difference between $x$ the four vector and $x$ the first space
coordinate should be clear from context. In momentum space we use:
\begin{equation}
p\equiv\left({E,\vec{p}}\right)=\left({E,{p_{x}},{p_{y}},{p_{z}}}\right)
\end{equation}

We have similar rules for clock time $\tau$ and its complementary
energy $f$:
\begin{equation}
\begin{array}{c}
{{\hat{g}}_{f}}=\frac{1}{{\sqrt{2\pi}}}\int\limits _{-\infty}^{\infty}d\tau{e^{\imath f\tau}}{g_{\tau}}\hfill\\
{g_{\tau}}=\frac{1}{{\sqrt{2\pi}}}\int\limits _{-\infty}^{\infty}df{e^{-\imath f\tau}}{{\hat{g}}_{f}}\hfill
\end{array}
\end{equation}

With these conventions when a Fourier transform is given by a convolution:

\begin{equation}
\hat{h}\left(p\right)=\int{dk\hat{f}\left({p-k}\right)\hat{g}\left(k\right)}
\end{equation}
the function in coordinate space is given by:
\begin{equation}
h\left(x\right)=\sqrt{2\pi}f\left(x\right)g\left(x\right)
\end{equation}

When it is obvious that a symbol represents a Fourier transform we
may drop the caret:
\begin{equation}
\hat{\varphi}\left(p\right)\to\varphi\left(p\right)
\end{equation}

\subsection{Gaussian test functions\label{subsec:app-conv-gtfs}}

As Gaussian test functions play a critical role in this investigation
it is useful to have a consistent notation with which to describe
them.

In general we define a Gaussian test function as a normalized Gaussian
function. It may be in position or momentum space. It is defined by
its expectations for position and momentum and by its dispersion in
either position or momentum. The most important single example here
is the Gaussian test function that describes the time part of the
wave function of a free particle:
\begin{equation}
{{\tilde{\varphi}}_{\tau}}\left(t\right)=F_{\tau}^{\left(t\right)}{e^{-\imath{E_{0}}t-\imath\frac{{E_{0}^{2}}}{{2m}}\tau-\frac{1}{{2\sigma_{t}^{2}f_{\tau}^{\left(t\right)}}}{{\left({t-{t_{0}}-\frac{{E_{0}}}{m}\tau}\right)}^{2}}}}
\end{equation}
with dispersion:
\begin{equation}
\sigma_{t}
\end{equation}
dispersion factor:
\begin{equation}
f_{\tau}^{\left(t\right)}\equiv1-\imath\frac{\tau}{{m\sigma_{t}^{2}}}
\end{equation}
and normalization factor:
\begin{equation}
F_{\tau}^{\left(t\right)}\equiv\sqrt[4]{\frac{1}{\pi\sigma_{t}^{2}}}\sqrt{\frac{1}{f_{\tau}^{\left(t\right)}}}
\end{equation}

The sign of the complex part of a dispersion factor is negative for time; positive for space:
\begin{equation}
f_{\tau}^{\left(x\right)}\equiv1+\imath\frac{\tau}{{m\sigma_{x}^{2}}}
\end{equation}

If we are dealing with multiple Gaussian test functions we may make
name changes $f\to g$,$F\to G$, $f\to h$,$F\to H$ to the dispersion
and normalization factors.

In general we can switch between time/energy and space/momentum forms
by taking the complex conjugate and interchanging variables $t\leftrightarrow x.$
This is our own small version of the $CPT$ transformations.

As expressions like $\sigma_{{p_{x}}}^{2}$ are cumbersome we sometimes
replace them with $\hat{\sigma}_{x}^{2}\equiv\sigma_{{p_{x}}}^{2}$
taking implicit advantage of the fact that with our conventions $\sigma_{x}^{2}=\frac{1}{{\sigma_{{p_{x}}}^{2}}}$.

If we need to tag various Gaussian wave functions we may assign each
specific letter $a,b,c$ as:
\begin{equation}
\tilde{\varphi}_{\tau}^{\left(a\right)}\left(t\right)=F_{\tau}^{\left(a\right)}{e^{-\imath{E_{a}}t-\imath\frac{{E_{a}^{2}}}{{2m}}\tau-\frac{1}{{2\sigma_{t}^{\left(a\right)2}f_{\tau}^{\left(a\right)}}}{{\left({t-{t_{a}}-\frac{{E_{a}}}{m}\tau}\right)}^{2}}}}
\end{equation}

We usually use $\varphi$ for Gaussian test functions but may use
a capital letter to reduce notational clutter, as $A\equiv\varphi^{\left(a\right)}$.

\subsection{Acronyms}
\begin{description}
\item [{CM}] Classical Mechanics: all four dimensions treated as parameters. 
\item [{SQM}] Standard Quantum Mechanics: quantum mechanics with the three
space dimensions treated as observables, time as a parameter. 
\item [{TQM}] Temporal Quantum Mechanics: SQM but with time treated as
an observable on the same basis as the three space dimensions. 
\end{description}

\section{Classical equations of motion\label{sec:app-classical}}

We verify that we get the classical equations of motion from the Lagrangian.

Broken out into time and space parts the Lagrangian is:
\begin{equation}
L\left(t,\vec{x},\dot{t},\dot{\vec{x}}\right)=-\frac{1}{2}m\dot{t}^{2}+\frac{1}{2}m\dot{\vec{x}}\cdot\dot{\vec{x}}-q\dot{t}\Phi\left(t,\vec{x}\right)+q\dot{x}_{j}A_{j}\left(t,\vec{x}\right)-\frac{1}{2}m
\end{equation}
The Euler-Lagrange equations are:
\begin{equation}
\frac{d}{d\tau}\frac{\delta L}{\delta\dot{x}^{\mu}}-\frac{\delta L}{\delta x^{\mu}}=0
\end{equation}
From the Euler-Lagrange equations we have:
\begin{equation}
m\ddot{t}=-q\dot{\Phi}+q\dot{t}\Phi_{,0}-q\dot{x}_{j}A_{j,0}=-q\dot{x}_{j}\left(\Phi_{,j}+A_{j,0}\right)
\end{equation}
\begin{equation}
m\ddot{x}_{i}=-q\dot{A}_{i}-q\dot{t}\Phi_{,i}+q\dot{x}_{j}A_{j,i}=-q\dot{t}A_{i,0}-q\dot{x}_{j}A_{i,j}-q\Phi_{,i}\dot{t}+q\dot{x}_{j}A_{j,i}
\end{equation}
Here the Roman indexes, $i$ and $j$, go from 1 to 3 and if present
in pairs are summed over. We use an overdot to indicate differentiation
by the laboratory time $\tau$.

By using:
\begin{equation}
\vec{E}=-\nabla\Phi-\frac{\partial\vec{A}}{\partial t}
\end{equation}
\begin{equation}
\vec{B}=\nabla\times\vec{A}
\end{equation}
we get:
\begin{equation}
m\ddot{t}=q\vec{E}\cdot\dot{\vec{x}}
\end{equation}
and:
\begin{equation}
m\ddot{\vec{x}}=q\dot{t}\vec{E}+q\dot{\vec{x}}\times\vec{B}
\end{equation}
which are the familiar equations of motion of a classical particle
in an electromagnetic field if we take $\tau$ as the proper time
of the particle. 

In the text we take $\tau$ as the laboratory time rather than the proper time of the particle. 
In the non-relativistic case, they are nearly the same.

\section{Unitarity\label{sec:app-unitarity}}

Since there is a chance of the wave function ``sneaking past'' the
plane of the present, we have to be particularly careful to confirm
unitarity.

To establish that the path integral kernel is unitary we need to establish
that it preserves the normalization of the wave function. The analysis
in the subsection \ref{subsec:Normalization} only established this
for the free case. We therefore need to confirm that the normalization
of the wave function is preserved in the general case. We use a proof
from Merzbacher \cite{Merzbacher:1998tc} but in four rather than
three dimensions.

We form the probability:
\begin{equation}
P\equiv\int d^{4}x\psi^{*}\left(x\right)\psi\left(x\right)
\end{equation}
We therefore have for the rate of change of probability in time:
\begin{equation}
\frac{dP}{d\tau}=\int d^{4}x\left(\psi^{*}\left(x\right)\frac{\partial\psi\left(x\right)}{\partial\tau}+\frac{\partial\psi^{*}\left(x\right)}{\partial\tau}\psi\left(x\right)\right)
\end{equation}
The Schr\"{o}dinger equations for the wave function and its complex conjugate
are:
\begin{equation}
\frac{\partial\psi}{\partial\tau}=-\frac{\imath}{2m}\partial^{\mu}\partial_{\mu}\psi+\frac{q}{m}\left(A^{\mu}\partial_{\mu}\right)\psi+\frac{q}{2m}\left(\partial^{\mu}A_{}\mu\right)\psi+\imath\frac{q^{2}}{2m}A^{\mu}A_{\mu}\psi-\imath\frac{m}{2}\psi
\end{equation}
\begin{equation}
\frac{\partial\psi^{*}}{\partial\tau}=\frac{\imath}{2m}\partial^{\mu}\partial_{\mu}\psi^{*}+\frac{q}{m}\left(A^{\mu}\partial_{\mu}\right)\psi^{*}+\frac{q}{2m}\left(\partial^{\mu}A_{}\mu\right)\psi^{*}-\imath\frac{q^{2}}{2m}A^{\mu}A_{\mu}\psi^{*}+\imath\frac{m}{2}\psi^{*}
\end{equation}
We rewrite $\frac{\partial\psi}{\partial\tau}$ and $\frac{\partial\psi^{*}}{\partial\tau}$
using these and throw out canceling terms. Since the probability density
is gauge independent, we choose the Lorentz gauge $\partial A=0$
to get:
\begin{equation}
\frac{dP}{d\tau}=\int d^{4}x\left(\psi^{*}\left(-\frac{\imath}{2m}\partial^{\mu}\partial_{\mu}\psi+\frac{q}{m}\left(A\partial\right)\psi\right)+\left(\frac{\imath}{2m}\partial^{\mu}\partial_{\mu}\psi^{*}+\frac{q}{m}\left(A\partial\right)\psi^{*}\right)\psi\right)
\end{equation}
We integrate by parts; we are left with zero on the right:
\begin{equation}
\frac{dP}{d\tau}=0
\end{equation}
Therefore the rate of change of probability is zero, as was to be
shown. And therefore the normalization is correct in the general case.

\section{Gauge transformations\label{sub:app-gauge}}

As noted in the text, the ambiguities in the phase of the normalization of the
wave function may be seen as representing a kind of gauge transformation.
We have all the usual possibilities for gauge transformations. And
we have in addition the possibility of gauge transformations which
are a function of the laboratory time.

To explore this, we write the wave function as a product of a gauge
function in coordinate time, space, and laboratory time and a gauged
wave function:
\begin{equation}
\psi'_{\tau}\left(t,\vec{x}\right)=e^{\imath q\Lambda_{\tau}\left(t,\vec{x}\right)}\psi_{\tau}\left(t,\vec{x}\right)
\end{equation}

%This clearly has no effect on the normalization of the wave function as we have by inspection:
%\begin{equation}
%{\psi '_\tau }{\left( {t,\vec x} \right)^*}{\psi '_\tau }\left( {t,\vec x} \right) = {\psi _\tau }{\left( {t,\vec x} \right)^*}{\psi _\tau }\left( {t,\vec x} \right)
%\end{equation}

If the original wave function satisfies a gauged Schr\"{o}dinger equation:
\begin{equation}
\left(\imath\frac{\partial}{\partial\tau}-q\mathcal{A}_{\tau}\left(x\right)\right)\psi_{\tau}\left(x\right)=-\frac{1}{2m}\left(\left(p-qA\right)^{2}-m^{2}\right)\psi_{\tau}\left(x\right)
\end{equation}
the gauged wave function also satisfies a gauged Schr\"{o}dinger equation:
\begin{equation}
\left(\imath\frac{\partial}{\partial\tau}-q\mathcal{A}'_{\tau}\left(x\right)\right)\psi'_{\tau}\left(x\right)=-\frac{1}{2m}\left(\left(p-qA'\right)^{2}-m^{2}\right)\psi'_{\tau}\left(x\right)
\end{equation}
provided we have:
\begin{equation}
\mathcal{A}'_{\tau}\left(x\right)=\mathcal{A}_{\tau}\left(x\right)-\frac{\partial\Lambda_{\tau}\left(x\right)}{\partial\tau}
\end{equation}
and the usual gauge transformations:
\begin{equation}
A'^{\mu}=A^{\mu}-\partial^{\mu}\Lambda_{\tau}\left(x\right)
\end{equation}
or:
\begin{equation}
\begin{array}{l}
\Phi'=\Phi-\frac{\partial\Lambda}{\partial t}\\
\vec{A}'=\vec{A}+\nabla\Lambda
\end{array}
\end{equation}
If the gauge function $\Lambda$ is not a function of the laboratory
time ($\Lambda=\Lambda\left(t,\vec{x}\right)$) then we recover the
usual gauge transformations for $\Phi$ and $\vec{A}$. On the other
hand, we could let the gauge depend on the laboratory time, perhaps
using different gauges for different parts of the problem in hand.

\section{Free wave functions and kernels\label{sec:app-free}}

We here assemble the solutions of the free Schr\"{o}dinger\textsc{'}s equation for reference.

In general the free wave functions and kernels can be written as a
coordinate time part times a familiar non-relativistic part. The division
into coordinate time, space, and -- occasionally -- clock time parts
is to some extent arbitrary.

\subsection{Plane waves\label{subsec:app-free-Plane-wave-functions}}

Plane wave in coordinate time:
\begin{equation}
{{\tilde{\phi}}_{\tau}}\left(t\right)=\frac{1}{{\sqrt{2\pi}}}\exp\left({-\imath E_{0}t+\imath\frac{{E_{0}^{2}}}{{2m}}\tau}\right)
\end{equation}

Plane wave in space:
\begin{equation}
{{\bar{\phi}}_{\tau}}\left({\vec{x}}\right)=\frac{1}{{{\sqrt{2\pi}}^{3}}}\exp\left({\imath\vec{p}_{0}\cdot\vec{x}-\imath\frac{{{\vec{p_{0}}}^{2}}}{{2m}}\tau}\right)
\end{equation}

The full plane wave is the product of coordinate time and space plane
waves:
\begin{equation}
{\phi_{\tau}}\left(x\right)={{\tilde{\phi}}_{\tau}}\left(t\right){{\bar{\phi}}_{\tau}}\left({\vec{x}}\right)\exp\left({\imath\frac{{m^{2}}}{{2m}}\tau}\right)=\frac{1}{{4{\pi^{2}}}}\exp\left({-\imath p_{0}x-\imath{f_{0}}\tau}\right)
\end{equation}
with definition of clock frequency:
\begin{equation}
{f_{0}}\equiv-\frac{{{E_{0}^{2}}-{{\vec{p_{0}}}^{2}}-{m^{2}}}}{{2m}}
\end{equation}

The equivalents in momentum space are $\delta$ functions with a clock
time dependent phase:
\begin{equation}
\begin{array}{c}
{{\hat{\tilde{\phi}}}_{\tau}}\left(E\right)=\delta\left({E-{E_{0}}}\right)\exp\left({\imath\frac{{E_{0}^{2}}}{{2m}}\tau}\right)\hfill\\
{{\hat{\bar{\phi}}}_{\tau}}\left({\vec{p}}\right)={\delta^{\left(3\right)}}\left({\vec{p}-{{\vec{p}}_{0}}}\right)\exp\left({-\imath\frac{{{\vec{p}_{0}}^{2}}}{{2m}}\tau}\right)\hfill\\
{{\hat{\phi}}_{\tau}}\left(p\right)={{\hat{\tilde{\phi}}}_{0}}\left(E\right){{\hat{\bar{\phi}}}_{0}}\left({\vec{p}}\right)\exp\left(-{\imath\frac{{m^{2}}}{{2m}}\tau}\right)={\delta^{\left(4\right)}}\left({p-{p_{0}}}\right)\exp\left({-\imath{f_{0}}\tau}\right)\hfill
\end{array}
\end{equation}

\subsection{Gaussian test functions\label{subsec:app-free-gtfs} }

By Morlet wavelet decomposition any normalizable wave function may
be written as a sum over Gaussian test functions. 
We have specified
the conventions we are using for Gaussian test functions above; here
we look specifically at Gaussian test functions as solutions of the
free Schr\"{o}dinger equation.

\subsubsection{Time and energy\label{subsubsec:app-free-Time-and-energy}}

Gaussian test function in coordinate time at clock time zero:
\begin{equation}
\tilde{\varphi}_{0}\left(t\right)\equiv\sqrt[4]{\frac{1}{\pi\sigma_{t}^{2}}}e^{-\imath E_{0}\left(t-t_{0}\right)-\frac{\left(t-t_{0}\right)^{2}}{2\sigma_{t}^{2}}}
\end{equation}
\begin{equation}
{{\hat{\tilde{\varphi}}}_{0}}\left(E\right)\equiv\sqrt[4]{{\frac{1}{{\pi\sigma_{E}^{2}}}}}{e^{\imath E{t_{0}}-\frac{{{\left({E-{E_{0}}}\right)}^{2}}}{{2\sigma_{E}^{2}}}}}
\end{equation}

With these conventions, the energy and coordinate time dispersions
are reciprocals:
\begin{equation}
{\sigma_{E}}=\frac{1}{{\sigma_{t}}}
\end{equation}

As noted, it is
often convenient to thread a letter through the wave function to label
the constants, e.g.:
\begin{equation}
\tilde{\varphi}_{a}\left(t\right)\equiv\sqrt[4]{\frac{1}{\pi\sigma_{t}^{(a)2}}}e^{-\imath E_{a}\left(t-t_{a}\right)-\frac{\left(t-t_{a}\right)^{2}}{2\sigma_{t}^{(a)2}}}
\end{equation}
\begin{equation}
{{\hat{\tilde{\varphi}}}_{a}}\left(E\right)\equiv\sqrt[4]{{\frac{1}{{\pi\sigma_{E}^{(a)2}}}}}{e^{\imath E{t_{a}}-\frac{{{\left({E-{E_{a}}}\right)}^{2}}}{{2\sigma_{E}^{(a)2}}}}}
\end{equation}

A typical case is where $a\to 0$, to represent the function at $\tau=0$.

Gaussian test function for coordinate time as a function of clock time:
\begin{equation}
\tilde{\varphi}_{\tau}\left(t\right)=\sqrt[4]{\frac{1}{\pi\sigma_{t}^{2}}}\sqrt{\frac{1}{f_{\tau}^{\left(t\right)}}}e^{-\imath E_{0}t+\imath\frac{E_{0}^{2}}{2m}\tau-\frac{1}{2\sigma_{t}^{2}f_{\tau}^{\left(t\right)}}\left(t-t_{0}-\frac{E_{0}}{m}\tau\right)^{2}}
\end{equation}
with dispersion factor $f_{\tau}^{\left(t\right)}\equiv1-\imath\frac{\tau}{m\sigma_{t}^{2}}$
and with expectation, probability density, and uncertainty:
\begin{equation}
\begin{array}{l}
\left\langle {t_{\tau}}\right\rangle ={t_{0}}+\frac{E}{m}\tau={t_{0}}+\gamma\tau\\
{{\tilde{\rho}}_{\tau}}\left(t\right)=\sqrt{\frac{1}{{\pi\sigma_{t}^{2}}}}\exp\left({-\frac{{{\left({t-\left\langle {t_{\tau}}\right\rangle}\right)}^{2}}}{{\sigma_{t}^{2}\left({1+\frac{{\tau^{2}}}{{{m^{2}}\sigma_{t}^{2}}}}\right)}}}\right)\\
{\left({\Delta t}\right)^{2}}\equiv\left\langle t^{2}\right\rangle -{\left\langle t\right\rangle ^{2}}=\frac{{\sigma_{t}^{2}}}{2}\left|{1+\frac{{\tau^{2}}}{{{m^{2}}\sigma_{t}^{4}}}}\right|
\end{array}
\end{equation}

In the non-relativistic case, if we start with $t_{0}=\tau_{0}$, then
we have $\left\langle t\right\rangle \approx\tau$ throughout.

For longer clock times the uncertainty in coordinate time is proportional
to the dispersion in the energy:
\begin{equation}
\Delta t\sim\frac{\tau}{{m{\sigma_{t}}}}=\frac{\tau}{m}{{\sigma}_{E}}
\end{equation}

Gaussian test function in energy:
\begin{equation}
{{\hat{\tilde{\varphi}}}_{\tau}}\left(E\right)\equiv\sqrt[4]{{\frac{1}{{\pi\sigma_{E}^{2}}}}}{e^{\imath E{t_{0}}-\frac{{{\left({E-{E_{0}}}\right)}^{2}}}{{2\sigma_{E}^{2}}}+\imath\frac{{E^{2}}}{{2m}}\tau}}
\end{equation}
with expectation, probability density, and uncertainty:
\begin{equation}
\begin{array}{c}
{\left\langle E\right\rangle }={E_{0}}\hfill\\
{{\hat{\tilde{\rho}}}_{\tau}}\left(E\right)={{\hat{\tilde{\rho}}}_{0}}\left(E\right)=\sqrt{\frac{1}{{\pi\sigma_{E}^{2}}}}\exp\left({-\frac{{{\left({E-{E_{0}}}\right)}^{2}}}{{\sigma_{E}^{2}}}}\right)\hfill\\
{\left({\Delta E}\right)^{2}}=\frac{{\sigma_{E}^{2}}}{2}\hfill
\end{array}
\end{equation}

\subsubsection{Single space/momentum dimension\label{subsecsec:app-free-Space-and-momentum-1D}}

Gaussian test function in one space dimension at clock time zero:
\begin{equation}
{{\bar{\varphi}}_{0}}\left(x\right)=\sqrt[4]{{\frac{1}{{\pi\sigma_{x}^{2}}}}}{e^{\imath p_{0}\left({x-{x_{0}}}\right)-\frac{{{\left({x-{x_{0}}}\right)}^{2}}}{{2\sigma_{x}^{2}}}}}
\end{equation}
and in momentum:
\begin{equation}
{{\hat{\bar{\varphi}}}_{0}}\left(p\right)=\sqrt[4]{{\frac{1}{{\pi\sigma_{p}^{2}}}}}{e^{-\imath p{x_{0}}-\frac{{{\left({p-{p_{0}}}\right)}^{2}}}{{2\sigma_{p}^{2}}}}}
\end{equation}

As with time and energy, the space and momentum dispersions are reciprocal:
\begin{equation}
{\sigma_{p}}\equiv\frac{1}{{\sigma_{x}}}
\end{equation}

When we have to consider the dispersion for all three space momentum
we may reduce the level of nesting by writing:
\begin{equation}
{{\hat{\sigma}}_{x}}\equiv{\sigma_{{p_{x}}}},{{\hat{\sigma}}_{y}}\equiv{\sigma_{{p_{y}}}},{{\hat{\sigma}}_{z}}\equiv{\sigma_{{p_{z}}}}
\end{equation}
Gaussian test function in one space dimension as a function of clock
time:
\begin{equation}
\bar{\varphi}_{\tau}\left(x\right)=\sqrt[4]{\frac{1}{\pi\sigma_{x}^{2}}}\sqrt{\frac{1}{f_{\tau}^{\left(x\right)}}}e^{\imath p_{0}x-\frac{1}{2\sigma_{x}^{2}f_{\tau}^{\left(x\right)}}\left(x-x_{0}-\frac{p_{0}}{m}\tau\right)^{2}-\imath\frac{p_{0}^{2}}{2m}\tau}
\end{equation}

The definition of the dispersion factor $f_{\tau}^{\left(x\right)}=1+\imath\frac{\tau}{m\sigma_{x}^{2}}_{\tau}$
is parallel to that for coordinate time (but with the opposite sign
for the imaginary part).

Expectation, probability density, and uncertainty for $x$:
\begin{equation}
\begin{array}{l}
\left\langle {x_{\tau}}\right\rangle ={x_{0}}+\frac{{p_{x}}}{m}\tau={x_{0}}+\gamma{v_{x}}\tau\\
{{\bar \rho }_\tau }\left( x \right) = \sqrt {\frac{1}{{\pi \sigma _x^2}}} \exp \left( { - \frac{{{{\left( {x - \left\langle {{x_\tau }} \right\rangle } \right)}^2}}}{{\sigma _x^2\left( {1 + \frac{{{\tau ^2}}}{{{m^2}\sigma _x^2}}} \right)}}} \right)\\
{\left({\Delta x}\right)^{2}}\equiv\left\langle {x_{\tau}^{2}}\right\rangle -{\left\langle {x_{\tau}}\right\rangle ^{2}}=\frac{{\sigma_{x}^{2}}}{2}\left|{1+\frac{{\tau^{2}}}{{{m^{2}}\sigma_{x}^{4}}}}\right|
\end{array}
\end{equation}
and similarly for $y$ and $z$.

As clock time goes to infinity, the uncertainty in space scales as:
\begin{equation}
\Delta x \sim \frac{\tau }{{m{\sigma _x}}} = \frac{\tau }{m}{{\hat \sigma }_x}
\end{equation}

Negative x-momentum is movement to the left, positive to the right.
As we require the most complete parallelism between time and space,
we therefore have that positive energy corresponds to movement into
the future, negative into the past. As most of our wave functions
have an energy of order:
\begin{equation}
E\sim m+\frac{{{\vec{p}}^{2}}}{{2m}}\gg0
\end{equation}
they are usually going into the future. As expected.

Gaussian test function for momentum in one dimension as a function
of clock time:
\begin{equation}
{{\hat{\bar{\varphi}}}_{\tau}}\left(p\right)=\sqrt[4]{{\frac{1}{{\pi\sigma_{p}^{2}}}}}{e^{-\imath p{x_{0}}-\frac{{{\left({p-{p_{0}}}\right)}^{2}}}{{2\sigma_{p}^{2}}}-\imath\frac{{p^{2}}}{{2m}}\tau}}
\end{equation}

The expectation, probability density, and uncertainty for $p$ are constant:
\begin{equation}
\begin{array}{c}
{\left\langle p\right\rangle _{\tau}}={\left\langle p\right\rangle _{0}}={p_{0}}\hfill\\
{{\hat{\bar{\rho}}}_{\tau}}\left(p\right)={{\hat{\bar{\rho}}}_{0}}\left(p\right)=\sqrt{\frac{1}{{\pi\sigma_{p}^{2}}}}\exp\left({-\frac{{{\left({p-{p_{0}}}\right)}^{2}}}{{\sigma_{p}^{2}}}}\right)\hfill\\
{\left({\Delta p}\right)^{2}}=\frac{{\sigma_{p}^{2}}}{2}\hfill
\end{array}
\end{equation}

\subsubsection{Covariant forms\label{subsubsec:app-free-Four-space}}

Usually we treat four dimensions wave functions as simple products of
one dimensional wave functions. 
But it is more appropriate in general to treat them as a single covariant object.

We define the four dimensional dispersion $\Sigma$ in position space
at clock time zero:
\begin{equation}
\Sigma_{0}^{\mu\nu}\equiv\left({\begin{array}{cccc}
{\sigma_{t}^{2}} & 0 & 0 & 0\\
0 & {\sigma_{x}^{2}} & 0 & 0\\
0 & 0 & {\sigma_{y}^{2}} & 0\\
0 & 0 & 0 & {\sigma_{z}^{2}}
\end{array}}\right)
\end{equation}

The determinant takes the simple form:
\begin{equation}
\det\Sigma_{0}=\sigma_{t}^{2}\sigma_{x}^{2}\sigma_{y}^{2}\sigma_{z}^{2}
\end{equation}

With this we have the wave function at clock time zero:
\begin{equation}
{\varphi_{0}}\left(x\right)=\sqrt[4]{{\frac{1}{{{\pi^{4}}{\det}\left({\Sigma_{0}}\right)}}}}{e^{-\imath p_{0}^{\mu}{{\left({x-{x_{0}}}\right)}_{\mu}}-\frac{1}{2}{{\left({x-{x_{0}}}\right)}_{_{\mu}}}\Sigma_{0}^{{-1}\mu\nu}{{\left({x-{x_{0}}}\right)}_{_{\nu}}}}}
\end{equation}

Four dimensional dispersion as a function of clock time $\tau$:
\begin{equation}
\Sigma_{\tau}^{\mu\nu}\equiv\left({\begin{array}{cccc}
{\sigma_{t}^{2}f_{\tau}^{\left(t\right)}=\sigma_{t}^{2}-\imath\frac{\tau}{m}} & 0 & 0 & 0\\
0 & {\sigma_{x}^{2}f_{\tau}^{\left(x\right)}=\sigma_{x}^{2}+\imath\frac{\tau}{m}} & 0 & 0\\
0 & 0 & {\sigma_{y}^{2}f_{\tau}^{\left(y\right)}=\sigma_{y}^{2}+\imath\frac{\tau}{m}} & 0\\
0 & 0 & 0 & {\sigma_{z}^{2}f_{\tau}^{\left(z\right)}=\sigma_{z}^{2}+\imath\frac{\tau}{m}}
\end{array}}\right)
\end{equation}

Four dimensional wave function as a function of clock time:
\begin{equation}
{\varphi_{\tau}}\left(x\right)=\sqrt[4]{{\frac{{\det\left({\Sigma_{0}}\right)}}{{{\pi^{4}}{{\det}^{2}}\left({\Sigma_{\tau}}\right)}}}}{e^{-\imath p_{0}^{\mu}{{\left({{x}-{x_{0}}-v\tau}\right)}_{\mu}}-\frac{1}{2}{{\left({x-{x_{0}}-v\tau}\right)}_{_{\mu}}}\Sigma_{\tau}^{-1\mu\nu}{{\left({x-{x_{0}}-v\tau}\right)}_{_{\nu}}{-\imath{f_{0}}\tau}}}}
\end{equation}
with the obvious definition of the four velocity:
\begin{equation}
{v_{\mu}}\equiv\frac{{p_{\mu}}}{m}
\end{equation}

In the general case, the wave function does not split
into coordinate time and space factors. But it may still be represented
as a sum over basis wave functions that do, via Morlet wavelet decomposition.

We have similar but simpler formulas in momentum space. In momentum
space at clock time zero:
\begin{equation}{{\hat \varphi }_0}\left( p \right) = \sqrt[4]{{\frac{1}{{{\pi ^4}\det \left( {\hat \Sigma } \right)}}}}{e^{ - \imath p{x_0} - \frac{1}{2}{{\left( {p - {p_0}} \right)}_\mu }{{\left( {{\Sigma ^{ - 1}}} \right)}^{\mu \nu }}{{\left( {p - {p_0}} \right)}_\nu }}}\end{equation}
and as a function of clock time:
\begin{equation}{{\hat \varphi }_\tau }\left( p \right) = \sqrt[4]{{\frac{1}{{{\pi ^4}\det \left( {\hat \Sigma } \right)}}}}{e^{ - \imath p{x_0} - \frac{1}{2}{{\left( {p - {p_0}} \right)}_\mu }{{\left( {{\Sigma ^{ - 1}}} \right)}^{\mu \nu }}{{\left( {p - {p_0}} \right)}_\nu } - \imath {f_0}\tau }} = {{\hat \varphi }_0}\left( p \right)\exp \left( { - \imath {f_0}\tau } \right)\end{equation}

We define the momentum dispersion matrix $\hat{\Sigma}$ as the reciprocal
of the coordinate space dispersion matrix $\Sigma_{0}$:

\begin{equation}
\hat{\Sigma}\equiv\left({\begin{array}{cccc}
{\hat{\sigma}_{t}^{2}} & 0 & 0 & 0\\
0 & {\hat{\sigma}_{x}^{2}} & 0 & 0\\
0 & 0 & {\hat{\sigma}_{y}^{2}} & 0\\
0 & 0 & 0 & {\hat{\sigma}_{z}^{2}}
\end{array}}\right)
\end{equation}

\subsection{Free kernels\label{subsec:app-free-Kernels}}

We list the kernels corresponding to the free Schr\"{o}dinger equation
in time. These are retarded kernels going from clock time zero to
clock time $\tau$, so include an implicit $\theta\left(\tau\right)$.

\subsubsection{Coordinate time and space\label{subsubsec:app-free-kernel-coordinate}}

Kernel in coordinate time:
\begin{equation}
{{\tilde{K}}_{\tau}}\left({{t^{\prime\prime}};{t^{\prime}}}\right)=\sqrt{\frac{{\imath m}}{{2\pi\tau}}}\exp\left({-\imath m\frac{{{\left({{t^{\prime\prime}}-{t^{\prime}}}\right)}^{2}}}{{2\tau}}}\right)
\end{equation}

In three space we have the familiar non-relativistic kernel (e.g.
Merzbacher \cite{Merzbacher:1998tc}):
\begin{equation}
{{\bar{K}}_{\tau}}\left({{{\vec{x}}^{\prime\prime}};{{\vec{x}}^{\prime}}}\right)={\sqrt{-\frac{\imath}{{2\pi\tau}}}^{3}}\exp\left({\imath m\frac{{{\left({\vec{x}''-\vec{x}'}\right)}^{2}}}{{2\tau}}}\right)
\end{equation}

The full kernel is:
\begin{equation}
{K_{\tau}}\left({x'';x'}\right)={\tilde{K}_{\tau}}\left({{t^{\prime\prime}};{t^{\prime}}}\right){\bar{K}_{\tau}}\left({{{\vec{x}}^{\prime\prime}};{{\vec{x}}^{\prime}}}\right)\exp\left(-\imath\frac{m}{2}\tau\right)
\end{equation}

Explicitly:
\begin{equation}
{K_{\tau}}\left({x'';x'}\right)=-\imath\frac{{m^{2}}}{{4{\pi^{2}}{\tau^{2}}}}{e^{-\imath m\frac{{{\left({{t^{\prime\prime}}-{t^{\prime}}}\right)}^{2}}}{{2\tau}}+\imath m\frac{{{\left({\vec{x}''-\vec{x}'}\right)}^{2}}}{{2\tau}}-\imath\frac{m}{2}\tau}}
\end{equation}

\subsubsection{Momentum space\label{subsubsec:app-free-kernel-momentum}}

Energy part:
\begin{equation}
{{\hat{\tilde{K}}}_{\tau}}\left({E'';E'}\right)=\delta\left({E''-E'}\right)\exp\left({\imath\frac{{{{E'}^{2}}-{m^{2}}}}{{2m}}\tau}\right)
\end{equation}

Three momentum part:
\begin{equation}
{{\hat{\vec{K}}}_{\tau}}\left({\vec{p}'';\vec{p}'}\right)={\delta^{\left(3\right)}}\left({\vec{p}''-\vec{p}'}\right)\exp\left({-\imath\frac{{{\left({\vec{p}'}\right)}^{2}}}{{2m}}\tau}\right)
\end{equation}

Again, the same as the usual non-relativistic kernel.

The full kernel is:
\begin{equation}
{{\hat{K}}_{\tau}}\left({p''-p'}\right)={{\hat{\tilde{K}}}_{\tau}}\left({E'';E'}\right){{\hat{\vec{K}}}_{\tau}}\left({\vec{p}'';\vec{p}'}\right)\label{eq:app-free-kernel-coord}
\end{equation}
Spelled out:
\begin{equation}
{{\hat{K}}_{\tau}}\left({p'';p'}\right)={\delta^{\left(4\right)}}\left({p''-p'}\right)\exp\left({\imath\frac{{{{E'}^{2}}-{{\left({\vec{p}'}\right)}^{2}}-{m^{2}}}}{{2m}}\tau}\right)\label{eq:app-free-kernel-mom}
\end{equation}
or:
\begin{equation}
{{\hat{K}}_{\tau}}\left({p'';p'}\right)={\delta^{\left(4\right)}}\left({p''-p'}\right)\exp\left({-\imath{f_{p'}}\tau}\right)
\end{equation}

With definition of clock frequency as above:
\begin{equation}
{f_{p}}\equiv-\frac{{{E^{2}}-{{\vec{p}}^{2}}-{m^{2}}}}{{2m}}
\end{equation}

\section{Checkpoint Copenhagen\label{sec:app-time}}

\begin{figure}[h]
\begin{centering}
\includegraphics[width=13cm]{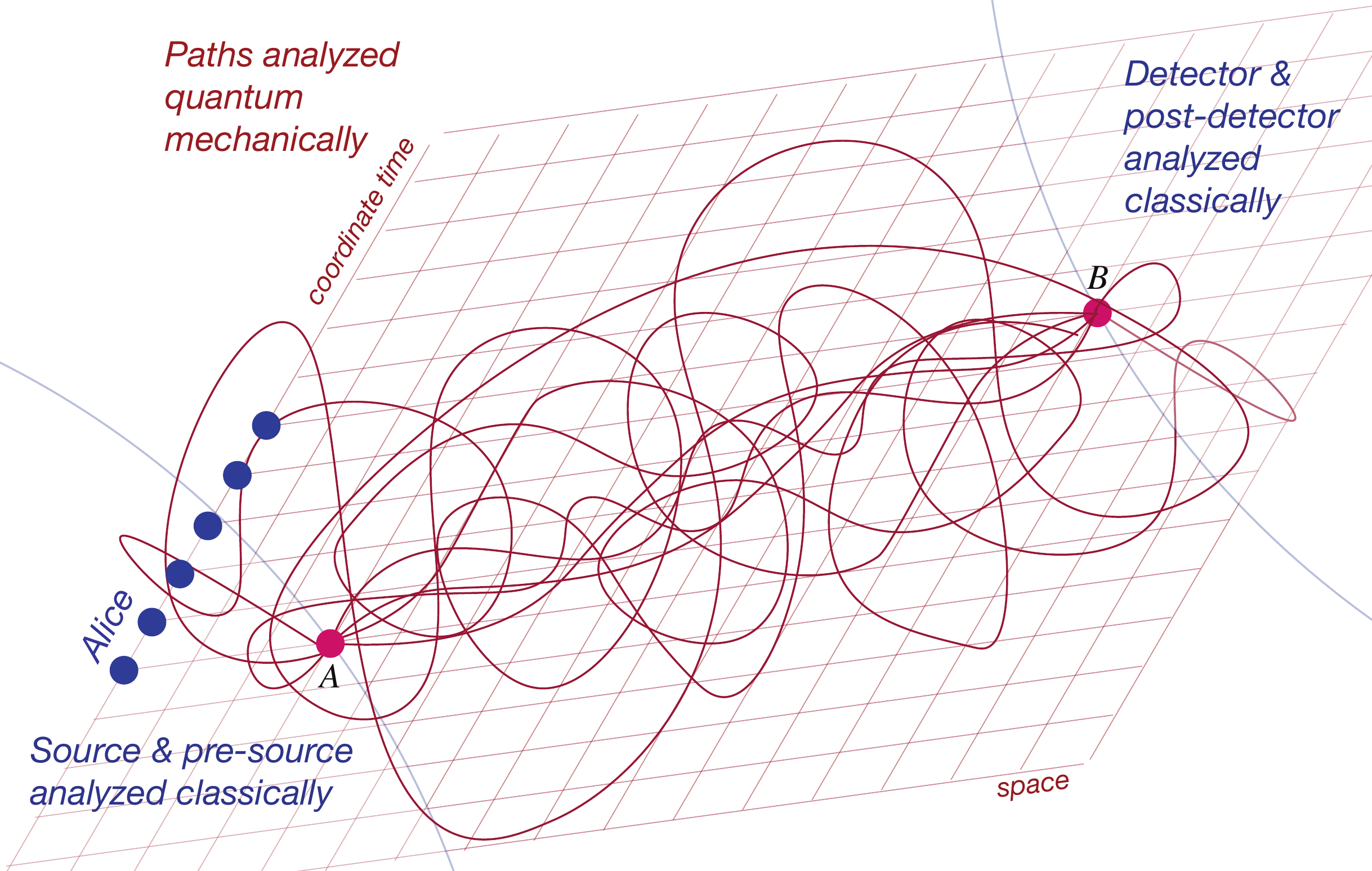}
\par\end{centering}
\caption{Clock time and coordinate time\label{fig:app-time-clock-coordinate}}
\end{figure}

\begin{quotation}
``Atoms are completely impossible from the classical point of view,
since the electrons would spiral into the nucleus.'' -- Richard
P. Feynman \cite{Feynman:1965ah} 
\end{quotation}

\subsection{Four dimensions and an approximation scheme}

Naively we might appear to have a five dimensional coordinate system
here: clock time, coordinate time, and the three space dimensions.
When this work was presented at the 2018 International Association
for Relativistic Dynamics (IARD) conference Dr. Asher Yahalom observed
that this is potentially a bit cumbersome.

Per subsequent discussion with Dr. Yahalom, it is more accurate to
describe it as four dimensions plus an approximation scheme.

Let\textsc{'}s start with a sheet of four dimensional graph paper
representing discretized space time. Alice is off to one side, drawing
paths on it. She is interested in the amplitude for a particle to
get from $A$ to $B$. And to compute this she draws all possible paths
from $A$ to $B$, planning to sum them using the rules in the text.

But Alice herself is a part of the universe she is observing.

Therefore we draw Alice on the left as a series of blue dots representing
her at successive clock ticks. At each clock tick she has a corresponding
three dimensional hyper-surface representing her rest frame. These
are the horizontal lines in the diagram. She is off to one side, because
she is after all observing the particle going from $A$ to $B$, but she
is on the same piece of four dimensional graph paper because she is
part of the same universe.

If she is using SQM to do the calculation then at each clock tick
each of her paths will slide side to side on the corresponding hyper-surface.
But if she is using TQM then the paths will also go forwards and backwards
in time, going off the current hyper-surface, often in quite elaborate
ways.
%\footnote{We are ignoring the complication that at each clock tick,
%the set of paths Alice is looking at is slightly different.  We are assuming implicitly that $\frac{{\imath \partial }\psi}{{\partial \tau }} =0$}

If Bob is also present, we can represent him by a series of green
dots on the right with his own coordinate system. (We have left this
off the illustration because it would make it too busy. Please imagine
Bob\textsc{'}s green dots and hyper-surfaces are present.) He will
be looking at the same set of paths, but slicing them up differently
because he has in general a different set of three dimensional hyper-surfaces.

And if we need to resolve the slight differences between Alice and
Bob\textsc{'}s descriptions -- per subsection
\ref{subsec:lsa-frame} -- we can call in Vera in the $V$ frame
to provide the definitive story.

So far so good, we have only one four dimensional coordinate system
with different observers. Each observer has his or her clock time,
but these clock times are present, tick by tick, on the same piece
of graph paper. And we have well-defined rules for going from one
observer\textsc{'}s frame to another\textsc{'}s.

The problem comes in when we try to reconcile the quantum descriptions
Alice uses for particles with the classical descriptions she uses
for detectors, emitters, or herself. We have an impedance mismatch
between quantum and classical descriptions. This is the
problem of measurement.

\subsection{Quantum descriptions and classical approximations}

The critical observation here is Feynman\textsc{'}s: there are no
classical atoms. Since the emitters, detectors, and observers are
all made of atoms all are quantum objects.

This means that there is no transition from a quantum to a classical
realm. Everything on both sides of the act of detection is a quantum
system.

So what is going on here?

Most of the systems we deal with can treated for all practical purposes
(FAPP) as if they were classical systems. It is only when we look
at certain parts of the system that we need to get down to the quantum
level. When we describe a particle being ejected from an atom we need
to use a quantum description. When we describe a particle in flight
we need to use a quantum description. When we describe its encounter
with a detector we need to use a quantum description. But once the
detector has registered a click, we can use classical approaches to
describe the counting and processing of those clicks.

The rules for detection and emission are ways to navigate the associated
approximations. In a detector, the wave function is not collapsing,
instead we are passing from a quantum to a classical description.
And in emission, the particle is not originally classical, it is just
that up to the starting gun, it can be treated as if it were.

In human terms, picture Alice going thru passport control. Bob, now
a customs official (he gets around), stamps her passport with a visa
stamp. She then heads to her ultimate destination. Both Alice and
Bob are -- per Feynman -- quantum systems. Their previous and subsequent
paths are unknown to official customs. But we have that visa stamp
and associated computer records. They are the measurement. They certify
that at time $T$ position $X$ with uncertainties $\Delta T$ and
$\Delta X$ Alice and Bob encountered each other.

Since Bob\textsc{'}s location is highly localized -- he is in a customs
booth, they are not big -- we treat this as a measurement of Alice\textsc{'}s
position. But the situation is in reality completely symmetrical.
We can think of it as Alice measuring Bob or Bob measuring Alice.
But for customs purposes only the visa stamp matters. The visa stamp
is the measurement.

Ultimately we must always be prepared to go down to the quantum level.
The quantum rules are decisive; the classical rules a mere useful
approximation. Even the visa stamp itself is made of atoms, of quantum
mechanical objects. But in some cases we can get away with a classical
analysis.

The division between quantum and classical is a division of analysis.
It is not part of the physical universe, it is part of how we describe
that universe -- allowing for the fact that we are a part of what
we are describing.

And we get our definition of clock time from the rest of the universe:
\begin{equation}
\tau=\left\langle {\text{rest of universe}}\left|t\right|{\text{rest of universe}}\right\rangle 
\end{equation}

This may be seen as a variation on Fanchi's ``historical time''\cite{Fanchi:1993ab}.
He used an outside system, $S2$, to define a reference clock. 
We extend his $S2$ to be the entire rest of the universe (fortunately there are no budget constraints on thought experiments).

\subsection{Where, when, and to what extent?}

But to take this point from the realm of philosophy to the realm of
science we need to throw some numbers into the mix, we need to answer
the questions: ``when, where, and to what extent does the classical
approximation break down?''

The rules in the Copenhagen interpretation are not specific. ``Somewhere''
between where the particle is being described by quantum mechanics
and where it is detected, the particle goes through a ``checkpoint
Copenhagen'', where its description collapses from the fuzziness
of quantum mechanics to the determinism and specificity of classical
mechanics.

This is both a strength and a weakness of the Copenhagen interpretation.

It is a strength because it let physicists get on with physics. And
because it does not try to ``explain'' what is going on.

But it is a weakness because the terms of the transition are not specified.
Schr\"{o}dinger\textsc{'}s cat experiment is the most striking illustration
of this point \cite{Schrodinger:1935}.

Decoherence provides part of the answer \cite{Omnes:1994ao,Giulini:1996vp,Heiss:2002pd,Joos:2003gf,Schlosshauer:2007rr}.
But decoherence is not yet as quantitative as one might like (although
see Venugopalan, Qureshi, and Mishra \cite{Venugopalan:2018sf}).

It is possible that new physics plays a role in the transition. The
Ghirardi, Rimini, and Weber ($GRW$) \cite{Ghirardi:1986yl} approach
and more generally the continuous spontaneous localization ($CSL$)
approaches (see Dickson \cite{Dickson:1998sw}) hypothesize additional physics.
These alternatives are helpful in parameterizing the transition, but
have not had any experimental confirmation.

Meanwhile, the quantum cats continue to get bigger and bigger, less
and less microscopic \cite{Norte:2015uq,Vinante:2016sf}. At some
point we will be able to see the transition itself, or show
that there is none.

TQM does \textit{not} specifically address this question. We have
taken the existing rules as given. However the temporal fluctuations
in $TQM$ (see especially the entropic estimate of the initial wave
function subsubsection \ref{subsubsec:free-initial-maximum-entropy}) provide a
source of ``internal decoherence'' so would affect estimates of
the size and rate of decoherence. (We owe the phrase ``internal decoherence''
to Dr. Daniel Braun at the 2007 Feynman Festival.)

\section*{References}

\bibliography{taqm} 
\bibliographystyle{iopart-num}

%\begin{thebibliography}{111}
%\end{thebibliography}

\end{document}